\documentclass[10pt]{iopart}

\usepackage{graphicx}
\usepackage{subfigure}
\usepackage{verbatim}
\usepackage{dcolumn}% Align table columns on decimal point
\usepackage{bm}% bold math
\usepackage{epsf}
\usepackage{color}
\usepackage[colorlinks=true,citecolor=blue,linkcolor=blue,urlcolor=blue]{hyperref}
\usepackage{hhline}
\usepackage{amssymb}
\usepackage{iopams}
\usepackage{cite}
\usepackage{tikz}
\usetikzlibrary{arrows,shapes,calc,matrix}

  \expandafter\let\csname equation*\endcsname\relax
  \expandafter\let\csname endequation*\endcsname\relax
\usepackage{amsmath}

\makeatletter
\newcommand\footnoteref[1]{\protected@xdef\@thefnmark{\ref{#1}}\@footnotemark}
\makeatother

\newcommand{\bra}[1]{\left\langle #1\right|}
\newcommand{\ket}[1]{\left|#1\right\rangle}
\newcommand{\braket}[2]{\left\langle #1|#2\right\rangle}

\newcommand{\trace}[1]{\mathrm{tr}\left\{#1\right\}}
\newcommand{\ptr}[2]{\mathrm{tr_{#1}}\left\{#2\right\}}

\newcommand{\la}{\left\langle}
\newcommand{\ra}{\right\rangle}
\newcommand{\pd}{\partial}

\newcommand{\td}{\mathrm{d}}
\newcommand{\ma}[1]{\max{\left\{#1\right\}}}
\newcommand{\mi}[1]{\min{\left\{#1\right\}}}
\newcommand{\etals}{\textit{et al. }}
\newcommand{\ex}[1]{\exp{\left(#1\right)}}
\newcommand{\loge}[1]{\ln{\left(#1\right)}}
\newcommand{\id}{\mathbb{I}}
\newcommand{\com}[2]{\left[#1,\,#2\right]}
\newcommand{\acom}[2]{\left\{#1,\,#2\right\}}
\newcommand{\co}[1]{\cos{\left(#1\right)}}
\newcommand{\si}[1]{\sin{\left(#1\right)}}

\newcommand{\bla}{bla\\bla\\bla\\bla\\bla}

\newcommand{\mb}[1]{\mbox{\boldmath$#1$}}
\newcommand{\mc}[1]{\mathcal{#1}}
\newcommand{\mbb}[1]{\mathbb{#1}}
\newcommand{\mf}[1]{\mathfrak{#1}}
\newcommand{\mrm}[1]{\mathrm{#1}}

\newcommand{\revised}[1]{{\color{black}#1}}

\newcommand{\draftmode}{1}    %to control draft colors below
\newcommand{\notetoself}[1]{\ifnum \draftmode=1 {\color[rgb]{0,0,0.8} [#1]} \fi}  %notes to self in blue when \draftmode==1.  invisible otherwise
\newcommand{\cuttext}[1]{\ifnum \draftmode=1 {\color[rgb]{0,0.5,0} [#1]} \fi}  %cut out text in green when \draftmode==1.  invisible otherwise
\newcommand{\warntext}[1]{\ifnum \draftmode=1 {\color[rgb]{0.9,0.6,0} #1} \else {#1} \color{black} \fi}
\begin{document} 

\topical[Quantum speed limits]{Quantum speed limits: from Heisenberg's uncertainty principle to optimal quantum control} 

\author{Sebastian Deffner$^1$ and Steve Campbell$^{2}$}
\address{$^1$Department of Physics, University of Maryland Baltimore County, Baltimore, MD 21250, USA}
\address{$^2$Istituto Nazionale di Fisica Nucleare, Sezione di Milano \& Dipartimento di Fisica, Universit{\`a} degli Studi di Milano, via Celoria 16, 20133 Milan, Italy}
\ead{$^1$\mailto{deffner@umbc.edu} and $^2$\mailto{steve.campbell@mi.infn.it}}
\date{\today}

\begin{abstract}
One of the most widely known building blocks of modern physics is Heisenberg's indeterminacy principle. Among the different statements of this fundamental property of the full quantum mechanical nature of physical reality, the uncertainty relation for energy and time has a special place. Its interpretation and its consequences have inspired continued research efforts for almost a century. In its modern formulation, the  uncertainty relation is understood as setting a fundamental bound on how fast any quantum system can evolve. In this Topical Review we describe important milestones, such as the Mandelstam-Tamm and the Margolus-Levitin bounds on the \emph{quantum speed limit}, and summarise recent applications in a variety of current research fields -- including quantum information theory, quantum computing, and quantum thermodynamics amongst several others. To bring order and to provide an access point into the many different notions and concepts, we have grouped the various approaches into the \emph{minimal time approach} and the \emph{geometric approach}, where the former relies on quantum control theory, and the latter arises from measuring the distinguishability of quantum states. Due to the volume of the  literature, this Topical Review can only present a snapshot of the current state-of-the-art and can never be fully comprehensive. Therefore, we highlight but a few works hoping that our selection can serve as a representative starting point for the interested reader.\\ \\
\textbf{Keywords:} Quantum speed limits, Heisenberg uncertainly principle, optimal control theory, shortcuts to adiabaticity, quantum information theory, quantum thermodynamics.
\end{abstract}
%\pacs{...}
%\submitto{\jpa}

%\tableofcontents

\newpage
%%%%%%%%%%%%%%%%%%%%%%%%%%%%%%%%%%%%%%%%%%%%%%%%%%%%%%%%%%%%%%%%%%%%%%%%%%%%%%%%%
%\input{chapters/intro.include.tex}

\section{Introduction\label{intro}}

It is a historic fact that Einstein never seemed quite comfortable with the probabilistic interpretation of quantum theory. In a letter to Born he once remarked \cite{Born1971}:
\begin{quote}
\emph{Quantum mechanics is certainly imposing. But an inner voice tells me that it is not yet the real thing. The theory says a lot, but does not really bring us any closer to the secret of the old one. I, at any rate, am convinced that He does not throw dice\footnote{English translation of the German original: Die Quantenmechanik ist sehr achtung-gebietend. Aber eine innere Stimme sagt mir, da\ss\,  das doch nicht der wahre Jakob ist. Die Theorie liefert viel, aber dem Geheimnis des Alten bringt sie uns kaum n\"aher. Jedenfalls bin ich überzeugt, da\ss\,  der nicht w\"urfelt.}.}
\end{quote}
Nevertheless, quantum mechanics is built on the very notion of indeterminacy, which is rooted in Heisenberg's uncertainty principles, and which can be expressed in terms of the famous inequalities \cite{Heisenberg1927,Heisenberg2008},
\begin{equation}
\label{intro:eq01}
\Delta p\,\Delta x\gtrsim \hbar\quad \mrm{and}\quad \Delta E\,\Delta t\gtrsim \hbar\,.
\end{equation}
Although physically insightful, these relations, Eq.~\eqref{intro:eq01}, were originally motivated only by plausibility arguments and by ``observing'' the commutation relations of canonical variables in first quantization\footnote{Without any mathematical justification Heisenberg explicitly writes \cite{Heisenberg1927}, $E t-t E=i\hbar$.}.

While the uncertainty relation for position and momentum was quickly put on solid grounds by Bohr \cite{Bohr1928} and Robertson \cite{Robertson1929}, the proper formulation and interpretation for the uncertainty relation of time and energy proved to be a significantly harder task. In its modern interpretation the uncertainty relation for position and momentum expresses the fact that the position and the momentum of a quantum particle cannot be measured simultaneously with infinite precision \cite{Dirac1958}.  However, if the uncertainty principle is a statement about \emph{simultaneous} events, the interpretation of an uncertainty in \emph{time} is far from obvious \cite{Hilgevoord1996}.

\begin{figure}[t]
\begin{centering}
\includegraphics[width=0.95\textwidth]{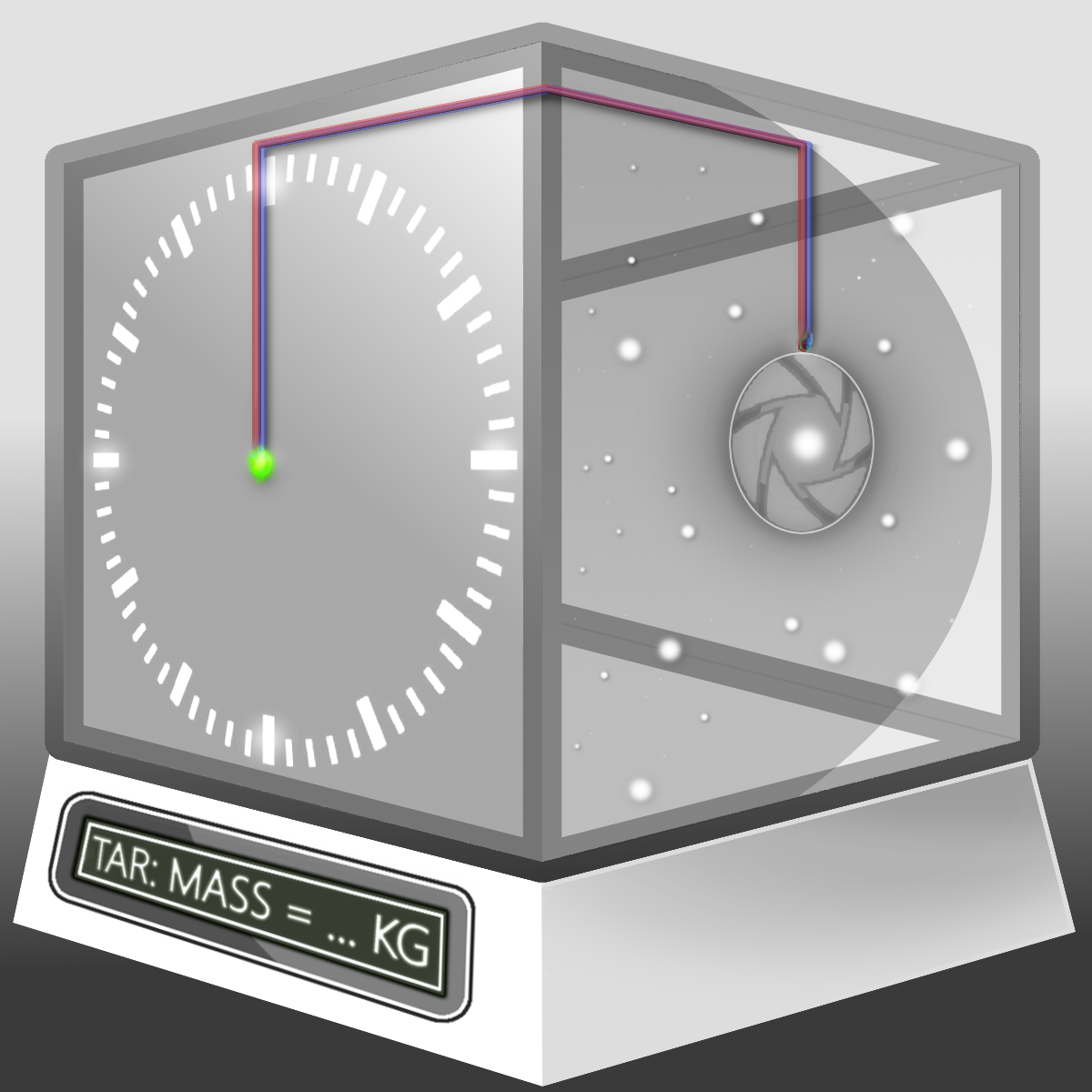}
\caption{Sketch of the gedankenexperiment envisaged by Einstein.}
\label{fig:intro}
\end{centering}
\end{figure}

Thus, only three years after its inception Einstein challenged the validity of the energy-time uncertainty relation with the following gedankenexperiment as depicted in Fig.~\ref{fig:intro} \cite{Hilgevoord1998}: Imagine a box containing photons, which has a hole in one of its walls. This hole can be opened and closed by a shutter controlled by a clock inside the box. At a preset time the shutter opens the hole for a short period and lets photons escape. Since the clock can be classical, the duration can be determined with infinite precision. From special relativity we know that energy and mass are equivalent, $E=mc^2$. Hence, by measuring the mass of the box in the gravitational field, the change in energy due to the loss of photons can also be determined with infinite precision. As a consequence, special relativity seems to negate the existence of an uncertainty principle for energy and time! 

Bohr's counterargument in essence states that in order to measure time, position and momentum of the hands of the clock have to be determined. In addition, accounting for time dilation due to motion in the gravitational field, the uncertainty relation for position and momentum implies an uncertainty relation for energy and time. Although insightful, Bohr's interpretation cannot be considered entirely satisfactory \cite{Hilgevoord1996,Hilgevoord1998,Hilgevoord2002a}, since it merely circumvents the problem of explaining the existence of the uncertainty principle in non-relativistic quantum mechanics.

A major breakthrough was achieved by Mandelstam and Tamm \cite{mandelstam45}, who realised that $\Delta E\,\Delta t\gtrsim \hbar$ is not a statement about simultaneous measurements, but rather about the \emph{intrinsic time scale} of unitary quantum dynamics. Hence, one should rather write $\Delta t \gtrsim \hbar/\Delta E$, where $\Delta t$ is interpreted as the time a quantum system needs to evolve from an initial to a final state. More specifically Mandelstam and Tamm derived the first expression of the quantum speed limit time $\tau_\mrm{QSL}=\pi\hbar/2\Delta H$, where $\Delta H^2$ is the variance of the Hamiltonian, $H$, of the quantum system \cite{mandelstam45}. As an application of their bound, they also argued that $\tau_\mrm{QSL}$ naturally quantifies the life time of quantum states \cite{mandelstam45}, which has found widespread prominence in the literature \cite{Gislason1985,Uffink1985,Hilgevoord1991,Luo2005,Boykin2007,Li2013}.

Nevertheless, the desire to formalise time as a proper quantum observable persisted \cite{Eberly1973,Kobe1994}. To make matters worse, it was further argued that the variance of an operator is not an adequate measure of quantum uncertainty \cite{Uffink1993,Hilgevoord2002}, which only highlighted that the uncertainty relation for time and energy needed to be put on even firmer grounds.

With the advent of quantum computing \cite{Feynman1982,Ladd2010} Mandelstam and Tamm's interpretation of the quantum speed limit as intrinsic time-scale experienced renewed prominence. Their interpretation was further solidified by Margolus and Levitin \cite{margolus98}, who derived an alternative expression for $\tau_\mrm{QSL}$ in terms of the (positive) expectation value of the Hamiltonian, $\tau_\mrm{QSL}=\pi\hbar/2 \la H\ra$. Eventually, it was also shown that the combined bound, 
\begin{equation}
\label{intro:eq02}
\tau_\mrm{QSL}=\ma{\frac{\pi}{2}\,\frac{\hbar}{\Delta H},\frac{\pi}{2}\,\frac{\hbar}{\la H\ra}}
\end{equation}
is tight \cite{Levitin2009}. This means Eq.~\eqref{intro:eq02} sets the fastest attainable time-scale over which a quantum system can evolve. In particular, Eq.~\eqref{intro:eq02} sets the maximal rate with which quantum information can be communicated \cite{Bekenstein1981}, the maximal rate with which quantum information can be processed \cite{lloyd00}, the maximal rate of quantum entropy production \cite{Deffner2010}, the shortest time-scale for quantum optimal control algorithms to converge \cite{Caneva2009}, \revised{the best precision in quantum metrology \cite{Giovannetti2011}, and determines the spectral form factor \cite{delCampo2017}}.

The next major milestone in the development of the field was achieved only relatively recently with the generalisation of the quantum speed limit to open systems. In 2013 three letters proposed, in quick succession, three independent approaches of how to quantify the maximal quantum speed of systems interacting with their environments. Taddei \etals \cite{taddei13} found an expression in terms of the quantum Fisher information, del Campo \etals \cite{delcampo13} bounded the rate of change of the relative purity, and Deffner and Lutz \cite{Deffner2013PRL} derived geometric generalizations of both, the Mandelstamm-Tamm bound as well as the Margolus-Levitin bound. These three contributions \cite{taddei13,delcampo13,Deffner2013PRL} effectively opened a new field of modern research, since for the first time it became obvious that the Heisenberg uncertainty principle for time and energy is not only of fundamental importance, but actually of quite practical relevance. For instance, it became clear that quantum processes in systems interacting with non-Markovian environments can evolve faster than in systems coupled to memory-less, Markovian baths -- which has been verified in a cavity QED experiment \cite{Cimmarusti2015}.

The purpose of this Topical Review is to take a step back and bring order into the plethora of novel ideas, concepts, and applications. Thus, in contrast to earlier reviews on quantum speed limits \cite{Busch1990,Busch1990a,Pfeifer1995,Muga2007,Muga2009,Sen2014,Dodonov2015,Frey2016}, we will focus less on mathematical and technical details, but rather emphasise the interplay between the various concepts, tradeoffs between speed and physical resources, and consequences in real-life applications. We will begin with a historical overview in Section~\ref{history}, where we will also summarise the original derivations by Mandelstam-Tamm and Margolus-Levitin. Section~\ref{timeindependent} is dedicated to quantum systems with time-independent generators, whereas Section~\ref{minimaltime} focuses on optimal control theory. In Section~\ref{geometric} we will discuss the geometric approach, which so far has been the most fruitful approach in the description of open quantum systems, and which has led to the most interesting insights. The review is rounded off with Section~\ref{nonstandard}, in which we briefly summarise generalisations to relativistic and non-linear quantum dynamics, and Section~\ref{future} which outlines the relation of quantum speed limits to other fundamental bounds. 

When writing this Topical Review, we strove for objectivity and completeness. However, the sheer volume of publications demanded to select but a few works to be discussed in detail. Our selection was motivated by accessibility, pedagogical value, and conceptual milestones. Therefore, this review can never be a fully complete discussion of the literature, but rather only serve as a starting point for further study and research on quantum speed limits.

%%%%%%%%%%%%%%%%%%%%%%%%%%%%%%%%%%%%%%%%%%%%%%%%%%%%%%%%%%%%%%%%%%%%%%%%%%%%%%%%%
%\input{chapters/hist.include.tex}

\section{Energy-time uncertainty: Emergence of the quantum speed limit}
\label{history}

\subsection{Heisenberg's uncertainty principle}

For classical observers, one of the most remarkable properties of quantum systems is the inherent indeterminism of physical states that have not been measured. This indeterminism originates from trying to resolve the apparent wave-particle duality, which necessitates a probabilistic theory to describe the behaviour of small objects \cite{Heisenberg1927,Heisenberg2008}. The most prominent hallmark of this insight is the indeterminacy principle, which is typically expressed as the uncertainly relation,
\begin{equation}
\label{hist:eq01}
\Delta x\,\Delta p \gtrsim \hbar\,.
\end{equation}
This relation reflects that the position and  momentum of a quantum particle cannot be measured \emph{simultaneously} with infinite precision\footnote{\revised{For the sake of accessibility we have chosen to work with a standard textbook interpretation of the uncertainty relations \cite{Messiah1966}. For a conceptually more correct interpretation we refer to the literature \cite{Hilgevoord1990}.}}. 

In introductory texts Eq.~\eqref{hist:eq01} is often motivated by analysing the Fourier modes of wave packets \cite{Messiah1966}, which then also allows to derive an additional uncertainty relation for time, $t$, and energy, $E$. To this end, one identifies \cite{Messiah1966}
\begin{equation}
\label{hist:eq02}
\Delta t \simeq \frac{\Delta x}{v}\quad \mrm{and}\quad \Delta E\simeq \frac{\pd E}{\pd p}\,\Delta p=v\, \Delta p\,,
\end{equation}
where $v$ denotes the group velocity of a wave packet. Therefore, one concludes
\begin{equation}
\label{hist:eq03}
\Delta t\,\Delta E\gtrsim \hbar\,,
\end{equation}
which now expresses that also time and energy cannot be known simultaneously with infinite precision. 

That something of this argumentation is a bit fishy  \cite{Hilgevoord1996,Hilgevoord1998} becomes clear once one realises that the uncertainty principle for position and momentum is actually a consequence of the canonical commutation relation,
\begin{equation}
\label{hist:eq04}
\com{x}{p}=i\hbar\,.
\end{equation}
It was shown by Roberston \cite{Robertson1929} by simply invoking the Cauchy-Schwarz inequality \cite{Messiah1966} that for any two operators $A$ and $B$ we have
\begin{equation}
\label{hist:eq05}
\Delta A\,\Delta B\geq \frac{1}{2}\,\left|\la\com{A}{B} \ra\right|
\end{equation}
where $\Delta O\equiv\sqrt{\la O^2\ra-\la O\ra^2}$ with $O=A,B$ and $\la O\ra=\bra{\psi}O\ket{\psi}$. \revised{Strictly speaking \eqref{hist:eq05} does not correspond to simultaneous measurements since the quantum back action of measurements is not considered \cite{Hilgevoord1990}, and see also Sec.~\ref{sec:learn}. Rather, Eq.~\eqref{hist:eq05} can be understood as a special case of the Cramer-Rao bound \cite{Kholevo1974,Frowis2015}, and thus the uncertainty relation rather must be interpreted as a statement about the preparation of states, see also Secs.~\ref{quantummetrology} and \ref{sec:CR}. Imagine that an observable $A$ is measured on the first half of an ensemble, and $B$ on the second half, then Eq.~\eqref{hist:eq05} sets a limit on the product of standard deviations.

Independent of the interpretation of Eq.~\eqref{hist:eq05} time can, generally, not be expressed as a Hermitian operator \cite{Hilgevoord1996,Hilgevoord1998,Hilgevoord2002a,Erker2016}, and hence Eq.~\eqref{hist:eq05} cannot be reduced to the Heisenberg energy-time uncertainty relation \eqref{hist:eq03}. To avoid any misconceptions we emphasize that other time-like variables such as arrival or tunneling times \cite{Muga2007,Muga2009} can  very well be expressed as operators.

In this Topical Review, ``time'' will always be understood as the quantity with which one commonly associates an uncertainty relation of the form \eqref{hist:eq03}.} From its first appearance \revised{of such a notion of time} in Heisenberg's paper in 1927 \cite{Heisenberg1927} it took almost twenty years before a mathematically sound and physically insightful treatment was proposed by Mandelstam and Tamm \cite{mandelstam45}.

\subsection{The uncertainty relation of Mandelstam and Tamm}

Mandelstam and Tamm's analysis \cite{mandelstam45} rests on the fact that for quantum systems evolving under Schr\"odinger dynamics the evolution of any observable $A$ is given by the Liouville-von-Neumann equation
\begin{equation}
\label{hist:eq06}
\frac{\pd A}{\pd t}=\frac{i}{\hbar}\,\com{H}{A}\,,
\end{equation}
where $H$ is the Hamiltonian of the system. Therefore, the general uncertainty relation in Eq.~\eqref{hist:eq05} implies for $B=H$
\begin{equation}
\label{hist:eq07}
\Delta H\,\Delta A\geq \frac{\hbar}{2}\,\left|\la\frac{\pd A}{\pd t}\ra\right|\,.
\end{equation}
The latter inequality can be further simplified, if we choose the observable $A$ as the projector onto the initial state $\ket{\psi(0)}$, and $A=\ket{\psi(0)}\bra{\psi(0)}$. Thus, we also have $\la A\ra_0=1$ and 
\begin{equation}
\label{hist:eq07a}
\Delta A=\sqrt{\la A^2\ra-\la A\ra^2}=\sqrt{\la A\ra-\la A\ra^2}\,,
\end{equation}
which allows to integrate Eq.~\eqref{hist:eq07} and we obtain,
\begin{equation}
\label{hist:eq08}
\frac{1}{\hbar}\,\Delta H\,t\geq\frac{\pi}{2}-\arcsin\sqrt{\la A\ra_t}\,.
\end{equation}
If we now consider only processes in which the final state is orthogonal to the initial state, i.e. $\braket{\psi(0)}{\psi(\tau)}=0$, then the minimal time for a quantum system to evolve between two orthogonal states is determined by
\begin{equation}
\label{hist:eq09}
\tau\geq\tau_\mrm{QSL}\equiv\frac{\pi}{2}\,\frac{\hbar}{\Delta H}\,.
\end{equation}
As a main breakthrough, Mandelstam and Tamm not only put Heisenberg's uncertainty principle for time and energy on solid physical grounds, but also proposed the first notion of a quantum speed limit time, $\tau_\mrm{QSL}$ \eqref{hist:eq09}. It is interesting to note that the proper interpretation of the energy-time uncertainty principle as a bound on the minimal time of quantum evolution was formalised by Aharonov and Bohm \cite{Aharonov1961}. They pointed out that Eq.~\eqref{hist:eq09} must not be interpreted as an uncertainty relation between the duration of a measurement and the energy transferred to the observed system. Rather, the quantum speed limit time, $\tau_\mrm{QSL}$, sets an \emph{intrinsic time scale} of any quantum evolution  \cite{Aharonov1961,Briggs2008}.

Over the next four decades $\tau_\mrm{QSL}$ \eqref{hist:eq09} was frequently studied and re-derived by Fleming \cite{Fleming1973},  Bhattacharyya \cite{Bhattacharyya1983}, Anandan and Aharonov \cite{Anandan1990}, and Vaidman \cite{Vaidman1992}. However, it was Uffink who realised \cite{Uffink1993} that in many situations $\Delta H$ gives a very unreasonable measure for the speed of a quantum evolution. The lower bound in Eq.~\eqref{hist:eq09} can be arbitrarily small, since the variance of the Hamiltonian, $\Delta H$, can diverge even if the average energy is finite \cite{margolus98}.

\subsection{The quantum speed limit of Margolus and Levitin}

To tackle this problem Margolus and Levitin proposed an alternative derivation of the quantum speed limit \cite{margolus98}. Their analysis starts with expanding the initial state, $\ket{\psi_0}$, in the energy eigenbasis
\begin{equation}
\label{hist:eq10}
\ket{\psi_0}=\sum_n c_n \ket{E_n}\,.
\end{equation}
Correspondingly, the solution to the time-dependent Schr\"odinger equation with constant Hamiltonian can be written as,
\begin{equation}
\label{hist:eq11}
\ket{\psi_t}=\sum_n c_n \ex{-i\,E_n t/\hbar}\,\ket{E_n}
\end{equation}
and we obtain for the time-dependent overlap with the initial state $S(t)\equiv \braket{\psi_0}{\psi_t}=\sum_n \left|c_n\right|^2 \ex{-i\,E_n t/\hbar}$.  The quantum speed limit is then obtained by estimating the real part of $S(t)$
\begin{equation}
\label{hist:eq12}
\begin{split}
\mf{R}(S)&=\sum_n \left|c_n\right|^2 \co{E_n t/\hbar}\\
&\geq \sum_n \left|c_n\right|^2 \left[1-\frac{2}{\pi}\left(\frac{E_n t}{\hbar}+\si{\frac{E_n t}{\hbar}}\right)\right]\\
&=1-\frac{2}{\pi}\frac{\la H\ra}{\hbar}\,t+\frac{2}{\pi}\,\mf{I}(S)\,,
\end{split}
\end{equation}
where we have used the trigonometric inequality, $\co{x}\geq1-2/\pi\,(x+\si{x})$, which is true $\forall x\geq 0$. Note that we here explicitly assume that the average energy, $\la H\ra$, is non-negative. Now, further noting that for orthogonal initial and final states we have $S(\tau)=0$, which also implies $\mf{R}(S)=0$ and $\mf{I}(S)=0$, and we obtain the minimal evolution time between two orthogonal states,
\begin{equation}
\label{hist:eq13}
\tau\geq\tau_\mrm{QSL}=\frac{\pi}{2}\,\frac{\hbar}{\la H\ra}\,.
\end{equation}
The Margolus-Levitin bound, Eq.~\eqref{hist:eq13}, does not suffer from the conceptual issues that plagued the Mandelstam-Tamm bound -- namely that the dynamical speed is determined by the variance of some quantum observable \cite{Uffink1993}. However, the discovery  of Eq.~\eqref{hist:eq13} also created the paradoxical situation that there seem to exist two apparently independent bounds based on two different physical properties of the same quantum state \cite{Levitin2009}. That both bounds, \eqref{hist:eq09} \emph{and} \eqref{hist:eq13}, hold true, however, was illustrated by further elementary proofs by, e.g., Uffink \cite{Uffink1993}, Brody \cite{Brody2003}, Andrecut and Ali \cite{Andrecut2004}, and Kosi\'{n}ski and Zych \cite{Zych2006}, and by more eleborated proofs for mixed and entangled states \cite{Giovannetti2003a,Svozil2005,Zander2007}.

\subsection{The unified bound is tight}

As a consequence it was simply assumed without much justification that the minimal time a quantum system needs to evolve between two orthogonal states is given by
\begin{equation}
\label{hist:eq14}
\tau_\mrm{QSL}=\ma{\frac{\pi}{2}\,\frac{\hbar}{\Delta H},\,\frac{\pi}{2}\,\frac{\hbar}{\la H\ra}}\,.
\end{equation}
However, it was Levitin and Toffoli \cite{Levitin2009}, who finally realised that the situation is not quite that simple. To this end, they proved the following theorem: 
\begin{quote}
Under the assumption that the ground state energy of a quantum systems is zero, the only state for which the Mandelstam-Tamm bound Eq.~\eqref{hist:eq09} as well as the Margolus-Levitin bound Eq.~\eqref{hist:eq10} are attained is given by
\begin{equation}
\label{hist:eq15}
\ket{\psi}=\frac{1}{\sqrt{2}}\,\left(\ket{E_0}+\ket{E_1}\right)\,,
\end{equation}
where $H\ket{E_k}=k E_1\,\ket{E_k}$ for all $k=0,1$, and $E_1$ is the energy of the first excited state. This state $\ket{\psi}$ Eq. \eqref{hist:eq15} is unique up to degeneracy of the first excited state and arbitrary phase factors.
\end{quote}

\paragraph{Margolus-Levitin bound.} That $\ket{\psi}$ is the only state to attain the Margolus-Levitin bound is easy to see from Eq.~\eqref{hist:eq12}. The trigonometric inequality $\co{x}\geq1-2/\pi\,(x+\si{x})$ becomes an equality for $x=0$ and $x=\pi$. Thus, in Eq.~\eqref{hist:eq12} we require that $E_nt/\hbar=0$ or $E_nt/\hbar=\pi$, which is possible if and only if the initial state is given by Eq.~\eqref{hist:eq15}, $\ket{\psi_0}=\ket{\psi}$.

\paragraph{Mandelstam-Tamm bound.} For the Mandelstam-Tamm bound we now consider the trigonometric inequality, $\co{x}\geq 1-4/\pi^2\,x\si{x}-2/\pi^2\,x^2$, which is again true $\forall x$. In complete analogy to above we again expand the initial state, $\ket{\psi_0}$, and the time-evolved state, $\ket{\psi_t}$, in the energy eigenbasis, Eqs.~\eqref{hist:eq10} and \eqref{hist:eq11}, respectively. Accordingly, we can write for the time-dependent overlap, $S(t)=\braket{\psi_0}{\psi_t}$,
\begin{equation}
\label{hist:eq16}
\begin{split}
\left|S(t)\right|^2&=\sum_{n,n'} \left|c_n\right|^2\,\left|c_{n'}\right|^2 \ex{-i\left(E_n-E_{n'}\right)\,t/\hbar}\\
&=\sum_{n,n'} \left|c_n\right|^2\,\left|c_{n'}\right|^2 \co{\left(E_n-E_{n'}\right)\,t/\hbar}\,,
\end{split}
\end{equation}
where we have used $\sum_n \left|c_n\right|^2=1$. The latter can be bounded from below
\begin{equation}
\label{hist:eq17}
\begin{split}
\left|S(t)\right|^2&\geq 1-\frac{4}{\pi^2}\sum_{n,n'}\left|c_{n'}\right|^2 \si{\left(E_n-E_{n'}\right)\,t/\hbar}\,\left(E_n-E_{n'}\right)\,t/\hbar \\
&-\frac{2}{\pi^2}\sum_{n,n'}\left|c_{n'}\right|^2 \left[\left(E_n-E_{n'}\right)\,t/\hbar\right]^2\,,
\end{split}
\end{equation}
which can be simplified to read
\begin{equation}
\label{hist:eq18}
\left|S(t)\right|^2\geq 1+\frac{4t}{\pi^2}\frac{d}{dt}\left|S(t)\right|^2-\frac{1}{\pi^2}\left(\frac{2 t}{\hbar}\,\Delta H\right)^2\,.
\end{equation}
Now further noting that $\left|S(t)\right|^2\geq 0$ we can choose a time $\tau$ such that we have
\begin{equation}
\label{hist:eq19}
0\geq 1-\frac{1}{\pi^2}\left(\frac{2 \tau}{\hbar}\,\Delta H\right)^2\,,
\end{equation}
which is nothing else but the Mandelstam-Tamm bound \eqref{hist:eq09}. This alternative proof, however, immediately allows us to determine when the inequality in Eq.~\eqref{hist:eq19} becomes an equality. One easily convinces oneself that this is the case if and only if $\ket{\psi_0}$ is given by Eq.~\eqref{hist:eq15}~ \cite{Levitin2009}. 

\paragraph{}As main results Levitin and Toffoli \cite{Levitin2009} showed that the unified bound, Eq.~\eqref{hist:eq14}, is tight and that it is attained only by states of the form Eq.~\eqref{hist:eq15}. From a few more rather technical considerations they further proved that no mixed state can have a larger speed, which means that Eq.~\eqref{hist:eq14} sets the ultimate speed limit for the evolution between orthogonal states and for time-independent Hamiltonians.

\subsection{Generalisations to arbitrary angles and driven \revised{dynamics}}

The natural question arises whether the expression for the quantum speed limit, Eq.~\eqref{hist:eq16}, can be further sharpened for arbitrary angles and for driven dynamics. \revised{Here and in the following ``driven'' refers to dynamics under parametrically varying Hamiltonians.}

Whereas for pure states addressing this question is rather straight forward \cite{Pfeifer1993}, for general mixed quantum states the situation is technically significantly more challenging. The difficulty arises from the fact that for pure states the angle is simply given by \cite{Wootters1981}
\begin{equation}
\label{hist:eq20}
\mc{L}\left(\ket{\psi_0},\ket{\psi_\tau}\right)=\arccos{\left(\left|\braket{\psi_0}{\psi_\tau}\right|\right)}\,,
\end{equation}
which \emph{ad hoc} has no unique generalisation to general, mixed quantum states. It was Uhlmann \cite{Uhlmann1976} who realised that the proper generalisation of the overlap of wavefunctions is given by the quantum fidelity,
\begin{equation}
\label{hist:eq21}
F\left(\rho_0,\rho_\tau\right)=\left[\trace{\sqrt{\sqrt{\rho_0}\rho_\tau\sqrt{\rho_0}}}\right]^2
\end{equation}
and Josza showed that the latter definition is the unique choice \cite{Jozsa1994}. Accordingly, the generalised angle between arbitrary quantum states is given by the Bures angle \cite{Kakutani1948,Bures1969}
\begin{equation}
\label{hist:eq22}
\mc{L}\left(\rho_0,\rho_\tau\right)=\arccos{\left(\sqrt{F\left(\rho_0,\rho_\tau\right)}\right)}\,.
\end{equation}
Equipped with the latter, Uhlmann was able to generalise the Mandelstam-Tamm bound to mixed states and driven dynamics \cite{Uhlmann1992}. \revised{Interestingly, without using its name Uhlmann already worked with the infinitesimal quantum Fisher information, on which we will expand shortly in Sec.~\ref{quantummetrology}.} However, Uhlmann's original treatment \cite{Uhlmann1992} is rather formal using parallel Hilbert-Schmidt operators, which is why we \revised{now} summarise the re-derivation by Deffner and Lutz \cite{Deffner2013}.

\paragraph{Mandelstam-Tamm bound for driven dynamics.}

Braunstein and Caves showed that the Bures angle for two infinitesimally close density operators, $\rho'=\rho+d\rho$ can be written as \cite{Braunstein1994},
\begin{equation}
\label{hist:eq23}
d\mathcal{L}^2=\trace{d\rho\, \mathcal{R}_\rho^{-1}(d\rho)}\ ,
\end{equation}
where the superoperator $\mathcal{R}^{-1}(O)$ reads in terms of the eigenvalues $p_i$ of $\rho$, $\rho=\sum_i\,p_i\ket{i}\bra{i}$,
\begin{equation}
\label{hist:eq24}
\mathcal{R}_\rho^{-1}(O)=\frac{1}{2}\sum_{j,k}\,\frac{\bra{j}O\ket{k}}{p_j+p_k}\ket{j}\bra{k} \ .
\end{equation}
Note that the superoperator $\mathcal{R}_\rho^{-1}$ is here defined as describing the infinitesimal Bures angle $\mathcal{L}$, and hence differs by a factor 4 from the one used in Ref.~\cite{Braunstein1994}, where $\mathcal{R}_\rho^{-1}$ is determined by the infinitesimal statistical distance.

Now we rewrite the von Neumann equation for the density operator of the system in the form,
\begin{equation}
\label{hist:eq25}
i\hbar\,\frac{d\rho_t}{dt} =\com{H_t}{\rho_t}=\com{H_t-\la H_t\ra}{\rho_t}=\com{\delta H_t}{\rho_t}\,,
\end{equation}
since the expectation value of the energy  $\la H_t\ra$ is a real number that can be included in the commutator. Combining Eqs.~\eqref{hist:eq23}-\eqref{hist:eq25}, we find,
\begin{equation}
\label{hist:eq26}
\begin{split}
\left(\frac{d\mathcal{L}}{dt}\right)^2&=\trace{\frac{d\rho_t}{dt}\, \mathcal{R}^{-1}_{\rho_t}\left(\frac{d\rho_t}{dt}\right)}=\frac{1}{2\hbar^2}\,\sum_{j,k}\,\frac{\left(p_j-p_k\right)^2}{p_j+p_k}\,\left|\bra{j}\delta H_t\ket{k}\right|^2 \\
&\leq\frac{1}{2\hbar^2}\,\sum_{j,k}\,\left(p_j+p_k \right)\,\left|\bra{j}\delta H_t\ket{k}\right|^2=\frac{\Delta H_t^2}{\hbar^2}\,,
\end{split}
\end{equation}
where the last line follows from a triangle-type inequality \cite{Braunstein1994}. The generalised energy-time uncertainty relation is now obtained by first taking the  positive root of   Eq.~\eqref{hist:eq26},
 \begin{equation}
\label{hist:eq27}
\frac{d\mathcal{L}}{dt}  \leq\left| \frac{d\mathcal{L}}{dt}\right|\leq\frac{\Delta H_t}{\hbar} \ ,
\end{equation}
 and then  performing the integral over both the Bures length and time,
\begin{equation}
\label{hist:eq28}
\int_0^{\mathcal{L}\left(\rho_0,\rho_\tau\right)} d\mathcal{L}\leq\frac{1}{\hbar}\,\int_0^\tau dt\,\Delta H_t\,.
\end{equation}
As a result, we obtain the inequality,
\begin{equation}
\label{hist:eq29}
\tau\geq\frac{\hbar}{\Delta E_\tau}\,\mathcal{L}\left(\rho_0,\rho_\tau\right)\,,
\end{equation}
where we have introduced the time averaged variance of the Hamiltonian,
\begin{equation}
\label{hist:eq30}
\Delta E_\tau\equiv\frac{1}{\tau}\,\int_0^\tau dt\,\Delta H_t=\frac{1}{\tau}\,\int_0^\tau dt\,\sqrt{\la H_t^2\ra -\la H_t \ra^2}\,.
\end{equation}
Equation  \eqref{hist:eq29} is the Mandelstam-Tamm uncertainty relation for arbitrary, initial and final mixed quantum states and arbitrary, driven Hamiltonians. Similar results were also found by Braunstein and Milburn \cite{Braunstein1995} and summarised by Braunstein \etals in Ref.~\cite{Braunstein1996}.

\paragraph{Margolus-Levitin bound from arbitrary angles.}

Similar to its original discovery, the generalisation of the Margolus-Levitin bound \eqref{hist:eq13} proved to be a significantly harder task. For time-independent Hamiltonians Giovannetti, Lloyd, and Maccone \cite{Giovannetti2003,Giovannetti2003b,Giovannetti2004} established numerically that the quantum speed limit time has to be given by
\begin{equation}
\label{hist:eq31}
\tau_\mrm{QSL}=\ma{\frac{\hbar}{\Delta H}\,\mc{L}\left(\rho_0,\rho_\tau\right),\,\frac{2\hbar}{\pi\la H\ra }\,\mc{L}^2\left(\rho_0,\rho_\tau\right)}\,.
\end{equation}
This result \eqref{hist:eq31} is particularly remarkable since Giovannetti, Lloyd, and Maccone \cite{Giovannetti2003,Giovannetti2003b,Giovannetti2004} numerically verified the analytical treatment of Pfeifer \cite{Pfeifer1993} and Uhlmann \cite{Uhlmann1992}. However, their study also highlighted that the Margolus-Levitin bound does not generalise as intuitively as one might hope.  \revised{Finally, it is interesting to note  that not all works were exclusively interested in lower bounds for the quantum speed limit time. For instance, in Ref.~\cite{Andrews2007} Andrews derived upper as well as lower bounds on the quantum speed, and hence minimal as well as maximal evolution times.}

Further attempts at generalisations of the \revised{bounds} to driven dynamics and arbitrary angles were undertaken by Jones and Kok \cite{Jones2010}, Zwierz \cite{Zwierz2012}, and Deffner and Lutz \cite{Deffner2013}. However, most of these earlier results lack the clarity and simplicity of the bound that was finally unveiled by Deffner and Lutz in Ref.~\cite{Deffner2013PRL} by considering a geometric approach to open system dynamics.

Before we move on to the case of open systems, however, there are many important consequences of the quantum speed limit for isolated dynamics to discuss first. Therefore, the next two sections will focus on the physical significance and conceptual insights from the quantum speed limit for time-independent dynamics, Eqs.~\eqref{hist:eq14} and \eqref{hist:eq31}, before we return to a more detailed discussion of driven systems and the geometric approach in Section~\ref{geometric}.

%%%%%%%%%%%%%%%%%%%%%%%%%%%%%%%%%%%%%%%%%%%%%%%%%%%%%%%%%%%%%%%%%%%%%%%%%%%%%%%%%
%\input{chapters/time_indep.include.tex}

\section{Quantum speed limits for time-independent generators}
\label{timeindependent}

\subsection{Bremermann-Bekenstein Bound}
A natural playground for exploring the ramifications of quantum speed limits are  information processing systems. One of the earliest explicit considerations was proposed by Bremermann~\cite{Bremermann1967},  who considered the physical limitations of any computational device. In particular, he argued that such a device must obey the fundamental laws of physics namely \emph{special relativity}, \emph{quantum mechanics}, and \emph{thermodynamics}. Thus the rate with which information is processed has to be bounded simultaneously by the {\it light barrier}, the {\it quantum barrier}, and the {\it thermodynamic barrier}. In an almost heuristic way, Bremermann invoked the quantum speed limit by first considering Shannon's seminal work on classical channel capacities and the associated noise energy, coupled with a maximum speed of propagation given by the speed of light, and then imposing the energy-time uncertainty principle. 

However, it was very quickly pointed out by Bekenstein~\cite{Bekenstein1981} that relating Shannon's noise energy to the energy uncertainty was, at the very least, dubious. Regardless, Bekenstein showed essentially the same fundamental bound on information transfer can be formulated from purely thermodynamic and causality considerations. 

His analysis starts with an upper bound on how much entropy can be stored in a given region of space, which can be expressed as the inequality
\begin{equation}
\frac{S}{\la H\ra} < \frac{2\pi k_B R}{\hbar c}\,,
\end{equation} 
where $R$ is the radius of a sphere enclosing the system and $\la H\ra$ is the mean energy. If the system's entropy is maximal then using all available internal states allows for up to $S/(k_B \ln2)$ bits of information to be stored, therefore our system (enclosed by the sphere) can store at most $I=(2\pi \la H\ra R)/(\hbar c \ln 2)$ bits.  

If we are now interested in learning about a system, information has to be exchanged between this system and an outside observer. An upper bound on the rate, $\dot{I}$, with which the information is communicated is given by the total information stored in the system divided by the minimal time it would take to erase all this information. Hence, using the Margolus-Levitin bound \eqref{hist:eq09} Bekenstein wrote 
\begin{equation}
\label{time:eq01}
\dot{I} < \frac{I}{\tau_\mrm{QSL}}< \frac{\pi \la H\ra}{\hbar \ln2},
\end{equation}
which can be equivalently expressed as
\begin{equation}
\frac{\la H\ra}{I} > \frac{\hbar \ln2}{\pi \tau_\mrm{QSL}}
\end{equation}
which simply gives the energy cost per bit for a message received in a time $\tau_\mrm{QSL}$. It is interesting that in these considerations the limits on transmission are imposed by the fundamental physical laws, such as the speed of light, rather than explicitly invoking Heisenberg's energy-time uncertainty relation. 

The Bremerman-Bekenstein bound \cite{Bekenstein1974,Bekenstein1990} is an important result in cosmology, since it gives an upper bound on how much can be learned about non-accessible objects in the Universe, such as black holes. It is further interesting to note that the Bekenstein-Hawking entropy of black holes saturates the bound \cite{Bekenstein1974,Bekenstein1990}. More recently, the Bremerman-Bekenstein bound was re-discovered in quantum thermodynamics \cite{Deffner2010}.

\subsection{Quantum thermodynamics}
Recent years have seen a surge of interest in exploring the thermodynamics of quantum systems~\cite{Goold2016}. It is apparent that the familiar laws of thermodynamics need to be adjusted to cope with situations when the working constituents are described by quantum mechanics. When dealing with quantum thermodynamics the notion of quantum speed limit times become fundamentally important, as is evidenced by two simple considerations. Firstly, irreversibility is a core aspect of thermodynamics and indeed understanding the emergence of this irreversibility will allow us to understand the arrow of time. From a practical point of view  however, controlling irreversibility is crucial to developing efficient devices. We must therefore define and quantify entropy production and entropy production rates in quantum systems. It is clear then that the ultimate bounds on the rate of entropy production must be intimately related to the quantum speed limit time. Secondly, if quantum systems are to be used, for example, as nano-scale engines~\cite{delCampo2013} then the time over which a given cycle is performed enters into the working description in a fundamental way. Clearly the quantum speed limit time allows us to define a maximally achievable efficiency and power. In the following we examine these two situations more closely.

\subsubsection{Entropy production rate and the quantum speed limit.}

\revised{Consider a closed quantum system with Hamiltonian $H_0$ initially in thermal equilibrium at inverse temperature $\beta$. If the system is driven by a time-dependent Hamiltonian, $H_\tau$, for a total elapsed time $\tau$, typically the system will be forced out-of-equilibrium and therefore lead to some degree of irreversible entropy production,
\begin{equation}
\la \Sigma\ra \equiv\beta \left(\la W\ra-\Delta F\right)\,,
\end{equation}
where $\la W \ra$ is the total work done on the system during time $\tau$ and $\Delta F$ is the free energy difference.

Typically, the quantum work distribution is  given by the difference  of  final and initial system energy eigenvalues, $E_{m}^{\tau }-E_{n}^{0}$, averaged over all initial states with thermal distribution $p_n^0 = \exp(-\beta E_n^0)/Z_0)$ and final states \cite{Talkner2007,Campisi2011,Deffner2016work},
\begin{equation}
\label{q03}
{\cal P}\left( W\right) =\sum_{m,n}\,\delta\left( W-\left( E_{m}^{\tau }-E_{n}^{0}\right) \right) \,p_{m,n}^{\tau}\, p^0_{n} \ ,
\end{equation}
where  $p_{m,n}^{\tau}=|\bra{m}U_\tau\ket{n}|^2$ are the unitary transition probabilities. Accordingly, we can write 
\begin{equation}
\label{q04}
\la W \ra=1/\beta\,\sum\limits_n\, p^0_n \,\ln{ p^0_n }-1/\beta\,\sum\limits_{m,n}\,p_n^0\, p_{m,n}^{\tau}\,\ln{ p_m^\tau }
-1/\beta\,\ln{\left(Z_\tau/Z_0\right) }\ .
\end{equation}
The last term on the right-hand side is equal to $\Delta F$, while the first two are $(1/\beta)$ times the quantum  
Kullback-Leibler divergence $S(\rho_\tau||\rho_\tau^\text{eq})$, or quantum relative entropy \cite{Umegaki1962}, between the actual density operator of the system $\rho_\tau$ at time $\tau$ and the corresponding equilibrium density operator $\rho_\tau^{\text{eq}}$. 

Therefore, $\la \Sigma\ra$ can be expressed as a relative entropy, which is always non-negative, and hence we have the Clausius inequality~\cite{Deffner2010,Deffner2013TL}
\begin{equation}
\la \Sigma\ra = S(\rho_{\tau} \| \rho_{\tau}^\text{eq}) =\trace{\rho_{\tau} \loge{\rho_{\tau} }}-\trace{\rho_{\tau} \loge{\rho_{\tau}^\text{eq} }}\geq 0\,.
\end{equation}}
A tighter bound can be derived by considering the geometric distance between these states~\cite{Deffner2010}
\begin{equation}
\la \Sigma\ra \geq \frac{8}{\pi^2} \mathcal{L}^2(\rho_{\tau}, \rho_{\tau}^\text{eq}). 
\end{equation}
Already the use of the Bures metric hints that some relation with the quantum speed limit might exist. This relation becomes more concrete when we consider an equally important quantity: the entropy production rate
\begin{equation}
\sigma = \frac{\la \Sigma\ra}{\tau}.
\end{equation}
Since the time for a state to evolve is bounded by the quantum speed limit time it allows us to establish an upper bound on the entropy production rate by replacing $\tau \to \tau_\text{QSL}$ as given by Eq.~\eqref{hist:eq31}. In the limit of large excitations, i.e. $\big< H_{\tau} \big> \gg \big< H_0 \big>$ the maximal entropy production rate is then given simply as
\begin{equation}
\label{ti:eq1}
\sigma_\mrm{max}=2\beta \la H_{\tau} \ra \mi{\frac{\Delta H_0}{ \hbar \mc{L}\left(\rho_0,\rho_{\tau} \right)},\,\frac{\pi\la H_0\ra }{2\hbar\mc{L}^2\left(\rho_0,\rho_\tau\right)}}\,.
\end{equation}
It is worth noting that if initial and final states are orthogonal and in the limit of high temperatures, Eq.~\eqref{ti:eq1} simplifies to the Bremermann-Bekenstein bound~\cite{Bekenstein1981}. While Eq.~\eqref{ti:eq1} uses the quantum speed limit time arising when assuming time-independent Hamiltonians, these bounds are readily generalisable to arbitrary processes by using a geometric approach to unambiguously define the quantum speed limit, and this will be discussed further in Sec.~\ref{geometric}.

\subsubsection{Efficiency and power of quantum machines.}
The study of thermal quantum engines has grown substantially in recent years and the quantum Otto cycle receiving particular focus, see for example Ref.~\cite{Kosloff2017}. The Otto cycle consists of four strokes: (i) isentropic compression where work $\la W_1 \ra$ is done, (ii) hot isochore where heat $\la Q_2 \ra$ is added, (iii) isentropic expansion where work $\la W_3 \ra$ is done, and (iv) cold isochore where heat $\la Q_4 \ra$ is removed. An interesting caveat associated with using quantum systems as the working substance is the expansion/compression strokes should be performed adiabatically, which according to the quantum adiabatic theorem requires them to be performed (infinitely) slowly, and thus render the considered engine useless as its output power would be zero. To circumvent this issue the use of ``shortcuts to adiabaticity" has been proposed to ensure the compression/expansion strokes are performed in a finite time, $\tau$~\cite{delCampo2013,Zheng2016}, see Ref.~\cite{Torrontegui2013} for a review of these techniques. The efficiency of such a ``superadiabatic engine" can be defined
\begin{equation}
\eta_{SA} = \frac{\text{Energy Output}}{\text{Energy Input}} = -\frac{\la W_1 \ra + \la W_3 \ra}{\la Q_2 \ra + \la H^{SA}_1 \ra + \la H_3^{SA} \ra}
\end{equation}
\revised{where $H^{SA}_i$ is the counterdiabatic Hamiltonian during the $i$th stroke, which can be written in terms of the instantaneous energy eigenstates, $\ket{n_t}$, as  \cite{Demirplak2003,Demirplak2005,Berry2009,Zheng2016,Campbell2017}
\begin{equation}
H^{SA}=i \hbar \com{\pd_t \ket{n_t}\!\!\bra{n_t}}{\ket{n_t}\!\!\bra{n_t}}\,.
\end{equation}
Accordingly its average $\la H^{SA}_i \ra$ can be understood as the} energetic cost of achieving the superadiabatic compression and expansion strokes~\cite{Abah2017,Zheng2016,Campbell2017}. The power is then
\begin{equation}
P_{SA} = -\frac{\la W_1 \ra + \la W_3 \ra}{\tau_{\text{cycle}}}
\end{equation}
where $\tau_{\text{cycle}}$ is the total time for the Otto cycle to be completed. Interestingly, a Margolus-Levitin-type quantum speed limit on the time required to achieve the transformations can be defined~\cite{Abah2017}
\begin{equation}
\label{eq:Abah}
\tau \geq \tau_\text{QSL} = \frac{\hbar \mathcal{L}(\rho_i,\rho_f)}{\la H_{SA} \ra}
\end{equation} 
This leads to bounds on the efficiency and power of the superadiabatic engines
\begin{equation}
\eta_{SA} \leq \eta_{SA}^\text{QSL} = -\frac{\la W_1 \ra + \la W_3 \ra}{\la Q_2 \ra + \hbar (\mathcal{L}_1 + \mathcal{L}_3)/(\tau_\text{QSL}^1+\tau_\text{QSL}^3)}
\end{equation}
\begin{equation}
P_{SA} \leq P_{SA}^\text{QSL} = -\frac{\la W_1 \ra + \la W_3 \ra}{\tau_\text{QSL}^1 + \tau_\text{QSL}^3 }.
\end{equation}
where $\tau_\text{QSL}^i$ is the quantum speed limit time given by Eq.~\eqref{eq:Abah} for the expansion/compression stroke and we have assumed that the thermalisation times during strokes 2 and 4 are much shorter than the expansion/compression stages~\cite{delCampo2013,Abah2017}. We remark that an alternative definition of the efficiency for such superadiabatic engines was proposed in Ref.~\cite{Kosloff2017}, however bounding this efficiency by using the quantum speed limit time can be done in essentially the same way. Finally, also the speed and efficiency of incoherent engines \cite{Mukherjee2016} and the effect of finite-sized clocks \cite{Woods2016} has been \revised{studied.}
 
\subsection{Quantum computation}
\label{quantumcomput}
It is interesting to note that in their original paper Margolus and Levitin make explicit reference to interpreting their result in the context of the number of gate operations a computing machine can achieve per second~\cite{margolus98}. Indeed, the original formulations of the quantum speed limits, where the evolutions are between orthogonal states, lends itself naturally to computational settings, in particular for a two-level system the situation clearly has a close analogy with bit erasure, which we will return to later in connection with Landauer's bound \revised{in Sec.~\ref{subsect:Landauer}}. 

Furthermore, Bremermann's work~\cite{Bremermann1967} explicitly used the energy-time uncertainty relation to discuss computational limitations, albeit the validity of this treatment has been largely disputed~\cite{lloyd00}. However, in Ref.~\cite{lloyd00} Lloyd re-examined this question by first assuming that we have a given amount of energy with which to perform a computation, which we denote $\la H \ra$ for consistency of notation. If $\Delta t_l$ denotes the number of logic operations that gate $l$ can perform per second, with each operation requiring an amount of energy $E_l$, then the total number of operations that a computer can perform per second is 
\begin{equation}
\label{eq:Lloyd}
\sum_l \frac{1}{\Delta t_l} = \sum_l \frac{2 E_l}{\pi \hbar} \leq \frac{2 \la H \ra}{\pi \hbar},
\end{equation}  
\revised{which, if the logic operation in question connects two orthogonal states, is exactly one over the Margolus-Levitin bound}. As noted by Lloyd, the rate at which a computer can process a computation is limited by the energy available. From Eq.~\eqref{eq:Lloyd} we clearly see that the more energy invested in a particular operation implies the faster it can be performed~\cite{lloyd00}. This natural conclusion has recently been shown more explicitly by Santos and Sarandy~\cite{SarandySciRep2015} - wherein by developing shortcuts to adiabaticity that achieve quantum gates, they showed that these operations can be performed faster, however this is accompanied by an increasing energetic cost. We will revisit the relation between employing shortcuts to adiabaticity and the quantum speed limit in Sec.~\ref{STAandQSL}.

Lloyd's discussion further puts into evidence that the time defined by the quantum speed limit is not always a physical evolution time but rather an intrinsic property of a given system. In the context of computation this is a very natural viewpoint as time resources are normally measured by number of gate operations rather than the absolute physically elapsed time.

More recently, Jordan \cite{Jordan2017} further pointed out that energy considerations alone are not sufficient to determine the computational speed. For realistic bounds additional assumptions about the information density and information transmission speed are necessary, see also Sec. \ref{Qucomm} on quantum communication.

\subsection{Quantum metrology}
\label{quantummetrology}
Quantum metrology deals with the use of techniques to achieve the best possible precision in estimating an unknown parameter, or parameters, of a given system. Imagine we wish to determine some unknown parameter, $\mu$ of a given quantum system, $\varrho(\mu)$. We choose a measurement strategy for our estimate $\hat \mu$ and repeat this $M$ times. Then through some smart data-processing we can arrive at an estimate for $\mu$. Under the assumption that our strategy is unbiased, i.e.  $\big< \hat \mu \big>=\mu$, the uncertainty in our estimate, is related to the variance of $\hat \mu$ and is lower bounded by the so-called quantum Cramer-Rao bound
\begin{equation}
\label{cramerrao}
\text{Var}(\hat \mu) \geq \frac{1}{M \mc{F}_Q},
\end{equation}
where $\mc{F}_Q$ is the quantum Fisher information (QFI)~\cite{Giovannetti2006,Giovannetti2011,Paris2009} \revised{which we have encountered above in its infinitesimal version in Eq.~\eqref{hist:eq23}. In this context it is also worth noting that the QFI is the convex roof of the variance~\cite{Toth2013,Yu2013}, and hence the relation to the Mandelstam-Tamm bound \eqref{hist:eq09} becomes immediately apparent.}

While from the outset the relation between quantum speed limits and quantum metrology may not be immediately obvious, considering Eq.~\eqref{cramerrao} is a type of uncertainty bound it seems wholly plausible that a strict relationship might exist. In particular, consider if the parameter we wish to estimate is the elapsed time from a given evolution governed by a time-independent Hamiltonian, $H$. Then Eq.~\eqref{cramerrao} bounds the time uncertainty by the QFI. It would then be sufficient to establish a relation between the QFI and the energy or variance of the Hamiltonian to arrive at a quantum speed limit. This was the approach explored in Ref.~\cite{Frowis2012}, which was one of the first to clearly elucidate the relationship between the QFI (and therefore metrology) and quantum speed limits, and did so by sharpening the Mandelstam-Tamm bound for mixed states.

Consider an isolated, in general mixed, initial state given by its spectral decomposition
\begin{equation}
\rho = \sum_i p_i \ket{i}\bra{i},
\end{equation}
the QFI is time-independent and can be shown to be bounded~\cite{Braunstein1994}
\begin{equation}
\mathcal{F}_Q = 2 \sum_{i,j} \frac{(p_i - p_j)^2}{p_i + p_j} \vert \bra{i} H/\hbar \ket{j} \vert^2 \leq \frac{4 (\Delta H)^2}{\hbar^2}.
\end{equation} 
Note that the latter equation is equivalent to Eq.~\eqref{hist:eq26}, from which we derived the quantum speed limit earlier. As before, we can rearrange this expression to show
\begin{equation}
\Delta H \geq \frac{\hbar \sqrt{\mc{F}_Q}}{2}.
\end{equation}
Using this relation in the Mandelstam-Tamm inequality \eqref{hist:eq09} we obtain
\begin{equation}
\tau_\text{QSL} \geq \frac{\pi}{\sqrt{\mc{F}_Q}}.
\end{equation}
While for pure states this expression is exactly equivalent to the Mandelstam-Tamm relation \eqref{hist:eq29}, it turns out to be a strictly tighter bound for generic mixed states. As remarked in Ref.~\cite{Frowis2012}, the similarity between this Mandelstam-Tamm bound and the Cramer-Rao bound alludes to the deep relation between metrology and quantum speed limits rooted in the fact that both are based on the distinguishability of the states at hand.

The relation between quantum metrology, the QFI, and quantum speed limits has been further explored in several works~\cite{Giovannetti2006,Giovannetti2011,Jones2010,Zwierz2010,Zwierz2012,taddei13,Zhang2016,Beau2017}. We will return to discuss some of these ideas in more detail in Sec.~\ref{geometric} and Sec.~\ref{sec:CR}.

%%%%%%%%%%%%%%%%%%%%%%%%%%%%%%%%%%%%%%%%%%%%%%%%%%%%%%%%%%%%%%%%%%%%%%%%%%%%%%%%%

%\input{chapters/mini.include.tex}

\section{Optimised quantum evolution  and the minimal time approach}
\label{minimaltime}
While we have stressed previously that the quantum speed limit time, $\tau_\text{QSL}$, is associated with the intrinsic properties of the system, it can of course correspond to a physically elapsed time. \revised{Indeed for a given time-independent Hamiltonian the quantum speed limit time implies that there exists a driven dynamic achieving this maximal speed.} The relation and importance of this first became evident in the seminal work of Caneva \etals \cite{Caneva2009} in the context of optimal control. In the following we explore the remarkable emergence of the quantum speed limit time as a fundamental limit in determining effective means to control quantum systems when we fix the physically allowed passage of time to be finite.

\subsection{Optimal control theory}
\label{OCtheory}
It is a well established fact that quantum systems are inherently fragile. Despite this drawback, the interest in exploiting single or many-body quantum systems to perform complex tasks, e.g. quantum gates, communication, or information processing, has steadily grown largely due to the perceived advantage that manipulating quantum systems can provide. All such endeavours then necessitate that the evolution of the quantum system is accurate, i.e. error-free, and in general this requires sophisticated control techniques. One such technique is optimal control theory~\cite{Krotov1996,Walmsley2003,Brody2003}, the aim of which is simple: we define an initial state, $\ket{\psi_0}$, and a desired target state, $\ket{\psi_T}$. Then using the tuneable parameters of the system's Hamiltonian we seek to maximise the final fidelity of the evolved state with $\ket{\psi_T}$. A particularly powerful tool in achieving this task is provided by the Krotov algorithm. Put simply, this method involves choosing an initial `guess pulse' for the functional form of the tuneable Hamiltonian parameter and then iteratively solving a Lagrange multiplier problem such that the fidelity, $F=|\braket{\psi_\tau}{\psi_T}|\to 1$. An interesting and important point is that the elapsed time over which this evolution is performed does not enter as a parameter to optimise explicitly and is typically pre-set before implementing Krotov's algorithm.

The very existence of the quantum speed limit implies that even these optimised evolutions cannot be performed in arbitrarily short times. In a remarkable work, Caneva \etals~\cite{Caneva2009} showed explicitly the connection between the quantum speed limit and optimal control. As a paradigmatic example they considered the Landau-Zener model (fixing units such that $\hbar=1$)
\begin{equation}
\label{eq:LZ}
H=\omega \sigma_x + \Gamma(t) \sigma_z,
\end{equation}
and set the initial and target states to be the ground state for $\Gamma(0)$ and $\Gamma(\tau)$ respectively. The Krotov algorithm was then performed taking various values for the elapsed time, $\tau$. \revised{Remarkably they found that there was a minimum value of $\tau$ given by Battacharyya's bound~\cite{Bhattacharyya1983} (i.e. the Mandelstam-Tamm bound for arbitrary angles), below which the algorithm failed to converge, while for values above this they consistently found $F\to1$.} 

As a particular example, by fixing $\Gamma(0)/\omega = -500$ and $\Gamma(\tau)/\omega=500$, it is easy to see that the fidelity between initial and final states is $F\sim0.002$. Therefore, the situation considered is very close to the original consideration by Mandelstam and Tamm as the aim is to evolve the initial state into an {\it almost} orthogonal state (which from the Mandelstam-Tamm bound we know would require an elapsed time of $\pi/(2\Delta H) \approx 1.5708 / \Delta H)$). From Battacharyya's bound~\cite{Bhattacharyya1983} it is easily shown
\begin{equation}
\label{eq:CanevaQSL}
\tau_\text{QSL} = \frac{\arccos\vert \braket{\psi_0}{\psi_T} \vert}{\Delta H} \approx 1.56881.
\end{equation}
Therefore, Caneva \etals  showed that using the Krotov algorithm the minimal elapsed time one can consider corresponds exactly to $\tau_\text{QSL}$. This is an important result as it shows that quantum speed limits are attainable and therefore provides a clear definition of `optimality' in the context of optimal control as a technique that achieves the quantum speed limit. In Ref.~\cite{Bason2012} this procedure, as well as a complementary approach using shortcuts to adiabaticity, was experimentally realised using a Bose-Einstein condensate. \revised{In addition, similar techniques have been employed to optimally charge a quantum battery \cite{Binder2015,Campaioli2017}.}

\subsubsection{Quantum communication.\label{Qucomm}}
\label{quantumcomm}
A key ingredient in virtually all proposed quantum technologies is the ability to transmit and read out information. This naturally leads to the further need for a manageable infrastructure on which these processes can be performed. One promising approach is to construct quantum channels consisting of  open ended one-dimensional chains of interacting qubits. The information is encoded in the first site, and then through the interaction this information is sent along the chain to the last site where it is then read out. This approach to quantum communication, first proposed by Bose~\cite{Bose2003}, has lead to a wide ranging field of study. 

\begin{figure}[t]
\begin{centering}
{\bf (a)}\\
\includegraphics[width=0.75\textwidth]{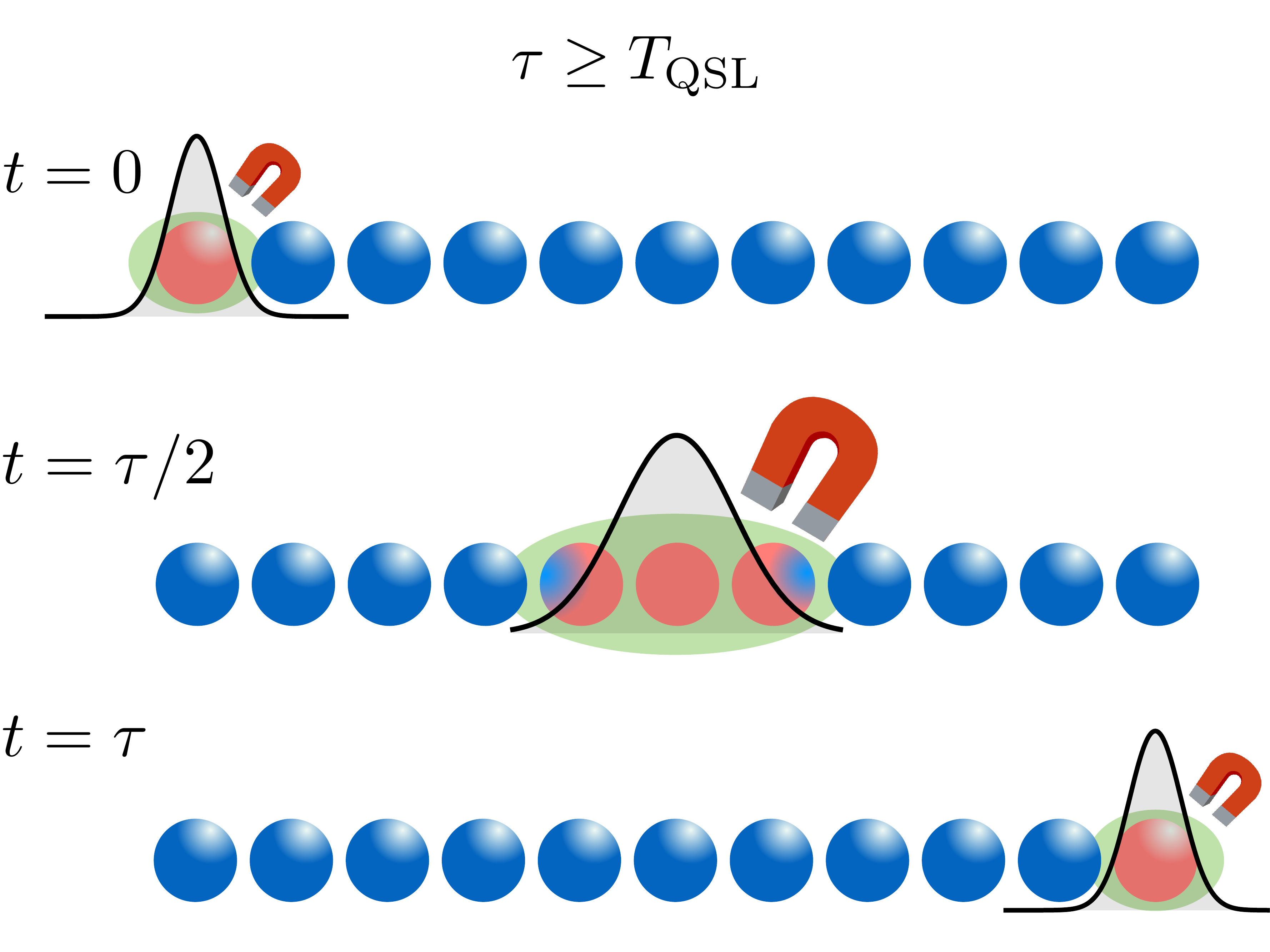}\\ \vspace{1em}
{\bf (b)}\\
\includegraphics[width=0.75\textwidth]{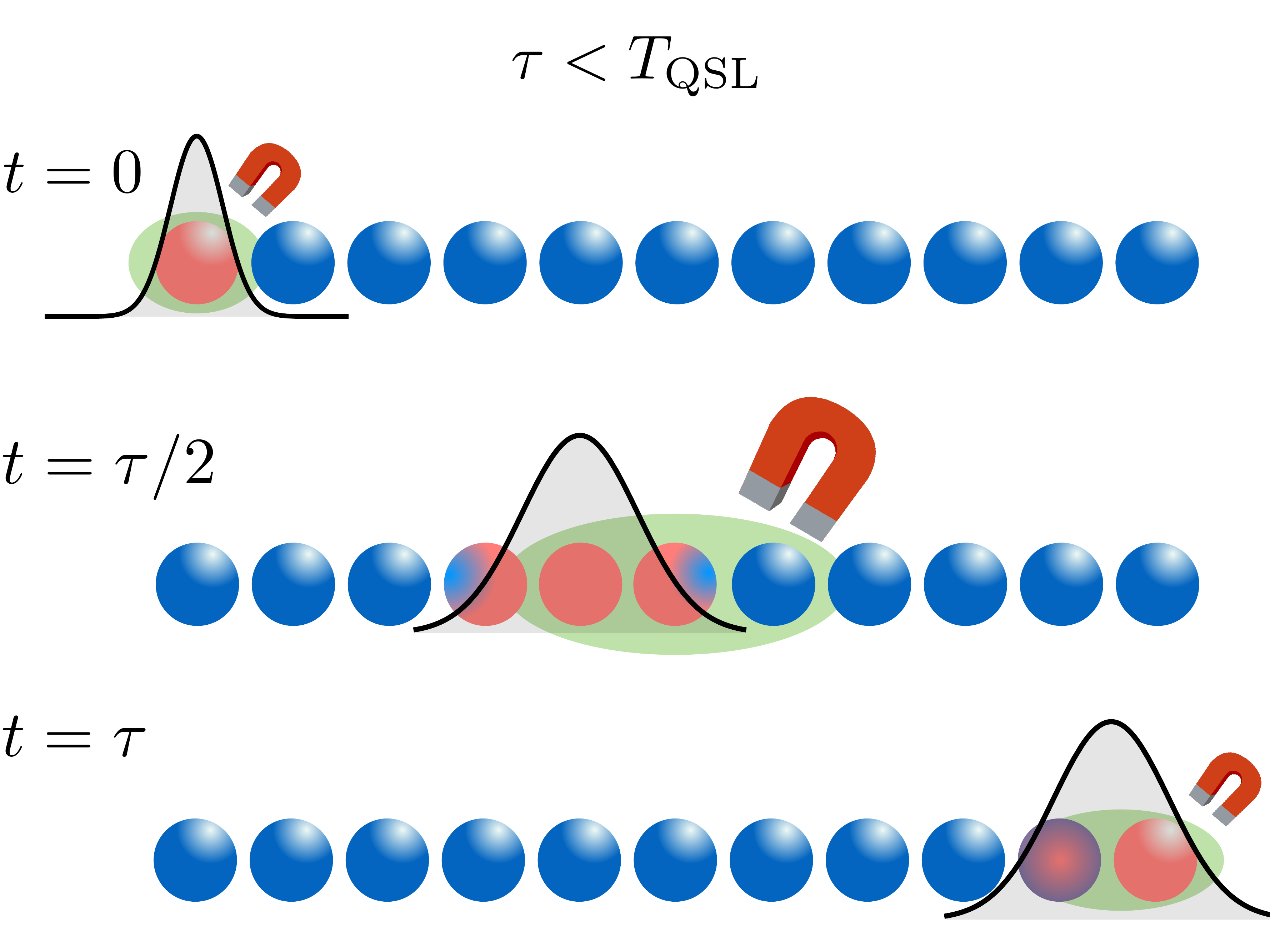}
\caption{Sketch of quantum communication along a spin chain. The first spin is encoded with the state which we wish to transmit. \revised{{\it Upper panel, (a):}} If the total duration of the protocol $\tau \geq T_\text{QSL}$ then this can be achieved using an optimised evolution. The initially localised state forms an excitation wave that is propagated along the chain. \revised{{\it Lower panel, (b):}} Conversely, if $\tau < T_\text{QSL}$, then evolution is too fast and the spin wave cannot keep up, resulting in only part of the state arriving at the final spin.}
\label{fig:qcommfigure}
\end{centering}
\end{figure}

The question of how fast this information can be propagated along the chain then becomes one of both fundamental and practical interest. In Ref.\cite{Murphy2010} Murphy \etals  explored this by studying the transmission of a single excitation, initially localised at the first site, along a spin chain, which is initialised in its ground state, with an isotropic Heisenberg interaction
\begin{equation}
H=-\frac{J}{2}\sum_n^{N-1} \left( \sigma_x^i \sigma_x^{i+1}+\sigma_y^i \sigma_y^{i+1}+\sigma_z^i \sigma_z^{i+1} \right) + \sum_n^N B_n(t) \sigma_z^n.
\end{equation}
By employing the Krotov algorithm to optimise the profile of the magnetic field, $B_n(t)$, they showed that there was a cutoff time $\tau_\text{QSL}$, below which the algorithm failed to converge, in close analogy to the discussions from Sec.~\ref{OCtheory}. Their analysis reveals an interesting aspect of the speed of evolution when dealing with interacting many-body systems: the presence of the interaction (and likely therefore entanglement) allows for faster communication. More specifically, Murphy \etals  established that for their system the quantum speed limit time to evolve the initial state into an orthogonal state is given by
\begin{equation}
\tau_\text{QSL} = \frac{\pi \hbar}{2 J}.
\end{equation}
For a chain of only 2 qubits, this is precisely the time it takes to perform a swap operation. Naively, we might assume then that the total quantum speed limit time to transmit the excitation along a chain of length $N$ will then simply be $(N-1)\tau_\text{QSL}$, and we would achieve this limit by performing sequential swap operations to neighbouring sites, which we call an ``orthogonal swap". However, by examining the dynamics we see that the optimal control pulse determined by the Krotov algorithm does not transmit the excitation completely to each site, but instead it forms an excitation wave which is spread across several sites at any time, see the schematic in Fig.~\ref{fig:qcommfigure} {\bf (a)}. Therefore, the optimised evolution performs a controlled propagation of the excitation wave, and can be understood as  a cascade of effective swaps, each of which has a duration shorter than the orthogonal swap. The total quantum speed limit time is then
\begin{equation}
T_\text{QSL} = \gamma (N-1) \tau_\text{QSL}.
\end{equation}
with $\gamma<1$ a dimensionless constant that quantifies the effective swap duration in terms of the orthogonal swap. Interestingly for $T<T_\text{QSL}$ Fig.~\ref{fig:qcommfigure} {\bf (b)} we see the excitation wave is unable to keep up with the pulse. The results show the interesting features that can emerge when dealing with many-body systems and puts into evidence the role that interactions and entanglement play in dictating the quantum speed limit.

Finally we note that the emergence of the quantum speed limit in quantum communication can be understood as a consequence of the Lieb-Robinson bound \cite{Lieb1972}, on which we will elaborate in Sec.~\ref{sec:LR}.

\subsubsection{Many-body systems.}
We have seen that for a many-body system, the speed of evolution of can be enhanced by allowing its constituents to interact. Such interacting many-body systems can exhibit remarkably interesting features, in particular the presence of distinct phases in the ground state. By varying an order-parameter, $\lambda$ (e.g. a magnetic field), the properties of the ground state of a large number of interacting quantum systems can exhibit sudden changes. In many systems these quantum phase transitions (QPTs) happen when the spectral gap between the ground and first excited state closes. As such, driving a critical system through its QPT typically requires timescales in the adiabatic limit in order to avoid generating defects. However, the adiabatic limit ensures that the system remains in its ground state at all times. If we are only interested in driving from the ground state in one phase, say for $\lambda_0$, to the ground state in another phase, $\lambda_1$, without requiring the system to always remain in its ground state, how fast can this transformation be achieved? 

Once again employing the minimal time approach, through the Krotov algorithm, this time was shown to be bounded by the quantum speed limit time~\cite{Caneva2011}. The analysis is closely related to that of Sec.~\ref{OCtheory} and \ref{quantumcomm}. However, an interesting additional aspect emerges when one recalls that the dynamics we are studying involves transitioning a critical point, which separates distinct static phases of the system. Defining the action, $s=\tau \Delta$, where $\Delta$ is the minimum gap between the ground and first excited state, it is possible to show that the quantum speed limit defines different dynamical regimes when driving through the critical point. In particular, if the driving is done linearly this action is shown to diverge as the system size is increased. Conversely, using the optimised pulses with duration $\tau=\tau_\text{QSL}$ we find $s\sim \pi$, and confirmed to occur in three distinct models, the Landau-Zener, Grover's search algorithm, and the Lipkin-Meshkov-Glick model, and therefore likely to hold in general. For $s\geq\pi$, which corresponds to durations $\tau\geq\tau_\text{QSL}$, the Krotov algorithm always converges and this is defined as a region in which adiabatic dynamics can be effectively achieved. For an action $s<\pi$, which means $\tau<\tau_\text{QSL}$ i.e. the driving time is less than the the quantum speed limit time, defects are produced~\cite{Caneva2011}.

\revised{\subsection{Parametric Hamiltonians and driven dynamics}}
The emergence of the quantum speed limit time as a fundamental limitation for optimal control methods outlined previously is indeed remarkable, however, it should be noted the somewhat special circumstances considered: namely the application of a particular algorithm to design the control pulses, and the form of the initial and final states. Returning to the Landau-Zener example Eq.~\eqref{eq:LZ}, the minimal time approach was revisited by Hegerfeldt~\cite{Hegerfeldt2013} and later by Poggi \etals \cite{Poggi2013}. The crucial difference in their approach is to revisit the notion of minimal times while also recalling that by engineering special control pulses for the Hamiltonian their system was no longer truly time-independent. This is particularly important considering the quantum speed limit time that emerged as the fundamental bound from Caneva \etals \cite{Caneva2009}, from Eq.~\eqref{eq:CanevaQSL}, actually assumes that the system's Hamiltonian is time-independent. 

Through a careful re-examination of the problem Hegerfeldt showed that the optimal control problem for a two level system is analytically treatable for arbitrary initial and final states. In particular, defining the initial and final target states as 
\begin{equation}
\begin{split}
\ket{\psi_0}&= i_0 \ket{0} + i_1 \ket{1},\\
\ket{\psi_T}&= f_0 \ket{0} + f_1 \ket{1},
\end{split}
\end{equation}
it was proven that the minimal time, $\tau_\text{min}$, satisfies
\begin{equation}
\label{eq:Hegerfeldt}
\cos(\omega\, \tau_\text{min}) = |f_0 i_0| +|f_1 i_1|.
\end{equation}
Interestingly, if we restrict $\ket{\psi_0}$ and $\ket{\psi_T}$ to be the respective ground states of the Landau-Zener Hamiltonian, Eq.~\eqref{eq:LZ}, at the start and end of the protocol, $\Gamma(0)=-\gamma$ and $\Gamma(\tau)=\gamma$, then $\tau_\text{min}=\tau_\text{QSL}$, precisely in line with Refs.~\cite{Caneva2009,Bason2012}. However, if the initial state, $\ket{\psi_0}$, is the ground state at $\Gamma(0)=-\gamma$ while the final target state, $\ket{\psi_T}$, is the excited state at $\Gamma(\tau)=\gamma$ it can be shown that
\begin{equation}
\label{eq:Tmin}
\tau_\text{min} = \frac{1}{\sqrt{\gamma^2+\omega^2}} < \frac{\pi}{2\omega} = \tau_\text{QSL}. 
\end{equation}
This indicates that care must be taken when applying the quantum speed limit for time-independent Hamiltonians {\it verbatim} to certain control problems. 

\subsection{Further reading on the minimal time approach}

The minimal time approach and associated techniques discussed in this section have been explored in a variety of other settings for which we refer the reader to Refs.~\cite{richerme14,delCampo2015,Khalil2015} on further analyses of critical and many-body systems systems, Refs.~\cite{Mukherjee2013,deffner14,Caneva2013,Mukherjee2015} on analyses concerning controlling quantum systems including the effects of noise, Ref.~\cite{Reich2013} on  a study of optimised molecular cooling, Ref.~\cite{Brody2015,Gajdacz2015} on studies of control in the presence of arbitrary external fields or potentials, Ref.~\cite{Baksic2016,Arenz2017} on accelerated quantum state transfer, Refs.~\cite{Barnes2013a,Boscain2012,Hegertfeldt2014} on further considerations of the control of two-level systems, and Ref.~\cite{Bukov2017} on quantum state preparation.

%%%%%%%%%%%%%%%%%%%%%%%%%%%%%%%%%%%%%%%%%%%%%%%%%%%%%%%%%%%%%%%%%%%%%%%%%%%%%%%%%

%\input{chapters/geo.include.tex}

\section{\label{geometric}Maximal quantum speed from the geometric approach}

In the preceding section we discussed the so-called \emph{minimal time approach} \cite{deffner14}. Within this paradigm one is interested in characterising the \revised{time optimal dynamics}, or more generally the optimal generator of the quantum dynamics that drives the quantum system from a particular initial state to a particular final state, in the shortest time allowed under the laws of quantum mechanics. In this section we will now \revised{slightly change the point of view, in that we are no longer interested in determining the shortest evolution time, but rather the maximal speed. This approach has become known as the \emph{geometric approach}.}

In the geometric approach one is interested to find an estimate for the \emph{maximal quantum speed} under a given quantum dynamics, which we will write as a quantum master equation,
\begin{equation}
\label{geo:eq01}
\dot{\rho}_t=L(\rho_t)\,.
\end{equation}
Here, $L(\rho_t)$ is an arbitrary, linear or non-linear, Liouvillian super-operator. At first glance, these two approaches appear to be equivalent, since they give the same results for time-independent generators \cite{Poggi2013,deffner14}. The fundamental difference, however, becomes obvious for parameterised, driven, and time-dependent $L(\rho_t)$ \cite{Jones2010}.

Above in Sec.~\ref{history} we already mentioned that for mixed quantum states defining an angle is rather involved. Finding such a measure of distinguishability, however, is necessary in order to be able to define the \emph{quantum speed}, which should be given by the derivative of some distance. For isolated systems it quickly became clear that the Bures angle \eqref{hist:eq22} would do the job \cite{Uhlmann1992,Deffner2013}, whereas for open systems the situation has been less obvious. For instance, del Campo \etals \cite{delcampo13} chose to work with the relative purity, Mondal and Pati \cite{Mondal2016PLA} saw the need to define a new metric, and Pires \etals made a strong case for the Wigner-Yanase information \cite{Pires2016}. Therefore, we continue with a brief summary of how to measure the distinguishability of quantum states, and of how to characterise the geometric quantum speed.

\subsection{Defining the geometric quantum speed}

In its standard interpretation \cite{Messiah1966} quantum mechanics is a probabilistic theory, in which the state of a physical system is described by a wave function $\psi(x)$. The modulus squared of $\psi(x)$ is the probability to find the quantum system at position $x$. More formally $\psi(x)$ is understood as a specific representation of a vector $\ket{\psi}$ in Hilbert space. Hence to be fully consistent, a proper measure of distinguishability of two wave functions $\psi_1(x)$ and $\psi_2(x)$ should be equivalent to the distance between $\ket{\psi_1}$ and $\ket{\psi_2}$.

This observation led Wootters \cite{Wootters1981} to carefully study the statistical distance $\ell$ induced by the Fisher-Rao metric, aka the Fisher information metric. For  a parametric path $\psi(x,t)$ with $\psi(x,0)=\psi_1(x)$ and $\psi(x,\tau)=\psi_2(x)$ we have,
\begin{equation}
\label{geo:eq02}
\ell(\psi_1,\psi_2)=\frac{1}{2}\,\int_0^\tau dt\, \sqrt{\int dx\,\frac{1}{p(x,t)}\left[\frac{dp(x,t)}{dt}\right]^2}\,,
\end{equation}
where $p(x,t)=|\psi(x,t)|^2$. \revised{\v{C}encov's theorem states that the Fisher-Rao metric is (up to normalisation) the unique metric whose geodesic distance is a monotonic function \cite{bengtsson2007}. Hence it is the only metric on the probability simplex that exhibits invariant properties under probabilistically natural mappings \cite{Campbell1986}.} Wootters \cite{Wootters1981} then showed that the geodesic, i.e., the shortest path connecting $\psi_1(x)$ and $\psi_2(x)$ is given by
\begin{equation}
\label{geo:eq03}
\ell(\psi_1,\psi_2)=\arccos{\left(\left|\braket{\psi_1}{\psi_2}\right|\right)}\,.
\end{equation}
As a main conclusion Wootters showed that the shortest path connecting vectors in Hilbert space, i.e. the angle between these vectors, is identical to the geodesic under the Fisher-Rao metric. Hence, measuring the distinguishability of probability distributions is identical to determining the angle between pure quantum  states.

Since $\ell$  constitutes the \emph{shortest} path connecting quantum states, the statistical distance \eqref{geo:eq03} serves as the natural choice to define the maximal quantum speed \cite{Pfeifer1993,Jones2010,Zwierz2012,Deffner2013,Poggi2013}. The geometric quantum speed limit is then found as an upper bound on the such defined speed,
\begin{equation}
\label{geo:eq04}
v\equiv\dot{\ell}(\psi(0),\psi(t))\leq\left|\dot{\ell}(\psi(0),\psi(t))\right|\leq v_\mrm{QSL}\,,
\end{equation}
and the quantum speed limit time is defined as one over the averaged speed \cite{Pfeifer1993,Jones2010,Zwierz2012,Deffner2013,Poggi2013},
\begin{equation}
\label{geo:eq05}
\tau_\mrm{QSL}\equiv\frac{\tau}{\int_0^\tau dt\,v_\mrm{QSL}}\,.
\end{equation}
It is worth emphasising that within the geometric approach $\tau_\mrm{QSL}$ \emph{no longer describes the actual evolution time} of the quantum system. In the following we will rather see that $ v_\mrm{QSL}$ and $\tau_\mrm{QSL}$ are characteristics of the dynamics and the geometry of the eigensystem of the generator of the dynamics.

Before we move on, however, we have to generalise the notion of quantum speed \eqref{geo:eq04} to mixed quantum states evolving under the general master equation \eqref{geo:eq01}. 

\paragraph{Quantum speed for mixed states.}

For pure states we have chosen the statistical distance \eqref{geo:eq03} as a natural basis for the definition of quantum speed. The situation is significantly more involved for mixed states, since mixed quantum states are no longer simply related to classical probability distributions\revised{, but rather are statistical mixtures of pure states. In other words, mixed states can be considered as \emph{distributions of probability distributions}. Nevertheless, it is still possible to express} mixed states as a partial trace over a pure state in a suitably enlarged Hilbert space \cite{bengtsson2007}. In a certain sense mixed states are the quantum analogues of marginal probability distributions. Hence, one would expect that one can find a generalisation of the statistical distance $\ell$ by taking the appropriate partial traces.

From rather formal arguments it can be shown \cite{Uhlmann1976,Jozsa1994,bengtsson2007} that the quantum statistical distance, aka the Bures angle is given by
\begin{equation}
\label{geo:eq06}
\mc{L}(\rho_1,\rho_2)=\arccos{\left(\sqrt{F(\rho_1,\rho_2)}\right)}\,,
\end{equation}
where the quantum fidelity $F(\rho_1,\rho_2)$ reads
\begin{equation}
\label{geo:eq07}
F(\rho_1,\rho_2)=\left[\trace{\sqrt{\sqrt{\rho_1}\rho_2\sqrt{\rho_1}}}\right]^2\,.
\end{equation}
It is easy to see that for pure states $\rho_1=\ket{\psi_1}\bra{\psi_1}$ and $\rho_2=\ket{\psi_2}\bra{\psi_2}$ the Bures angle $\mc{L}$ reduces to the statistical distance $\ell$ \eqref{geo:eq05}. Moreover, the Bures angle is a Riemannian distance and monotonically decreasing under stochastic maps \cite{bengtsson2007}. The latter property is very desirable since coarse graining means that discarded information cannot increase the distinguishability of quantum states. Finally, the Bures angle \eqref{geo:eq07} is the Fisher-Rao distance \eqref{geo:eq03} maximised over all possible purifications.

In conclusion, it appears natural to define the geometric quantum speed as derivation of the time-dependent Bures angle,
\begin{equation}
\label{geo:eq08}
v\equiv\dot{\mc{L}}(\rho_0,\rho_t)\,.
\end{equation}
However, already in Sec.~\ref{history} we alluded to the fact the derivative of $\mc{L}$ and the infinitesimal Bures angle, $\delta\mc{L}$, are far from trivial to handle, and thus several distinct approaches to determine the quantum speed limit $v_\mrm{QSL}$ have been developed.

\subsection{Differential geometry}

An approach inspired by differential geometry on density operator space was proposed by Taddei \etals \cite{taddei13}. Consider the quantum dynamics described by the master equation \eqref{geo:eq01} and two infinitesimally close states $\rho(t)$ and $\rho(t+\delta t)$ for some infinitesimal time step $\delta t$. Then the infinitesimal Bures angle $\delta \mc{L}$ can be expanded in powers of $\delta t$,
\begin{equation}
\label{geo:eq09}
\delta \mc{L}=1-\frac{\mc{F}_Q(t)}{4}\,\delta t^2+\mc{O}(\delta t^3)\,,
\end{equation}
where $\mc{F}_Q(t)$ is again the quantum Fisher information, i.e., the quantum generalisation of the Fisher-Rao metric. It may be defined as \cite{Braunstein1994,bengtsson2007},
\begin{equation}
\label{geo:eq10}
\mc{F}_Q(t)=\trace{\rho(t) G^2(t)}\quad\mrm{with}\quad\rho(t+\delta t)-\rho(t)=\rho(t)G(t)+G(t)\rho(t)\,.
\end{equation}
Hence, we immediately see that the square root of the quantum Fisher information, $\mc{F}_Q(t)$, is proportional to the instantaneous speed between  $\rho(t+\delta t)$ and $\rho(t)$. Taddei \etals \cite{taddei13} then showed that by discretising the evolution of $\rho(t)$ and integrating the right-hand side of Eq.~\eqref{geo:eq09} the Bures angle can be bounded from above by
\begin{equation}
\label{geo:eq11}
\mc{L}(\rho_0,\rho_\tau)=\arccos{\left(\sqrt{F(\rho_0,\rho_\tau)}\right)}\leq \frac{1}{2}\,\int_0^\tau dt\,\sqrt{\mc{F}_Q(t)}\,.
\end{equation}
Accordingly we can identify the quantum speed limit, $v_\mrm{QSL}$, as
\begin{equation}
\label{geo:eq12}
v_\mrm{QSL}=\frac{1}{2}\,\sqrt{\mc{F}_Q(t)}\,.
\end{equation}
Moreover, if the Liouvillian $L(\rho_t)$ simply gives the von-Neumann equation, i.e.,  $i\hbar\,\dot{\rho_t}=\com{H_t}{\rho_t}$, Eq.~\eqref{geo:eq05} reduces to the Mandelstam-Tamm bound \eqref{hist:eq29}.

Similar results  were obtained by Andersson and Heydari with more mathematical rigor, who put forward a geometric construction \cite{Andersson2014b} to study the relations between Hamiltonian dynamics and Riemannian structures \cite{Andersson2014}, and to characterise time-optimal Hamiltonians \cite{Andersson2014a}.

We conclude this section with an observation: When expanding the Bures angle in terms of infinitesimal time-steps \eqref{geo:eq09} the first non-vanishing order is quadratic in time. Therefore, Taddei \etals \cite{taddei13} only obtained a Mandelstam-Tamm type bound \eqref{hist:eq29}, whereas a Margolus-Levitin type bound \eqref{hist:eq13} is beyond the scope of this approach. How to resolve this issue is the starting point of Ref.~\cite{Deffner2013PRL}.

\subsection{Open systems and non-Markovian dynamics}

Generally, the Bures angle $\mc{L}$ \eqref{geo:eq06} is a mathematically rather involved quantity, since it is defined in terms of square roots of operators. The situation significantly simplifies for initially pure states, and we then have
\begin{equation}
\label{geo:eq13}
\mc{L}(\rho_0,\rho_\tau)=\arccos{\left(\sqrt{\bra{\psi_0}\rho_\tau\ket{\psi_0}}\right)}\,,
\end{equation}
where $\rho_0=\ket{\psi_0}\bra{\psi_0}$. Note that for general quantum dynamics \eqref{geo:eq01} the time-dependent state $\rho_\tau$ will be mixed, even if the initial state is pure.

In complete analogy to above \eqref{geo:eq04}, we then obtain 
\begin{equation}
\label{geo:eq14}
2 \co{\mc{L}}\si{\mc{L}}\,\dot{\mc{L}}\leq \left|\bra{\psi_0}\dot{\rho}_t\ket{\psi_0}\right|\,.
\end{equation}
From the latter both a Mandelstam-Tamm type as well as a Margolus-Levitin type bound are easily derived.

\paragraph{Margolus-Levitin type bound.}

To proceed, we introduce  the von Neumann  trace inequality for operators  which reads \cite{Mirsky1975},
\begin{equation}
\label{geo:eq15}
\left|\trace{A_1 A_2} \right|\leq \sum\limits_{i=1}^n \sigma_{1,i} \sigma_{2,i}.
\end{equation}
Inequality \eqref{geo:eq05} holds for any complex $n\times n$ matrices $A_1$ and $A_2$ with descending singular values, $\sigma_{1,1}\geq...\geq\sigma_{1,n}$ and $\sigma_{2,1}\geq...\geq\sigma_{2,n}$. The singular values of an operator $A$  are defined as the eigenvalues of $\sqrt{A^\dagger A}$.  For a Hermitian operator, they are given by the absolute value of the eigenvalues of $A$, and are positive real numbers. If  $A_1$ and $A_2$ are simple (positive) functions of density operators acting on the same Hilbert space, Eq.~\eqref{geo:eq05} remains true for arbitrary dimensions  \cite{Grigorieff1991}. The singular values of the operators $A$ and $A^\dagger$ are moreover identical. By taking $A_1=L(\rho_t)$ and $A_2=\rho_0$, we thus find,
\begin{equation}
\label{geo:eq16}
2\co{\mc{L}}\si{\mc{L}}\,\dot{\mc{L}}\leq \sum_i \sigma_i p_i=\sigma_1,
\end{equation}
where $\sigma_i$ are the singular values of $L(\rho_t)$ and  $p_i=\delta_{i,1}$ for the initially pure state $\rho_0$. Now finally noting that the largest singular value is identical to the operator norm we can write,
\begin{equation}
\label{geo:eq17}
v_\mrm{QSL}=\frac{\| L(\rho_t)\|_\mrm{op}}{2\co{\mc{L}}\si{\mc{L}}}
\end{equation}
and accordingly
\begin{equation}
\label{geo:eq18}
\tau_\mrm{QSL}=\frac{\left[\si{\mc{L}(\rho_0,\rho_\tau}\right]^2}{1/\tau\,\int_0^\tau dt\,\| L(\rho_t)\|_\mrm{op}}\,.
\end{equation}
The interpretation of Eq.~\eqref{geo:eq18} as a Margolus-Levitin type bound becomes more apparent by noting that for any operator $A$ the operator norm is bounded from above by the trace norm,
\begin{equation}
\label{geo:eq18a}
\| A\|_\mrm{op}=\sigma_1\leq\sum_i \sigma_i=\| A\|_\mrm{tr}\,.
\end{equation}
One easily convinces oneself that for unitary dynamics induced by time-independent, positive semi-definite Hamiltonians the time-averaged trace norm of the generator is identical to the average energy, $\| L(\rho_t)\|_\mrm{tr}=\la H\ra$ \cite{Deffner2013PRL}. This limit might appear rather restrictive, however this is exactly the dynamics considered in the original work by Margolus and Levitin \cite{margolus98}.

\paragraph{Mandelstam-Tamm type bound.}

Next, we also derive a Madelstam-Tamm type bound.  To this end, we rewrite Eq.~\eqref{geo:eq14} as
\begin{equation}
\label{geo:eq19}
2\co{\mc{L}}\si{\mc{L}}\,\dot{\mc{L}}\leq\left|\trace{L_t(\rho_t)\,\rho_0}\right|.
\end{equation}
The latter can be estimated from above with the help of the Cauchy-Schwarz inequality for operators:
\begin{equation}
\label{geo:eq20}
2\co{\mc{L}}\si{\mc{L}}\,\dot{\mc{L}}\leq\sqrt{\trace{{L(\rho_t)\,L(\rho_t)^\dagger}}\,\trace{\rho_0^2}}.
\end{equation}
Since  $\rho_0$ is a pure state, $\trace{\rho_0^2}=1$, we obtain,
\begin{equation}
\label{geo:eq21}
2\co{\mc{L}}\si{\mc{L}}\,\dot{\mc{L}}\leq\sqrt{\trace{{L(\rho_t)\,L(\rho_t)^\dagger}}}=\|L(\rho_t)\|_\mrm{hs},
\end{equation}
where $\|A\|_\mrm{hs}= \sqrt{\trace{A^\dagger A}}=\sqrt{\sum_i \sigma_i^2}$ is the Hilbert-Schmidt norm. Thus we now have,
\begin{equation}
\label{geo:eq22}
v_\mrm{QSL}=\frac{\| L(\rho_t)\|_\mrm{hs}}{2\co{\mc{L}}\si{\mc{L}}}
\end{equation}
and accordingly
\begin{equation}
\label{geo:eq23}
\tau_\mrm{QSL}=\frac{\left[\si{\mc{L}(\rho_0,\rho_\tau}\right]^2}{1/\tau\,\int_0^\tau dt\,\| L(\rho_t)\|_\mrm{hs}}\,.
\end{equation}
That Eq.~\eqref{geo:eq23} is a bound of the Mandelstam-Tamm type is obvious, since the Hilbert-Schmidt norm reduces for Hamiltonian dynamics to the variance of the energy \cite{Deffner2013PRL}.

\paragraph{The unified bound.}

Similarly to the case of isolated dynamics, see Sec.~\ref{history}, we obtain seemingly independent expressions for the quantum speed limit, which we can write together as \cite{Deffner2013PRL}
\begin{equation}
\label{geo:eq24}
\tau_\mrm{QSL}=\left[\si{\mc{L}(\rho_0,\rho_\tau}\right]^2 \ma{\frac{1}{\Lambda_\mrm{op}},\frac{1}{\Lambda_\mrm{tr}},\frac{1}{\Lambda_\mrm{hs}}}\,,
\end{equation}
where $\Lambda=1/\tau\,\int_0^\tau dt\,\| L(\rho_t)\|$. However, in the approach chosen in Ref.~\cite{Deffner2013PRL} the bounds are not independent. Rather noting that for trace class operators  we have (see Ref.~\cite{Simon1979}, Theorem 1.16),
\begin{equation}
\label{geo:eq26}
\|A\|_\mrm{op} \leq \|A\|_\mrm{hs}\leq\|A\|_\mrm{tr}\,,
\end{equation}
we conclude that the Margolus-Levitin type bound \eqref{geo:eq18} in terms of the operator norm is the sharpest bound. However, for general quantum dynamics determining the Hilbert-Schmidt norm is computationally cheaper, and hence the Mandelstam-Tamm type  bound \eqref{geo:eq23} could be considered more practical.

\subsubsection{Non-Markovian dynamics.}

As a first application of these new bounds on the maximal quantum speed \eqref{geo:eq24} Deffner and Lutz then proposed to investigate the influence of non-Markovian dynamics on the quantum speed limit time, $\tau_\mrm{QSL}$. To this end, they considered the damped Jaynes-Cummings model \cite{Garraway1997,Breuer1999}, whose dynamics \revised{can be written as
\begin{equation}
\label{geo:eq27}
\dot{\rho}_t=-\frac{i}{\hbar} \com{H_\mrm{qubit}}{\rho_t}-\frac{i}{2\hbar} \com{\lambda_t\, \sigma_+\,\sigma_-}{\rho_t}+\gamma_t\,\left(\sigma_-\rho_t\sigma_+-\frac{1}{2}\acom{\sigma_+\sigma_-}{\rho_t}\right)\,,
\end{equation}
where $H_\mrm{qubit}=\hbar \omega_0\,\sigma_+\sigma_-$ and $\sigma_\pm=\sigma_x\pm i\sigma_y$ are the Pauli operators. The time-dependent decay rate, $\gamma_t$, and the time-dependent Lamb shift, $\lambda_t$, are fully determined by the spectral density, $J(\omega)$, of the cavity mode. We have
\begin{equation}
\label{q13}
\lambda_t=-2\,\mrm{Im}\left\{\frac{\dot{c}_t}{c_t}\right\} \quad \mathrm{and} \quad \gamma_t=-2\,\mrm{Re}\left\{\frac{\dot{c}_t}{c_t}\right\}\,
\end{equation}
where $c_t$ is a solution of
\begin{equation}
\label{q14}
\dot{c}_t=-\int_0^t\td s\int \td \omega\,J(\omega)\,e^{i\hbar\left(\omega-\omega_0\right)(t-s)}\,c_s\,.
\end{equation}}
For a Lorentzian spectral density,
\begin{equation}
\label{geo:eq28}
J(\omega)=\frac{1}{2\pi}\frac{\gamma_0\lambda}{\left(\omega_0-\omega\right)^2+\lambda^2}\,
\end{equation}
the dynamics is analytically solvable and the decay rate can be written as \cite{Deffner2013PRL}
\begin{equation}
\label{geo:eq29}
\gamma_t=\frac{2\gamma_0\lambda\,\sinh\left(d t/2\right)}{d\cosh\left(d t/2\right) +\lambda\sinh\left(d t/2\right)}\,,
\end{equation}
where $d=\sqrt{\lambda^2-2\gamma_0\lambda}$. Deffner and Lutz \cite{Deffner2013PRL} then found that $\tau_\mrm{QSL}$ is a monotonically decreasing function of $\gamma_0/\omega_0$, which suggests that the non-Markovian backflow of information from the environment into the system can accelerate its dynamics.
\begin{figure}
\centering
{\bf (a)}\\
\includegraphics[width=.75\textwidth]{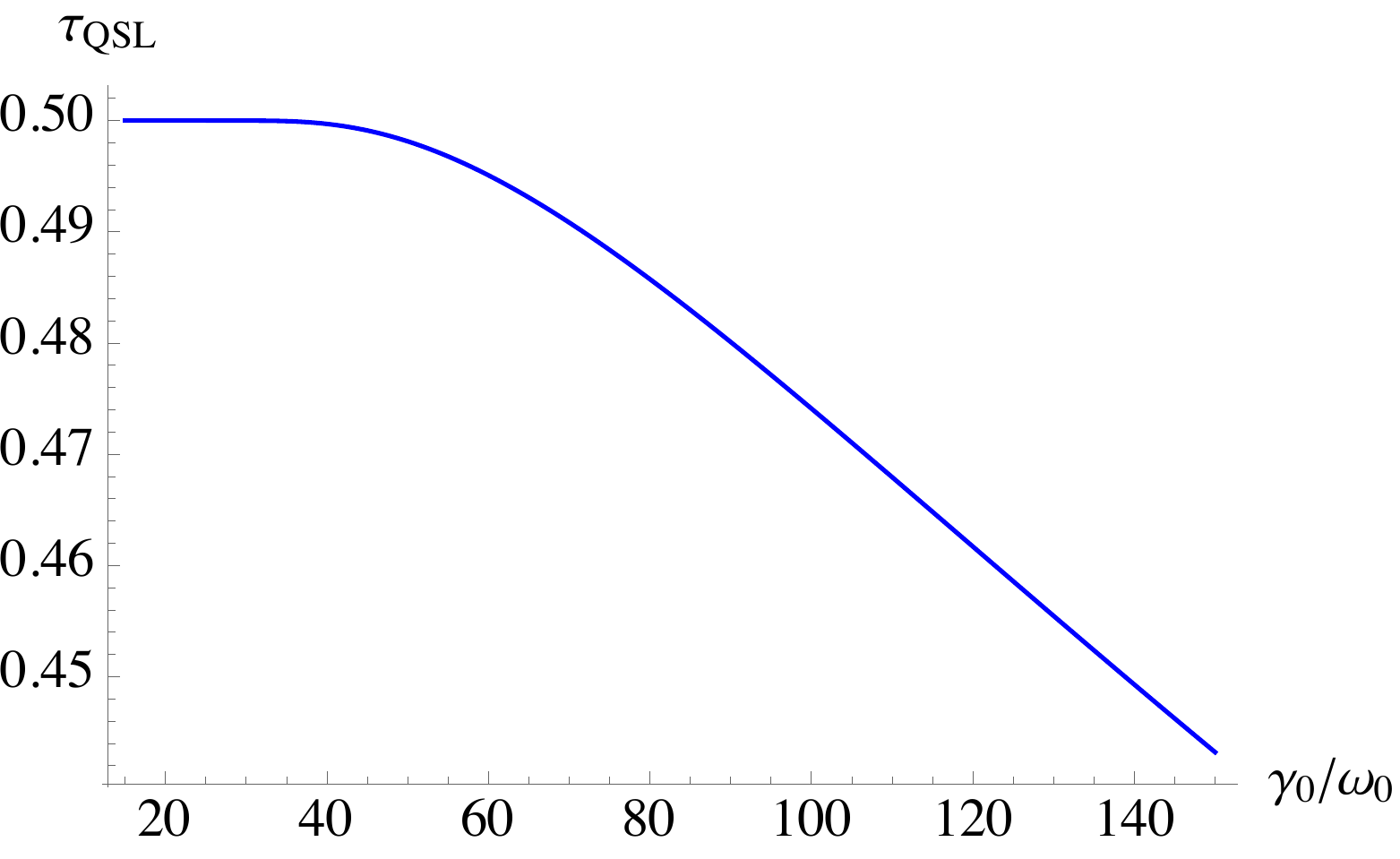}\\
{\bf (b)}\\
\includegraphics[width=.75\textwidth]{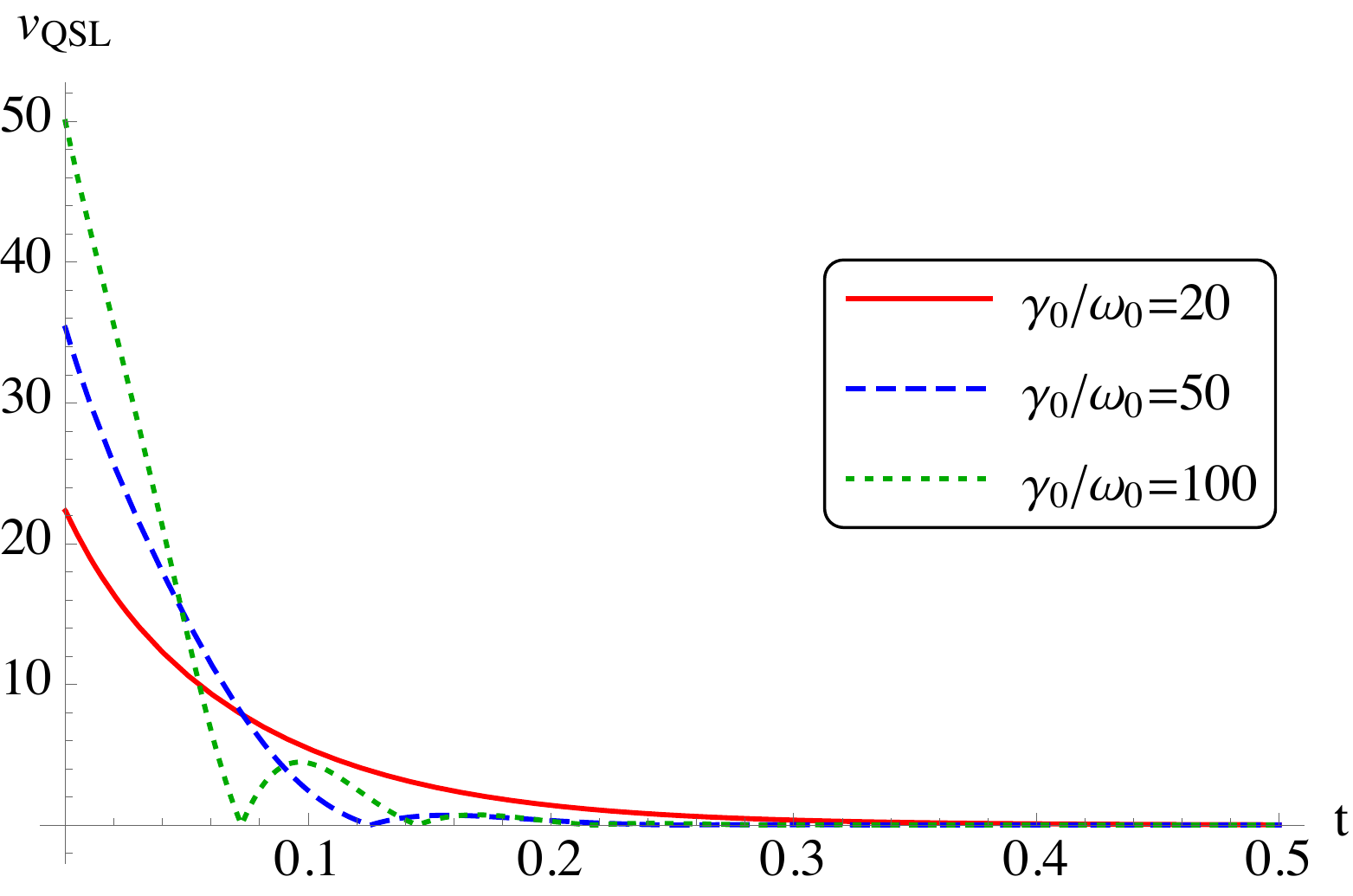}\\
\caption{\label{fig:non-Markov} Quantum speed limit time $\tau_\mrm{QSL}$ \eqref{geo:eq18} \revised{(upper panel, (a))} and quantum speed limit $v_\mrm{QSL}$ \eqref{geo:eq17} \revised{(lower panel, (b))} for the damped-Jaynes Cummings model \eqref{geo:eq27} and the initial state $\rho_0$ \eqref{geo:eq29a}. Parameters are $\tau=0.5$, $\hbar=1$ and $\lambda=50$.}
\end{figure}
In Fig.~\ref{fig:non-Markov} {\bf (a)} we plot $\tau_\mrm{QSL}$ in terms of the operator norm Eq.~\eqref{geo:eq18} as a function of the coupling strength $\gamma_0$ for the initial state 
\begin{equation}
\label{geo:eq29a}
\rho_0=\begin{pmatrix}
1 &0\\
0&0
\end{pmatrix}\,.
\end{equation}
As noted in Ref.~\cite{Deffner2013PRL} we observe that $\tau_\mrm{QSL}$ is indeed monotonically decreasing as a function of the coupling strength, $\gamma_0$. In the lower panel, we show the quantum speed limit $v_\mrm{QSL}$ \eqref{geo:eq17} for different values of $\gamma_0$. Notice that for strong coupling $v_\mrm{QSL}$ exhibits non-monotonic behaviour, which is a characteristic of non-Markovian dynamics.

This observation was made more precise by Xu \etals\cite{Xu2014}. Without loss of generality they chose the qubit to be initially prepared in its excited state, $\rho_0=\ket{1}\bra{1}$. Denoting by $P_t$ the population of the excited state one can then write,
\begin{equation}
\label{geo:eq30}
\tau_\mrm{QSL}=\frac{1-P_\tau}{1/\tau\,\int_0^\tau dt\, |\dot{P}_t|}\,.
\end{equation}
This representation of the quantum speed limit time is particularly elucidating since it is closely related to a measure of non-Markovianity introduced by Breuer \etals\cite{Breuer2009}.

Generally, quantifying non-Markovianity is complicated \cite{Brune1996,Madsen2011}. However, for simple quantum systems such as the Jaynes-Cummings model, Eq.~\eqref{geo:eq27}, the situation drastically simplifies. Breuer \etals\cite{Breuer2009} defined
\begin{equation}
\label{geo:eq31}
\mc{N}\equiv\max_{\rho_{1,2}(0)}\left\{\int_{\sigma>0} dt\, \sigma(t,\rho_{1,2}(0))\right\}
\end{equation}
where $\rho_{1,2}(0)$ are two initial states and
\begin{equation}
\label{geo:eq32}
\sigma(t,\rho_{1,2}(0))=\frac{1}{2}\frac{d}{dt}\|\rho_1(t)-\rho_2(t)\|_\mrm{tr}\,.
\end{equation}
It has been seen \cite{Breuer2009} that for Markovian dynamics all initial states monotonically converge towards a unique stationary state. Thus, $\sigma(t,\rho_{1,2}(0))$ in Eq.~\eqref{geo:eq32} is strictly negative and $\mc{N}=0$, Eq.~\eqref{geo:eq31}. Non-Markovian dynamics are characterised by an information backflow from the environment, and the convergence of $\rho(t)$ towards the stationary state is accompanied by oscillations, cf non-monotonic behaviour of $v_\mrm{QSL}$ in Fig.~\ref{fig:non-Markov} {\bf (b)}. Hence, $\sigma(t,\rho_{1,2}(0))$ in Eq.~\eqref{geo:eq32} can become positive, which amounts to finite values of $\mc{N}$.

It is then easy to see \cite{Xu2014} that the quantum speed limit time, Eq.~\eqref{geo:eq30} reads,
\begin{equation}
\label{geo:eq33}
\tau_\mrm{QSL}=\frac{\tau\,\left(1-P_\tau\right)}{2\,\mc{N}+\left(1-P_\tau\right)}\,.
\end{equation}
Equation~\eqref{geo:eq33} clearly demonstrates that the more non-Markovian the dynamics, the faster a quantum system can evolve. Similar conclusions where found by Zhang \etals for Ohmic spectral densities \cite{Zhang2014} \revised{and nonequilibrium environments \cite{Cai2017}}. However, it was also pointed out that the expression for $\tau_\mrm{QSL}$, and hence its behaviour sensitively depends on the choice of the initial state \cite{Xu2014a}, which was confirmed by Zhu and Xu \cite{Zhu2014}. Moreover, quantum speed-ups can also be achieved by a judicious choice of the external driving protocol \cite{Zhang2015}. Finally, Liu \etals\cite{Liu2016} showed that in addition to non-Markovianity, the preparation of the initial state, and the choice of the driving protocol, the characteristics of the energy spectrum govern the maximal speed of quantum evolution. \revised{Finally, it is worth noting that the quantum speed limit time also sets the time-scale over which decoherence is effective \cite{delCampo2017,Chenu2017,Beau2017arXiv}}. 

\subsubsection{Environment assisted speed-up in cavity QED.}

All this theoretical work posed the natural question whether \emph{environment assisted} speed-ups could be observed in an experiment. Motivated by the theoretical study of the Jaynes-Cummings model \cite{Deffner2013PRL}, Cimmarusti \etals\cite{Cimmarusti2015} set out to test whether a similar effect could be observed in cavity QED. To accomplish this feat, however, Cimmarusti \etals\cite{Cimmarusti2015} had to deviate from the conventional view on cavity QED systems. More specifically, Cimmarusti \etals\cite{Cimmarusti2015} considered the cavity field as the ``system'', and treated the atomic number that generates the atomic polarisation as a tuneable environment with a range of coupling constants. As a main result they found that varying the interaction strength of the optical cavity field with the environment by tuning the number of atoms modifies the time-dependent, nonclassical intensity correlation function. Moreover, the rate of evolution of the cavity field enhances -- speeds up -- with increasing the coupling. Finally, it was also shown \cite{Cimmarusti2015} that the cavity field undergoes non-Markovian dynamics, and that the environment assisted speed-up is an effect of the non-Markovianity. Hence, Cimmarusti \etals\cite{Cimmarusti2015} experimentally verified the theoretical predictions.

This experiment has inspired several other studies including the work by Xu \cite{Xu2016} on how to detect speed-ups in arbitrary systems and by Mo \etals on engineering multiple environments \cite{Mo2017}.

\subsection{Dynamics of multi-particle systems}

Non-Markovianity \revised{can arise} from collaborative excitations and correlations in the environment \cite{Hou2015} \revised{(see Ref. \cite{Vacchini2016} for a recent review of non-Markovian dynamics)}. Thus from studying the effect of correlations in its environment on the speed of a quantum system it is only natural to also analyse the influence of collaborative effects in the system itself. For instance, Liu \etals\cite{Liu2015} found that a open multi-qubit systems still exhibits the non-Markovian speed-up, but that also correlations intrinsic to the system can have the same effect. Similar effects were found in many level systems with avoided crossings by Poggi \etals\cite{Poggi2015,Poggi2015a}, and by Hou \etals for environments with star geometry \cite{Hou2015JPA} and finite XY spin chains \cite{Hou2017}. Moreover, Song \etals\cite{Song2016} found that these cooperative effects can be enhanced by a judicious choice of the external driving. Other speed-up mechanisms include criticality \cite{Wei2016,Heyl2017} and initial correlations between system and environment \cite{Yu2016}.

\paragraph{The Lipkin-Meshkov-Glick bath.}

When studying multi-particle quantum systems it becomes inevitable to also consider quantum phase transitions. Generically, quantum phase transitions are accompanied by a critical slowing down close to the critical point \cite{Wu2015,Francuz2016}. Therefore, the behavior of the quantum speed limit $v_\mrm{QSL}$ should be a good indicator of criticality. To make this observation more precise Hou \etals\cite{Hou2016} computed the quantum speed limit for a spin coupled to a Lipkin-Meshkov-Glick (LMG) bath. The LMG model was originally introduced to describe the tunneling of bosons between degenerate levels in nuclei \cite{Lipkin1965}, but it recently attracted attention as a testbed for shortcuts to adiabaticity \cite{Campbell2015LMG} and for quantum thermodynamics \cite{Campbell2016LMG}.

\begin{figure}
\centering
\includegraphics[width=.75\textwidth]{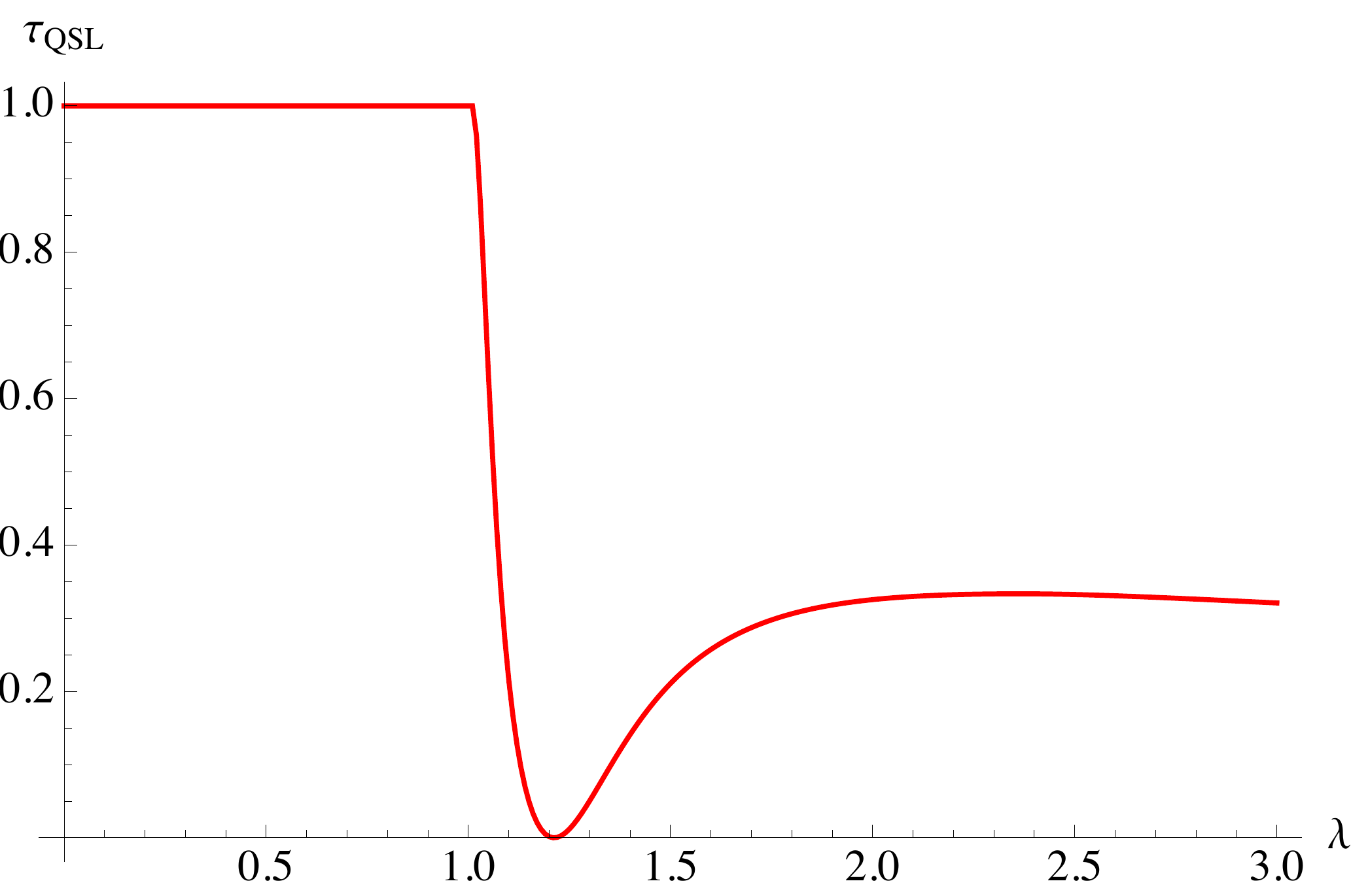}
\caption{Quantum speed limit time for a single qubit, initially in its excited state, weakly coupled to an LMG bath as given by Eq.~\eqref{geo:eq35}. We fix $\gamma=0.05$, $N=100$, and $\tau=1$. In the symmetry broken phase the probe evolves at the quantum speed limit, while in the symmetric phase the dynamics are significantly slower. Remarkably, near the critical point $\tau_\text{QSL} \to 0$ clearly spotlighting the quantum phase transition occurring in the bath.}
\label{fig:LMG} 
\end{figure}

In Hou \etals's \cite{Hou2016} analysis the total Hamiltonian is given by
\begin{equation}
\label{geo:eq34}
H=H_S+H_B+H_{SB}
\end{equation}
with
\begin{equation}
\label{geo:eq35}
\begin{split}
H_S&=-\sigma^z\,,\\
H_B&=-\frac{\lambda}{N}\sum^N_{i<j} \left(\sigma_i^x\sigma_j^x+\sigma_i^y\sigma_j^y\right)-\sum_{i=1}^N \sigma_i^z\,,\\
H_{SB}&=-\gamma\sum_{i=1}^N \left(\sigma_i^x\sigma^x+\sigma_i^y\sigma^y\right)\,,
\end{split}
\end{equation}
where $\sigma^\alpha$ and $\sigma_i^\alpha$, $\alpha=x,y,z$ are the Pauli matrices of the central spin and the $i$th spin of the bath. Note that $H_S$ can be understood as a probe system, which interacts through $H_{SB}$ with the LMG-model $H_B$. It is further easy to see \cite{Hou2016} that $H_B$ undergoes a quantum phase transition for $\lambda=1$, where the symmetry is broken for $0<\lambda<1$. The question is now, if the quantum speed limit for the probe system only, $\rho_S(t)=\ptr{B}{\rho_{tot}(t)}$, exhibits features of the phase transition in the environment. Hou \etals\cite{Hou2016} showed that this is indeed the case. Initialising the probe qubit in its excited state Hou \etals examined the corresponding quantum speed limit time, as defined by Deffner and Lutz~\cite{Deffner2013PRL}, fixing the evolution time $\tau=1$. By weakly coupling to the LMG environment and determining the corresponding $\tau_\text{QSL}$ they showed that if the bath is in the symmetry broken phase, the probe evolves at (or close to for larger systems) the quantum speed limit. This reflects the large spectral gap present in this phase, see for example Ref.~\cite{Campbell2016LMG}. Conversely, when the bath is characterised by the symmetric phase, where the spectral gap becomes vanishingly small with system size, $\tau_\text{QSL}$ witnesses a significant drop. Therefore we can conclude that the dynamics of the probe is now significantly slower than $\tau_\text{QSL}$. Interestingly we see $\tau_\text{QSL}\to 0$ in the vicinity of the critical point, thus reflecting the characteristic critical slowing down mentioned previously. These features are clearly shown in Fig.~\ref{fig:LMG}. Hou \etals further explored the relationship between the behaviour of the quantum speed limit and the non-Markovian features induced by the bath~\cite{Hou2016}.  

\subsection{Quantum speed limits for other quantities}

Conventionally, geometric quantum speed limits are derived as upper bounds on the rate of change of a geometric measure of distinguishability. In the preceding subsections we presented the case for the Bures angle, which is the generalised geometric angle between density operators. However, for many applications one is not necessarily interested in how exactly the quantum states evolve, but one rather needs a quick estimate of, for instance, the typical flow of entropy \cite{Hutter2012} or the rate of decoherence \cite{Gardas2016}. Thus, a plethora of other speed limits have been discussed in the literature. 

A particularly insightful treatment was proposed by Uzdin and Kosloff \cite{Uzdin2016}. Imagine that a system is initially prepared in a pure state, and then left to evolve under open system dynamics, for which the general master equation \eqref{geo:eq01} becomes
\begin{equation}
\label{geo:eq36}
L(\rho_t)=\frac{i}{\hbar}\,\com{H}{\rho_t}+\ex{i\phi(t)}\sum_k\left(A_k\,\rho_t\,A_k^\dagger-\frac{1}{2}\,\acom{A_k^\dagger A_k}{\rho_t}\right)\,,
\end{equation}
with $\phi(t)=0$ for Markovian systems. Then its rate of decoherence, or more generally its rate of thermalisation, is characterised by the dynamics of the purity 
\begin{equation}
\label{geo:eq37}
\mc{P}(t)=\trace{\rho_t^2}\,.
\end{equation}
Using the triangle inequality and the Cauchy-Schwarz inequality Uzdin and Kosloff \cite{Uzdin2016} then find 
\begin{equation}
\label{geo:eq38}
\left|\loge{\mc{P}(\tau)}-\loge{\mc{P}(0)}\right|\leq 4\int_0^\tau dt \sum_k\|A_k\|_\mrm{hs}\,.
\end{equation}
The latter can be re-written to define a quantum speed limit time,
\begin{equation}
\label{geo:eq39}
\tau_\mrm{QSL}^\mc{P}=\frac{\left|\loge{\mc{P}(\tau)}-\loge{\mc{P}(0)}\right|}{4/\tau\,\int_0^\tau dt \sum_k\|A_k\|_\mrm{hs}}.
\end{equation}
Equation~\eqref{geo:eq39} can be interpreted as a Mandelstam-Tamm type bound for the rate with which the purity, $\mc{P}(t)$, decays in open systems. \revised{Note that the denominator in Eq.~\eqref{geo:eq39} only depends on the Lindblad operators, and thus the resulting quantum speed limit time, $\tau_\mrm{QSL}^\mc{P}$, is independent of the time-dependent quantum states $\rho_t$.}

A Margolus-Levitin type bound follows from re-writing the density operator in Liouville space, i.e., $\rho_t$ is re-shaped as a vector in $\ket{\rho_t}\in\mbb{C}^{1\times N^2}$, and we have $\ket{\dot{\rho_t}}=\mc{H}\ket{\rho_t}$, where the Hamiltonian superoperator is given by \cite{Machnes2014}
\begin{equation}
\label{geo:eq40}
\mc{H}=\frac{i}{\hbar}\left(H\odot\mbb{I}-\mbb{I}\odot H\right)+\sum_k \left(A_k\odot A^\dagger-\frac{1}{2}\mbb{I}\odot A_k^\dagger A_k-\frac{1}{2} A_k^\dagger A_k\odot\mbb{I}\right)\,.
\end{equation}
Here the $\odot$-product is defined as \cite{Machnes2014},
\begin{equation}
\label{geo:eq41}
\left(A\odot B\right)\ket{\rho}\equiv \left(A\otimes \left(B^T\right)\right)\ket{\rho}\,.
\end{equation}
It then follows in complete analogy to above (see Eqs.~\eqref{geo:eq16}-\eqref{geo:eq18a}) that we have,
\begin{equation}
\label{geo:eq42}
\left|\loge{\mc{P}(\tau)}-\loge{\mc{P}(0)}\right|\leq \int_0^\tau dt \,\|\mc{H}-\mc{H}^\dagger\|_\mrm{op}
\end{equation}
and hence we obtain for the quantum speed limit time,
\begin{equation}
\label{geo:eq43}
\tau_\mrm{QSL}^\mc{P}=\frac{\left|\loge{\mc{P}(\tau)}-\loge{\mc{P}(0)}\right|}{1/\tau\,\int_0^\tau dt \,\|\mc{H}-\mc{H}^\dagger\|_\mrm{op}}\,.
\end{equation}
Accordingly a unified quantum speed limit time based on the dynamics of the purity becomes,
\begin{equation}
\label{geo:eq44}
\tau_\mrm{QSL}^\mc{P}\equiv \left|\loge{\mc{P}(\tau)}-\loge{\mc{P}(0)}\right|\ma{\frac{1}{4/\tau\,\int_0^\tau dt \sum_k\|A_k\|_\mrm{hs}},\frac{1}{1/\tau\,\int_0^\tau dt \,\|\mc{H}-\mc{H}^\dagger\|_\mrm{op}}}.
\end{equation}
Similarly to the case of speed limits based on the Bures angle \eqref{geo:eq24} Uzdin and Kosloff \cite{Uzdin2016} again find that the Margolus-Levitin type bound \eqref{geo:eq43} is tighter, but that the Mandelstam-Tamm type bound \eqref{geo:eq39} is much easier to compute. This is illustrated by Uzdin and Kosloff \cite{Uzdin2016} among other examples for dephasing channels and erasure of classical and quantum correlations.

The quantum speed limit based on the dynamics of the purity \eqref{geo:eq44} serves as an important and illustrative example of various studies. Other examples include: Dehdashti \etals\cite{Dehdashti2015} used the quantum speed limit based on the relative purity \cite{delcampo13} to study decoherence in the spin-deformed boson model, whereas Jing \etals\cite{Jing2016} quantified the generation of quantumness. On the other hand, Mondal \etals\cite{Mondal2016}, as well as Pires \etals\cite{Pires2015}, Marvian and Lidar \cite{Marvian2015}, and Marvian \etals\cite{Marvian2016} focused on the loss of quantum coherence in isolated and open dynamics. Finally, Sun \etals\cite{Sun2015} focused on computable bounds. 

\subsection{Universal, geometric quantum speed limits}

Comparing the expressions for the quantum speed limit time, $\tau_\mrm{QSL}$, based on the Bures angle \eqref{geo:eq24} and based on the purity Eq.~\eqref{geo:eq44}, as well as the expressions for the relative purity \cite{delcampo13} and the Winger-Yanase information \cite{Pires2016} we notice that the choice of the measure of distinguishability only determines the numerator.  The actual speed limit is set in all of these cases by a time-averaged  Schatten-$p$-norm of the generator of the dynamics. This observation can be made more precise, if one starts the derivation of the speed limit directly with such a norm \cite{Deffner2017}. 

To this end, consider the Schatten-$p$-distance
\begin{equation}
\label{geo:eq45}
\ell_p(\rho_t,\rho_0)=\|\,\rho_t-\rho_0\,\|_p\equiv\left(\trace{\left|\rho_t-\rho_0\right|^p}\right)^{1/p}\,,
\end{equation}
where $p$ is an arbitrary, positive, real number.  For $p=2$ we have the Hilbert-Schmidt distance, for $p=1$ the trace distance, and for $p=\infty$ the operator norm.

The universal geometric speed \eqref{geo:eq08} can be written as
\begin{equation}
\label{geo:eq46}
\dot{\ell}_p(\rho_t,\rho_0)=\left(\trace{\left|\rho_t-\rho_0\right|^p}\right)^{\frac{1}{p}-1}\,\trace{\left[\left(\rho_t-\rho_0\right)^2\right]^{\frac{p}{2}-1}\,\left(\rho_t-\rho_0\right)\,\dot{\rho}_t}\,.
\end{equation}
The latter expression looks rather involved, but it can be simplified by using, $\dot{\ell}_p(\rho_t,\rho_0)\leq |\dot{\ell}_p(\rho_t,\rho_0)| $ and employing the triangle inequality for operators, $|\trace{A}|\leq \trace{|A|}$, to read
\begin{equation}
\label{geo:eq47}
\dot{\ell}_p(\rho_t,\rho_0)\leq \left(\trace{\left|\rho_t-\rho_0\right|^p}\right)^{\frac{1}{p}-1} \trace{\left|\rho_t-\rho_0\right|^{p-1} \left|\dot{\rho}_t\right|}\,.
\end{equation}
Equation~\eqref{geo:eq47} can be further simplified with the help of H\"older's inequality \cite{Holder1889}
\begin{equation}
\label{geo:eq48}
\trace{\left|A B\right|}\leq \left(\trace{\left|A\right|^{q_1}}\right)^{1/q_1}\,\left(\trace{\left|B\right|^{q_2}}\right)^{1/q_2}
\end{equation}
which is true for all $1/q_1+1/q_2=1$. Now choosing $B=\dot{\rho}_t$ and $q_1=p/(p-1)$, for which $q_2=p$, we obtain
\begin{equation}
\label{geo:eq49}
\dot{\ell}_p(\rho_t,\rho_0)\leq \|\,L(\rho_t)\,\|_p\equiv v_\mrm{QSL}^p\,.
\end{equation}
where we defined a universal quantum speed limit, $v_\mrm{QSL}^p $ -- independent of the physical situation and considered quantities.

However, computing most of the Schatten-$p$-norms is rather involved, since the singular values of the generator of the dynamics have to be determined.  An equivalent, and much more tractable quantum speed limit can be derived, if one considers the quantum state in its Wigner representation instead,
\begin{equation}
\label{geo:eq50}
W(x,p)=\frac{1}{\pi\hbar}\,\int dy\,\bra{x+y}\rho\ket{x-y}\,\ex{-\frac{2 i p \,y}{\hbar}}\,.
\end{equation}
It was shown in Ref.~\cite{Deffner2017} that if one considers the rate of Wasserstein-$p$-distance 
\begin{equation}
\label{geo:eq51}
\mc{D}_p(W_t,W_0)=\|\,W_t-W_0\,\|_p\equiv\left(\int d\Gamma\,\left|W(\Gamma,t)-W_0(\Gamma)\right|^p\right)^{1/p}\,,
\end{equation}
one can derive
\begin{equation}
\label{geo:eq52}
\dot{\mc{D}}_p(W_t,W_0)\leq \|\,\dot{W}_t\,\|_p=v^W_\mrm{QSL}\,.
\end{equation}
The two expressions \eqref{geo:eq49} and \eqref{geo:eq52} are fully equivalent up to normalization \cite{Deffner2017}.

In conclusion, it was shown in Ref.~\cite{Deffner2017} that, independent of the choice of measure, the quantum speed is universally characterised by the Schatten-$p$-norm of the generator of the dynamics. Moreover, this Schatten-$p$-norm can be computed from the mathematically tractable Wasserstein-$p$-norm of the corresponding Wigner function.

\subsection{Applications: Shortcuts to adiabaticity\label{STAandQSL}}

We conclude this section on the geometric approach by highlighting an important application that was recognised only very recently. In the area of quantum control so-called shortcuts to adiabaticity have become a prominent and active topic \cite{Torrontegui2013}. A shortcut to adiabaticity is a fast process with the same final state that would result from infinitely slow driving. Within only a few years a huge variety of techniques has been developed \cite{Chen2010,Campo2012,Masuda2010,Masuda2011,Torrontegui2012,Torrontegui2012a,Masuda2014a,Kiely2015,Deffner2016,Demirplak2003,Demirplak2005,Berry2009,
Deffner2014,Jarzysnki2013,Patra2016,Chen2011a,Stefanatos2013,Campbell2015LMG,Wu2015,Saberi2014,PolkovnikovArXiv,Xiao2014,
Masuda2014,Torrontegui2014,Acconcia2015,Garaot2015,Kiely2016,Chen2010PRL}. 

Among all of these techniques transitionless quantum driving is unique. In its original formulation \cite{Demirplak2003,Demirplak2005,Berry2009} one considers a time-dependent Hamiltonian $H_0(t)$ and constructs an additional counterdiabatic field, $H_1(t)$, such that the joint Hamiltonian $H(t)=H_0(t)+H_1(t)$ drives the dynamics precisely through the adiabatic manifold of $H_0(t)$. Moreover, $H_1(t)$ vanishes by construction in the beginning, $t=0$, and and the end, $t=\tau$, of the finite time process. At first glance, it seems that such an energetically free shortcut to adiabaticity could be implemented for any arbitrarily fast process of arbitrarily short duration $\tau$. That this is not the case and how the quantum speed limit enters the picture was formalised by Campbell and Deffner \cite{Campbell2017}.

One can construct $H(t)$ such that the adiabatic solution of $H_0(t)$ is an exact solution of the dynamics generated by $H(t)$, and we have \cite{Demirplak2003,Demirplak2005,Berry2009,Zheng2016}
\begin{equation}
\label{geo:eq53}
H_1(t)= i \hbar \com{\pd_t \ket{n_t}\!\!\bra{n_t}}{\ket{n_t}\!\!\bra{n_t}}\,,
\end{equation}
where we denote the instantaneous eigenstates of $H_0(t)$ by $\ket{n_t}$. In Ref.~\cite{Zheng2016} a family of functionals has been proposed to quantify the cost associated with implementing $H_1(t)$. The simplest member of the family is given by the trace norm, $\|\cdot\|_\mrm{tr}$ \cite{Zheng2016,SarandySciRep2015,Santos2016,Coulamy2016},
\begin{equation}
\label{geo:eq54}
C^1_t \equiv C=\int_0^\tau\, dt\, \|H_1(t)\|_\mrm{tr},
\end{equation}
It is easy to see that for a single 2-level spin, $\partial_t C$ is proportional to the average power input~\cite{Zheng2016}, i.e., $H_1(t)$ reduces to an orthogonal, magnetic field. More generally, $C$ can be interpreted as the additional action arising from the counterdiabatic driving. Hence, the relation to the quantum speed limit becomes apparent, since loosely speaking the quantum speed limit time sets a lower bound on the action $E\,\tau_\mrm{QSL}\simeq \hbar/2$ \cite{Deffner2013}.

It is easy to see \cite{Campbell2017}, that in the case of transitionless quantum driving the instantaneous cost, i.e., the trace norm of the counterdiabatic Hamiltonian, $H_1(t)$, reduces to
\begin{equation}
\label{geo:eq56}
\pd_t C=\| H_1(t)\|_\mrm{tr}=\sqrt{\braket{\pd_t n_t}{\pd_t n_t}}\,,
\end{equation}
where we used that $\braket{\pd_t n_t}{n_t}=0$, which is true for all Hamiltonians with entirely discrete eigenvalue spectrum. By further noting that $\ket{\psi_t}=\ket{n_t}$ and $H(t)=H_0(t)+H_1(t)$ with $H_1(t)$ as given in Eq.~\eqref{geo:eq53}. To determine the speed we require
\begin{equation}
\label{geo:eq57}
\epsilon_t = \| H(t) \rho \| =\sqrt{\varepsilon_n^2(t)+\braket{\pd_t n_t}{\pd_t n_t}}\,,
\end{equation}
where we employed again $\braket{\pd_t n_t}{n_t}=0$  and where $\varepsilon_n(t)$ are the instantaneous eigenvalues.

Collecting Eqs.~\eqref{geo:eq54}-\eqref{geo:eq57} the maximal speed becomes
\begin{equation}
\label{geo:eq58}
v_\mrm{QSL}=\frac{\sqrt{\varepsilon_n^2(t)+\left(\pd_t C\right)^2}}{\hbar\,\co{\mc{L}_t} \si{\mc{L}_t}}\,,
\end{equation}
and the corresponding quantum speed limit time becomes
\begin{equation}
\label{geo:eq59}
\tau_\mrm{QSL}=\frac{\hbar\tau\,\left[\si{\mc{L}_\tau}\right]^2}{2\int_0^\tau{ dt\,\sqrt{\varepsilon_n^2(t)+\left(\pd_t C\right)^2}}}\,.
\end{equation}
Equations~\eqref{geo:eq58} and \eqref{geo:eq59} express, in a transparent and immediate way, the trade-off of speed and cost of a shortcut to adiabaticity. The faster a quantum system evolves along its adiabatic manifold, the higher is the cost of implementing the shortcut.

Related results were found by Santos and Sarandy \cite{SarandySciRep2015,Santos2016} in the context of quantum computation and by Funo \etals\cite{Funo2017} from the work fluctuations during the finite-time process. 
 
%%%%%%%%%%%%%%%%%%%%%%%%%%%%%%%%%%%%%%%%%%%%%%%%%%%%%%%%%%%%%%%%%%%%%%%%%%%%%%%%%
%\input{chapters/non.include.tex}

\section{Quantum speed limits in non-Schr\"odinger quantum mechanics\label{nonstandard}}

In the above sections we have been aiming to provide an overview over the various notions of a quantum speed limit in standard quantum mechanics. So far, we have always assumed that the dynamics of an isolated systems is described by the von-Neumann equation \eqref{hist:eq06}, and that open systems can be described by a quantum master equation \eqref{geo:eq01}. 

In this section we will broaden the scope of the analysis, and discuss generalisations and applications of the quantum speed limit for dynamics that are beyond the applicability of the Schr\"odinger equation. 

\subsection{Relativistic systems and Dirac dynamics}

As a first example we are interested in relativistic quantum dynamics described by the Dirac equation,
\begin{equation}
\label{non:eq01}
i\hbar\, \dot{\Psi}(\mb{x},t)=\left(-i\hbar c\, \mb{\alpha}\cdot\nabla+ e c\,\mb{\alpha}\cdot\mb{A}(\mb{x})+\alpha_0\,mc^2\right)\,\Psi(\mb{x},t)\,.
\end{equation}
Here, $\Psi(\mb{x},t)$ is the wave function of an electron with rest mass $m$ and elementary charge $e$ at position $\mb{x}=(x_1,x_2,x_3)$, and $c$ is the speed of light. Finally, $\mb{A}(\mb{x})$ is the vector-potential, with $\mb{B}(\mb{x})=\nabla \times \mb{A}(\mb{x})$. In covariant form the matrices $\mb{\alpha}=(\alpha_1,\alpha_2,\alpha_3)$ and $\alpha_0$ can be expressed as \cite{Peskin1995,Thaller1956},
\begin{equation}
\label{non:eq02}
\alpha^0=\gamma^0\quad\mathrm{and}\quad\gamma^0\, \alpha^k=\gamma^k\,.
\end{equation}
The $\gamma$-matrices are commonly written in terms of $2\times 2$ sub-matrices with the Pauli-matrices $\sigma_x, \sigma_y, \sigma_z$ and the identity $\id_2$ as,
\begin{eqnarray}
\label{non:eq03}
\fl
\gamma^0=\begin{pmatrix}\id_2&0\\0&-\id_2 \end{pmatrix}\quad&\gamma^1=\begin{pmatrix}0&\sigma_x\\-\sigma_x&0 \end{pmatrix}\quad
\gamma^2=\begin{pmatrix}0&\sigma_y\\-\sigma_y&0 \end{pmatrix}\quad&\gamma^3=\begin{pmatrix}0&\sigma_z\\-\sigma_z&0 \end{pmatrix}\,.
\end{eqnarray}
The solution of Eq.~\eqref{non:eq01}, the Dirac wave function $\Psi(\mb{x},t)$, is a bispinor, which can be interpreted as a superposition of a spin-up electron, a spin-down electron, a spin-up positron, and a spin-down positron \cite{Thaller1956,Peskin1995}.

Since its inception \cite{Dirac1928} the Dirac equation \eqref{non:eq01} has proven to be one of the most versatile results of theoretical physics.  Originally it was designed as a relativistic wave equation to describe massive spin-$1/2$ particles, such as electrons and quarks \cite{Thaller1956,Peskin1995}. Up to today, the Dirac equation enjoys plenty of attention  \cite{Pickl2008,Fillion2012,Fillion2013,Fillion2013a}, since it also found application in opto-mechanics \cite{Schmidt2015}, in quantum thermodynamics \cite{Deffner2015}, and in condensed matter physics to describe so-called Dirac materials \cite{Wehling2014,Fillion2015}.

It is only natural to ask whether and how the quantum speed limit generalises to relativistic Dirac dynamics. To this end, Villamizar and Duzzioni \cite{Villamizar2015} further asked two specific questions: (i) which quantum states allow for the greatest spatial displacement in the shortest time, and (ii) can the relativistic quantum speed limit be used to (in-)validate a model for quantum dynamics.

To keep things as simple as possible Villamizar and Duzzioni \cite{Villamizar2015} then considered systems in the $x$-$y$ plane with $\mb{B}(\mb{x})=B\,\hat{z}$, and an initial state that is is a homogeneous superposition of two radial eigenstates in different Landau energy levels. Thus, the initial states is of the form Eq.~\eqref{hist:eq15}, for which the Mandelstam-Tamm and the Margolus-Levitin bound become identical and tight, and $\tau_\mrm{QSL}$ \eqref{hist:eq14} is the minimal time for the initial state to evolve to an orthogonal state. 

If the dynamics was described by a Schr\"odinger equation we would have \cite{Villamizar2015}
\begin{equation}
\label{non:eq04}
\tau^S_\mrm{QSL}=\frac{\pi\hbar}{2(\la H\ra-E_0)}=\frac{\pi m}{e B}\,,
\end{equation}
where $E_0$ denotes the ground state energy. For the relativistic Dirac dynamics \eqref{non:eq01} we obtain \cite{Villamizar2015}
\begin{equation}
\label{non:eq05}
\tau^D_\mrm{QSL}=\frac{\pi\hbar}{\sqrt{(m c^2)^2+4\hbar c^2e\,B}-\sqrt{(m c^2)^2+2\hbar c^2e\,B}}\simeq \frac{\pi}{2 c\beta\left(\sqrt{2}-1\right)}
\end{equation}
where the second equality holds for strong magnetic fields, $\hbar c^2e\,B\gg(m c^2)^2$, and where $\beta=\sqrt{eB/2\hbar}$. \revised{Note that in the non-relativistic limit $\tau^S_\mrm{QSL}$ and $\tau^D_\mrm{QSL}$ become equivalent.}

It is then a simple exercise \cite{Villamizar2015} to also compute the mean displacement, which can be written for Schr\"odinger dynamics as
\begin{equation}
\label{non:eq06}
\la \Delta r^S\ra=\sqrt{\frac{\pi\hbar}{2e\, B}}
\end{equation}
and for Dirac dynamics \revised{in the ultra-relativistic limit}
\begin{equation}
\label{non:eq07}
\la \Delta r^D\ra=\frac{\sqrt{\pi}}{4\beta}\left(1+\frac{3}{2\sqrt{2}}\right)\,.
\end{equation}
Combining the expressions for the quantum speed limit time Eqs.~\eqref{non:eq04}-\eqref{non:eq05} with the average displacements \eqref{non:eq06} and \eqref{non:eq06}, we can define the average speed of an electron
\begin{equation}
\label{non:eq08}
\la v^S\ra_S\equiv\frac{\la \Delta r^S\ra}{\tau^S_\mrm{QSL}}=\frac{1}{m}\sqrt{\frac{e B\hbar}{2\pi}}\quad\mrm{and}\quad \la v^D\ra\equiv\frac{\la \Delta r^D\ra}{\tau^D_\mrm{QSL}}=\frac{c}{4\sqrt{2\pi}}\left(1+\sqrt{2}\right)\,.
\end{equation}
We immediately see that for strong magnetic fields, $B\gg1$, the average speed computed from the Schr\"odinger dynamics can surpass the speed of light, $c$. As expected, this is not the case for the Dirac equation, and even for arbitrarily strong fields $B$ we always have $\la v^D\ra<c$. Thus, Villamizar and Duzzioni \cite{Villamizar2015} conclude that the quantum speed limit can be used to check the physical consistency of a physical model. Villamizar and Duzzioni 's\cite{Villamizar2015} work has been followed by several further analyses for which we refer the reader to the literature \cite{Tian2015,Khan2015,Wang2017,Wang2017a}

Finally, it is interesting to note that in passing Villamizar and Duzzioni \cite{Villamizar2015} resolved the longstanding debate of Einstein and Bohr (see Sec.~\ref{intro}) by bringing  the energy-time uncertainty principle in full agreement with special relativity.

\subsection{PT-symmetric quantum mechanics}

Another area of research that has attracted considerable attention are non-Hermitian \cite{Mostafazadeh2010,Berry2011JO,Moiseyev2011,Gardas2016Coriolis,Gardas2016thermo} and $\mc{PT}$-symmetric quantum systems \cite{Bender1998,Bender2007,Deffner2015PRL,Gardas2016,Bender2017}. Since its originial development as a rather mathematical theory \cite{Bender1998,Bender2013}, $\mc{PT}$-symmetric quantum mechanics has found experimental realisation  as systems with balanced loss and gain \cite{Bender2003,Rushhaupt2005,Klaiman2008,Makris2008,Musslimani2008,Ruter2010,Bender2013}. 

In $\mc{PT}$-symmetric quantum mechanics the Hamiltonian $H$ is non-Hermitian, but commutes with $\mc{P}$ and $\mc{T}$, i.e., $\left[ \mc{PT},H\right]=0$. Here, $\mc{P}$ is the space reflection (parity) operator, and $\mc{T}$ is the time-reflection operator \cite{Bender2007},
\begin{equation}
\label{non:eq09}
\begin{split}
&\mc{P}\,x\,\mc{P}=-x\quad \mathrm{and}\quad \mc{P}\,p\,\mc{P}=-p\\
&\mc{T}\,x\,\mc{T}=x,\quad \mc{T}\,p\,\mc{T}=-p\quad \mathrm{and}\quad \mc{T}\,i\,\mc{T}=-i\,
\end{split}
\end{equation}
where $x$ and $p$ are position and momentum operators, respectively.  Since $\mc{T}$ also changes the sign of the imaginary unit $i$, canonical commutation relations such as $\com{x}{p}=i\hbar$ are invariant under $\mc{PT}$.

The major difference between Hermitian and $\mc{PT}$-symmetric quantum mechanics is the definition of the inner product \cite{Bender2002,Bender2007}. For Hermitian Hamiltonians we have,
\begin{equation}
\label{non:eq10}
\braket{\psi_1}{\psi_2}=\psi_1^\dagger \cdot \psi_2\,,
\end{equation}
where, as usual, $\dagger$ denotes conjugate transpose. This inner product, however, results in indefinite norms for the non-Hermitian, but $\mc{PT}$-symmetric case \cite{Bender2007}. This observation will become crucial below, since this means that for the geometric approach the angle Eq.~\eqref{geo:eq03} will have to be re-defined.

In $\mc{PT}$-symmetric quantum mechanics the inner product is defined in terms of the metric operator $\mc{C}$ as \cite{Bender2007}
\begin{equation}
\label{non:eq11}
\braket{\psi_1}{\psi_2}_\mc{CPT}=\left(\mc{CPT}\psi_1\right) \cdot \psi_2\,.
\end{equation}
In the unbroken regime, i.e., if the eigenvalues of $H$ are real, $\mc{C}$ can be determined from \cite{Bender2007},
\begin{equation}
\label{non:eq12}
\left[\mc{C},H\right]=0 \quad\mathrm{and}\quad \mc{C}^2=\id\,.
\end{equation}
Note that the time evolution induced by a time-independent Hamiltonian with real eigenvalues is unitary. Thus, one would naively expect that both the Mandelstam-Tamm \eqref{hist:eq09} as well as the Margolus-Levitin bound \eqref{hist:eq13} for time-dependent generators to remain valid.

To analyse this claim Bender \etals \cite{Bender2007a} considered a simple case study. Imagine a qubit, which is initially prepared in $\ket{\psi_0}$ and we want to drive it to the final state $\ket{\psi_T}$ with
\begin{equation}
\label{non:eq13}
\ket{\psi_0}=\begin{pmatrix} 1 \\0 \end{pmatrix}\quad \mrm{and}\quad \ket{\psi_T}= \begin{pmatrix} a \\b \end{pmatrix}=e^{i H t/\hbar}\,\begin{pmatrix} 1 \\0 \end{pmatrix}\,.
\end{equation}
The most general, but $\mc{PT}$-symmetric Hamiltonian can be written as
\begin{equation}
\label{non:eq14}
H=\begin{pmatrix}
r\,e^{i\theta}&s\\
s&r\, e^{-i\theta}
\end{pmatrix}
\end{equation}
and thus the final state, $\ket{\psi_T}$, reads \cite{Bender2007a}
\begin{equation}
\label{non:eq15}
\ket{\psi_T}=\frac{e^{-itr\co{\theta}/\hbar}}{\co{\alpha}}\,\begin{pmatrix} \co{\frac{\omega t}{2\hbar}-\alpha} \\-i \si{\frac{\omega t}{2\hbar}} \end{pmatrix}\,,
\end{equation}
where $\alpha$ and $\omega$ are determined by $\si{\alpha}=r \si{\theta}/s$ and $\omega^2=4s^2-4 r^2\sin^2{\left(\theta\right)}$, respectively. 

Bender \etals \cite{Bender2007a} then argued that $\mc{PT}$-symmetric Hamiltonians provide means to accelerate quantum mechanics. For instance, if we chose $a=0$ and $b=1$ in Eq.~\eqref{non:eq13} then Eq.~\eqref{non:eq15} suggests that the quantum system could evolve from $\ket{\psi_0}$ to $\ket{\psi_T} $ in arbitrarily short times. However, Bender \etals \cite{Bender2007a} also note that such $\ket{\psi_0}$ and $\ket{\psi_T} $ are no longer orthogonal, since under $\mc{PT}$-symmetric Hamiltonians the inner product, and hence the geometry of Hilbert space has to be modified \eqref{non:eq12}.

Thus, Uzdin \etals \cite{Uzdin2012} revisited the geometric approach to the quantum speed limit. As a starting point of their analysis they consider the angle $\mc{L}$ between two arbitrary, complex vectors $\ket{\psi_1}$ and $\ket{\psi_2}$
\begin{equation}
\label{non:eq16}
\co{\mc{L}}=\frac{|\braket{\psi_1}{\psi_2}|}{\sqrt{\braket{\psi_1}{\psi_1}}\sqrt{\braket{\psi_2}{\psi_2}}}\,.
\end{equation}
Note that generally $\ket{\psi_1}$ and $\ket{\psi_2}$ are not normalised. In complete analogy to above (see Eqs.~\eqref{hist:eq23}-\eqref{hist:eq27}) it can then be shown that
\begin{equation}
\label{non:eq17}
\left(\frac{d\mc{L}}{dt}\right)^2=\bra{\Psi}H^\dagger(t)H(t)\ket{\Psi}-\bra{\Psi}H^\dagger(t)\ket{\Psi}\bra{\Psi}H(t)\ket{\Psi}\,,
\end{equation}
where $\ket{\Psi}\equiv\ket{\psi(t)}/\sqrt{\braket{\psi(t)}{\psi(t)}}$. Repeating the by now familiar steps Uzdin \etals \cite{Uzdin2012} find that a quantum speed limit for non-Hermitian systems can be written in complete analogy to Hermitian systems as
\begin{equation}
\label{non:eq18}
\left| \frac{d\mc{L}}{dt}\right|\leq v_\mrm{QSL}=\|\mc{H}\|_\mrm{op}\,,
\end{equation}
where we here have
\begin{equation}
\label{non:eq19}
\mc{H}=H-\mu\, \id\quad\mrm{and}\quad \mu=\trace{H}/N\,,
\end{equation}
and $N$ is the dimension of the Hilbert space of $H$.  In their analysis Uzdin \etals \cite{Uzdin2012} then use Eq.~\eqref{non:eq18} to identify time-optimal Hamiltonians, for which the fastest evolution between specific initial and final states can be achieved. 

Identifying time-optimal Hamiltonians is sometimes also dubbed the \emph{quantum brachistochrone problem}, which has been further elaborated in the literature for Hermitian \cite{Frydryszak2008,Borras2008,Kuzmak2013,Kuzmak2015,Russel2014,Russel2015,Villamizar2017} as well as non-Hermitian systems \cite{Mostafazadeh2007,Assis2008,Giri2008,Gunther2008,Gunther2008PRL,Mostafazadeh2009,Bender2009}.

\subsection{Non-linear systems}

Finally, the notion of quantum speed limits has also been generalised to non-linear systems. Non-linear Schr\"odinger equations arise from effective descriptions of many-particle systems, such as Bose-Einstein condensates (BEC) \cite{Huang2009} or in non-linear optics \cite{Shen1984}. Consequently, Chen \etals \cite{Xi2016} generalised Hergerfeldt's earlier work \cite{Hegerfeldt2013,Hegertfeldt2014} to two models, namely one describing standard BECs,
\begin{equation}
\label{non:eq20}
i\hbar\, \frac{d}{dt}\ket{\psi}=\left[\left(\Gamma(t)+\kappa \left(|\psi_1|^2-|\psi_2|^2\right)\right)\,\sigma_z+\omega(t)\,\sigma_x\right]\ket{\psi}
\end{equation}
where where $\psi_1$ and $\psi_2$ are probability amplitudes of two possible states, $\kappa$ describes the interaction between atoms, and $\Gamma(t)$ and $\omega(t)$ energy bias and coupling strength. The second model is more involved and describes oscillations between atomic and molecular BECs. 

It had been claimed by Dou \etals \cite{Dou2014} that nonlinear interactions strongly affect the minimal time a quantum system needs to evolve between distinct states. This claim was refuted by Chen \etals \cite{Xi2016}, and they showed explicitly that the maximal quantum speed of nonlinear systems for unconstrained controls is independent of the strength of the non-linearity, and that non-linear systems at maximal speed evolve as fast as equivalent linear systems. In particular, they found for unconstrained control protocols $\Gamma(t)$,
\begin{equation}
\label{non:eq21}
\tau_\mrm{min}=\frac{\left|\theta_i-\theta_f\right|}{2 \omega_0}\,,
\end{equation}
where $\theta_i$ and $\theta_f$ are initial and final azimuth on the Bloch sphere. Note that $\tau_\mrm{min}$ is independent of $\kappa$ and that $\tau_\mrm{min}$ is similar to the result for linear qubits with more than one control parameter \eqref{eq:Tmin}.

%%%%%%%%%%%%%%%%%%%%%%%%%%%%%%%%%%%%%%%%%%%%%%%%%%%%%%%%%%%%%%%%%%%%%%%%%%%%%%%%%

%\input{chapters/future.include.tex}

\section{Relation to other fundamental bounds}
\label{future}

\subsection{Cramer-Rao bound and beyond\label{sec:CR}}
From Sec.~\ref{quantummetrology} we have already established that a strict relationship exists between certain formulations of the quantum speed limit and the quantum Fisher information, and therefore with the quantum Cramer-Rao bound. As alluded to earlier, the existence of a tight relationship is natural: Heisenberg's uncertainty principles limits the accuracy with which we can measure conjugate variables, while the Cramer-Rao bound tells us that no procedure for estimating the value of a given quantity can have a precision scaling better than the inverse of the standard deviation of a conjugate quantity. Therefore, while the quantum speed limit arises by virtue of the energy-time uncertainty relation, in principle we can use the same reasoning for any pair of conjugate variables. This approach was  recently followed in order to re-examine the measurement precision bounds by Giovannetti, Lloyd, and Maccone~\cite{Giovannetti2012}. Interestingly, while the Cramer-Rao bound is asymptotically tight (in the number of measurements performed), for moderate numbers of measurements this line of reasoning leads to a tighter bound.

If the parameter to estimate, $\mu$, is encoded in some state, the quantum fluctuations associated with this can be connected to a conjugate operator, $H$, that generates translations of $\mu$, $U_\mu = e^{-i \mu H}$. We remark, $H$ can be, although is not necessarily, the Hamiltonian of the system at hand. Assuming there are $M$ copies of the state, along with pure probe states and a unitary encoding scheme, we can establish the quantum Cramer-Rao bound takes the form
\begin{equation}
\label{eq:CR}
\Delta \mu \geq \frac{1}{2\sqrt{M} \Delta H}.
\end{equation}
In Ref.~\cite{Giovannetti2012} the authors derive an alternative bound on the root mean squared error for $\mu$
\begin{equation}
\label{eq:GLM}
\Delta \mu \geq \frac{\kappa}{M(\langle H \rangle - E_0)},
\end{equation}
where $E_0$ is the minimum eigenvalue of $H$ and $\kappa$ is a constant. To establish Eq.~\eqref{eq:GLM} the authors use the fact that a ``quantum speed limit" in this instance, which links the fidelity, $F$, of the states for values of the parameter $\mu$ and $\mu'$ to the difference $\mu'-\mu$ through the mapping $U_\mu$, is given by
\begin{equation}
| \mu' - \mu | \geq \frac{\pi}{2} \text{max} \left[ \frac{\alpha(F)}{M(\langle H \rangle - E_0)} , \frac{\beta(F)}{\sqrt{M} \Delta H} \right]
\end{equation}
where $\alpha(F)\approx \beta^2(F) = 4\arccos^2(\sqrt{F})/\pi^2$ from Ref.~\cite{Giovannetti2003}. These functions can then be bounded and optimised to show Eq.~\eqref{eq:GLM}~\cite{Giovannetti2012}. 

Of course, in the situation where the parameter to estimate is time and $H$ corresponds to the system Hamiltonian, then the analysis reduces to the ``standard" quantum speed limit. However, the remarkable insight of Ref.~\cite{Giovannetti2012} is that for any pair of conjugate variables this reasoning can be used to establish tight bounds on the precision with which they can be estimated.
 
\subsection{Landauer's bound}
\label{subsect:Landauer}
It is widely accepted that information is not an abstract entity, but rather a physical quantity. In particular, Landauer established that in order to erase a single bit of information there is an associated minimum energetic cost, given by the dissipated heat~\cite{Landauer1961,Deffner2013PRX}
\begin{equation}
\label{eq:landauer}
\mathcal{Q} = k_B T \ln 2.
\end{equation}
While the ramifications of this have lead to many significant insights regarding computational costs, the physical nature of information and a host of others, our interest lies assessing its recent experimental verification. The first experiment was performed by B\'erut \etals~\cite{Berut2012}, where they used a single colloidal particle trapped in a time-dependent double well potential. By modulating and tilting the potential slowly enough they were able to achieve Landauer limited bit erasure. Since then several other experiments have tested and verified Eq.~\eqref{eq:landauer} in a variety of settings both classical and quantum~\cite{Jun2014,PetersonPRSA,Ciampini2017,Gaudenzi2017}. However, most of these experiments shared one common aspect, namely that the erasure processes was performed slowly in order to avoid generating excess excitations and therefore costing more energy. Considering the discussions of Sec.~\ref{quantumcomput}, since Landauer's bound sets a minimal energy to transition between two orthogonal states, it therefore seems to naturally lend itself to exploring the quantum speed limit, provided the information is encoded into a quantum system. In fact, as shown in Ref.~\cite{Berut2012} where the erasure protocol was classical, they approach the Landauer limit only asymptotically with increasing cycle duration.

The recent experiment by Gaudenzi \etals~\cite{Gaudenzi2017}, however, explicitly considered a quantum set-up. In this case, Landauer's bound sets the minimal energy and the quantum speed limit, essentially independently, sets the minimal time to achieve the erasure. Using the energy-time uncertainty relation they established that the quantum limit is given by
\begin{equation}
\mathcal{Q} \cdot \tau_\text{QSL} = \frac{\pi \hbar}{2}
\end{equation}
By exploiting quantum tunnelling to realise a carefully controlled evolution, they were able to achieve  Landauer limited bit erasure several orders of magnitude faster (although not yet at the quantum speed limit) than any previous implementations.

\subsection{Lieb-Robinson bound \label{sec:LR}}
We saw in Sec.~\ref{quantumcomm} that the quantum speed limit fundamentally bounds the rate at which a quantum state can be communicated along a quantum channel. Indeed, while the emergence of $\tau_\text{QSL}$ is remarkable, the fact that information cannot propagate arbitrarily fast when the interaction range is limited is quite intuitive. In fact, this notion is formalised in the form of Lieb-Robinson bounds~\cite{Lieb1972}. In essence, these bounds set constraints on the speed with which local effects can propagate through a quantum system by rigorously showing that correlations between the expectation values of local observables are exponentially suppressed outside an effective ``space-time cone". We refer to Ref.~\cite{Kliesch2014} for an introduction to the more formal aspects of Lieb-Robinson bounds and some of their more far reaching implications. 

If we consider only nearest-neighbour interactions, then the speed with which correlations can travel along a one-dimensional chain is bounded linearly by the so-called Lieb-Robinson velocity, $v_\text{LR}$. This behaviour was recently verified experimentally in Ref.~\cite{Cheneau2012}, while the analysis for long range interactions, where such bounds can be violated is found in Ref.~\cite{richerme14}. To the best of our knowledge, it is currently an open question regarding a more stringent relation between $v_\text{LR}$ and the speed arising from the quantum speed limit, $v_\text{QSL}$.

\subsection{Maximal rate of quantum learning and Holevo's information\label{sec:learn}}

As a final example we return to the Bremerman-Bekenstein bound \eqref{time:eq01}. In its original formulation it gives an upper bound on the rate with which information can be extracted from a quantum system. However, the bound is rather weak, since it assumes that the total information stored in the quantum system is accessible. Generally this is not the case \cite{Groenewold1971} due to the quantum back action of measurements \cite{Nielsen2002}. 

Imagine that we have an observable $A=\sum_n a_n \Pi_n$, where $a_n$ are the measurement outcomes, and $\Pi_n$ are the projectors into the eigenspaces corresponding to $a_n$. Generally, the post-measurement quantum state $\rho$ will suffer from a back-action, i.e, information about the quantum system will be lost in the measurement. How much information is lost is quantified by Holevo's information,
\begin{equation}
\label{holevo:eq01}
\chi=S(\rho)-\sum_n \mf{p}_n S(\rho_n)
\end{equation}
where $S(\rho)=-\trace{\rho \loge{\rho}}$ is the von-Neumann entropy, and $\rho_n=\Pi_n \rho \Pi_n$ is the post-measurement state. Further, $\mf{p}_n=\trace{\Pi_n \rho}$ denotes the probability to obtain the $n$th measurement outcome. 

Very recently, Acconcia and Deffner \cite{Acconcia2017} then used $\chi$ to define the maximal rate of quantum learning. If $\chi$ is the accessible information, then a ``rate of learning'' quantifies the change of $\chi$ under a time-dependent perturbation. In complete analogy to the Bremerman-Bekenstein bound one can then find an upper bound on the rate with which the accessible information changes by
\begin{equation}
\label{holevo:eq02}
\dot{\chi}\leq \frac{\Delta\chi}{\tau_\mrm{QSL}}\,.
\end{equation}
In Ref.~\cite{Acconcia2017} this bound was studied for the harmonic oscillator and for the P\"oschl-Teller potential by means of time-dependent perturbation theory.

%%%%%%%%%%%%%%%%%%%%%%%%%%%%%%%%%%%%%%%%%%%%%%%%%%%%%%%%%%%%%%%%%%%%%%%%%%%%%%%%%

%\input{chapters/con.include.tex}

\section{Final remarks}
Throughout this Topical Review we have striven to focus on the practical applications of the quantum speed limit. In particular, with the rapid development of new quantum technologies and the recent surge of interest in exploring the thermodynamic working principles of quantum systems it is important to understand the limits on controlling such systems. Arguably the most basic question one can ask is: how fast can we achieve a desired outcome? To which one then naturally would attempt to reduce the time or, seemingly equivalently, speed up the process. Interestingly, these two approaches actually allow us to formulate and study the quantum speed limit in very different settings. In the former approach, while the establishment of the quantum speed limit time arises from the energy-time uncertainty principle, it is quite remarkable that precisely this time emerges as the minimal evolution time for optimised quantum systems. Employing the latter approach, based on the geometry of a quantum evolution, we were able to identify new and unexpected control mechanisms. In most set-ups for quantum computation one works hard to isolate the system against noise and decoherence from the environment. The geometric approach to quantum speed limits, however, indicated environmental correlations as a valuable resource to speed-up quantum processes. The implications are far reaching, in particular considering the development of quantum computers and quantum enhanced thermal devices, for which the quantum speed limit is crucial in characterising the associated limitations. 

While of fundamental interest, arguably it is the practical ramifications of the energy-time uncertainty principle that has lead to the renewed interest. The diverse range of techniques to control quantum systems has allowed us to exploit the quantum speed limit to understand the limits of metrology, features of criticality, define the efficiency of quantum devices, and understand the cost of controlled quantum evolutions. Undoubtedly there are many more revelations to come, for instance in further developing the notion in non-standard quantum mechanics or a clearer understanding for complex multipartite systems. As we have tried to evidence in this Topical Review, the seeming ubiquity of the quantum speed limit in controlling quantum systems indicates more is to be learned.

We close with a more philosophical remark: Quantum speed limits are an inherently quantum phenomenon with no classical analogue. The ramifications of such counter-intuitive, yet fundamental bounds of physical reality have puzzled the greatest among us -- as highlighted by the long debates by Einstein and Bohr in the XX century. Over the last century, however, quantum theory has undoubtedly become \emph{the} cornerstone of physics and its ``peculiarities'' are what make the Universe work. To say it in Zurek's words \cite{Lockwood1996}:
\begin{quote}
\emph{The only `failure' of quantum theory is its inability to provide a natural framework for our prejudices about the workings of the Universe.}
\end{quote}

%%%%%%%%%%%%%%%%%%%%%%%%%%%%%%%%%%%%%%%%%%%%%%%%%%%%%%%%%%%%%%%%%%%%%%%%%%%%%%%%%
\ack{We are grateful to Marta Paczy\'nska for creating the visual representation of Einstein's gedankenexperiment, Fig.~\ref{fig:intro}, and Lu (Lucy) Hou for providing the resources for Fig.~\ref{fig:LMG}. SD would like to thank Eric Lutz for many years of insightful discussions and supporting mentorship, and in particular for inciting our interest in quantum speed limits. This work was supported by the U.S. National Science Foundation under Grant No. CHE-1648973.}

\section*{References}

\bibliographystyle{unsrt}
\bibliography{qsl}

\begin{thebibliography}{100}

\bibitem{Born1971}
I.~Born, editor.
\newblock {\em The Born-Einstein Letters}.
\newblock Walker and Company, New York, 1971.
\newblock Letter to Max Born (4 December 1926).

\bibitem{Heisenberg1927}
W.~Heisenberg.
\newblock {\"U}ber den anschaulichen {I}nhalt der quantentheoretischen
  {K}inematik und {M}echanik.
\newblock {\em Z. Phys.}, 43:172, 1927.

\bibitem{Heisenberg2008}
W.~Heisenberg.
\newblock {\em "Die physikalischen Prinzipien der Quantumtheorie"}.
\newblock Hirzel, Stuttgart, Germany, 5th edition, 1944.

\bibitem{Bohr1928}
N.~Bohr.
\newblock The quantum postulate and recent developments of atomic theory.
\newblock {\em Nature}, 121:580, 1928.

\bibitem{Robertson1929}
H.~P. Robertson.
\newblock The uncertainty principle.
\newblock {\em Phys. Rev.}, 34:163, 1929.

\bibitem{Dirac1958}
P.~M.~A. Dirac.
\newblock {\em The principles of quantum mechanics}.
\newblock Oxford Science Publications, Oxford, UK, 4th edition, 1958.

\bibitem{Hilgevoord1996}
J.~Hilgevoord.
\newblock The uncertainty principle for energy and time.
\newblock {\em Am. J. Phys.}, 64:1451, 1996.

\bibitem{Hilgevoord1998}
J.~Hilgevoord.
\newblock The uncertainty principle for energy and time. {II}.
\newblock {\em Am. J. Phys.}, 66:396, 1998.

\bibitem{Hilgevoord2002a}
J.~Hilgevoord.
\newblock Time in quantum mechanics.
\newblock {\em Am. J. Phys.}, 70:301, 2002.

\bibitem{mandelstam45}
L.~Mandelstam and I.~Tamm.
\newblock The uncertainty relation between energy and time in nonrelativistic
  quantum mechanics.
\newblock {\em J. Phys.}, 9:249, 1945.

\bibitem{Gislason1985}
E.~A. Gislason, N.~H. Sabelli, and J.~W. Wood.
\newblock New form of the time-energy uncertainty relation.
\newblock {\em Phys. Rev. A}, 31:2078, 1985.

\bibitem{Uffink1985}
J.~B.~M. Uffink and J.~Hilgevoord.
\newblock Uncertainty principle and uncertainty relations.
\newblock {\em Found. Phys.}, 15:925, 1985.

\bibitem{Hilgevoord1991}
J.~Hilgevoord and J.~Uffink.
\newblock Uncertainty in prediction and in inference.
\newblock {\em Found. Phys.}, 21:323, 1991.

\bibitem{Luo2005}
S.~Luo.
\newblock On survival probability of quantum states.
\newblock {\em J. Phys. A: Math. Gen.}, 38:2991, 2005.

\bibitem{Boykin2007}
T.~B. Boykin, N.~Kharche, and G.~Klimeck.
\newblock Evolution time and energy uncertainty.
\newblock {\em Euro. J. Phys.}, 28:673, 2007.

\bibitem{Li2013}
J.~Li, M.~P. Silveri, K.~S. Kumar, J.-M. Pirkkalainen, A.~Veps\"{a}l\"{a}inen,
  W.~C. Chien, J.~Tuorila, M.~A. Sillanp\"{a}\"{a}, P.~J. Hakonen, E.~V.
  Thuneberg, and G.~S. Paraoanu.
\newblock {Motional averaging in a superconducting qubit.}
\newblock {\em Nat. Commun.}, 4:1420, 2013.

\bibitem{Eberly1973}
J.~H. Eberly and L.~P.~S. Singh.
\newblock Time operators, partial stationarity, and the energy-time uncertainty
  relation.
\newblock {\em Phys. Rev. D}, 7:359, 1973.

\bibitem{Kobe1994}
D.~H. Kobe and V.~C. Aguilera-Navarro.
\newblock Derivation of the energy-time uncertainty relation.
\newblock {\em Phys. Rev. A}, 50:933, 1994.

\bibitem{Uffink1993}
J.~Uffink.
\newblock The rate of evolution of a quantum state.
\newblock {\em Am. J. Phys.}, 61:935, 1993.

\bibitem{Hilgevoord2002}
J.~Hilgevoord.
\newblock The standard deviation is not an adequate measure of quantum
  uncertainty.
\newblock {\em Am. J. Phys.}, 70:983, 2002.

\bibitem{Feynman1982}
R.~P. Feynman.
\newblock Simulating physics with computers.
\newblock {\em Int. J. Theo. Phys.}, 21:467, 1982.

\bibitem{Ladd2010}
T.~D. {Ladd}, F.~{Jelezko}, R.~{Laflamme}, Y.~{Nakamura}, C.~{Monroe}, and
  J.~L. {O'Brien}.
\newblock {Quantum computers}.
\newblock {\em Nature}, 464:45, 2010.

\bibitem{margolus98}
N.~Margolus and L.~B. Levitin.
\newblock The maximum speed of dynamical evolution.
\newblock {\em Physica D}, 120:188, 1998.

\bibitem{Levitin2009}
L.~B. Levitin and Y.~Toffoli.
\newblock Fundamental limit on the rate of quantum dynamics: {T}he unified
  bound is tight.
\newblock {\em Phys. Rev. Lett.}, 103:160502, 2009.

\bibitem{Bekenstein1981}
J.~D. Bekenstein.
\newblock {Energy Cost of Information Transfer}.
\newblock {\em Phys. Rev. Lett.}, 46:623, 1981.

\bibitem{lloyd00}
S.~Lloyd.
\newblock Ultimate physical limits to computation.
\newblock {\em Nature}, 406:1047, 2000.

\bibitem{Deffner2010}
S.~Deffner and E.~Lutz.
\newblock {Generalized Clausius inequality for nonequilibrium quantum
  processes}.
\newblock {\em Phys. Rev. Lett.}, 105:170402, 2010.

\bibitem{Caneva2009}
T.~Caneva, M.~Murphy, T.~Calarco, R.~Fazio, S.~Montangero, V.~Giovannetti, and
  G.~E. Santoro.
\newblock Optimal control at the quantum speed limit.
\newblock {\em Phys. Rev. Lett.}, 103:240501, 2009.

\bibitem{Giovannetti2011}
V.~Giovannetti, S.~Lloyd, and L.~Maccone.
\newblock {Advances in quantum metrology}.
\newblock {\em Nat. Photonics}, 5:222--229, 2011.

\bibitem{delCampo2017}
A.~del Campo, J.~Molina-Vilaplana, and J.~Sonner.
\newblock Scrambling the spectral form factor: Unitarity constraints and exact
  results.
\newblock {\em Phys. Rev. D}, 95:126008, 2017.

\bibitem{taddei13}
M.~M. Taddei, B.~M. Escher, L.~Davidovich, and R.~L. de~Matos~Filho.
\newblock Quantum speed limit for physical processes.
\newblock {\em Phys. Rev. Lett.}, 110:050402, 2013.

\bibitem{delcampo13}
A.~del Campo, I.~L. Egusquiza, M.~B. Plenio, and S.~F. Huelga.
\newblock Quantum speed limits in open system dynamics.
\newblock {\em Phys. Rev. Lett.}, 110:050403, 2013.

\bibitem{Deffner2013PRL}
S.~Deffner and E.~Lutz.
\newblock Quantum speed limit for non-{Markovian} dynamics.
\newblock {\em Phys. Rev. Lett.}, 111:010402, 2013.

\bibitem{Cimmarusti2015}
A.~D. {Cimmarusti}, Z.~{Yan}, B.~D. {Patterson}, L.~P. {Corcos}, L.~A.
  {Orozco}, and S.~{Deffner}.
\newblock Environment-assisted speed-up of the field evolution in cavity
  quantum electrodynamics.
\newblock {\em Phys. Rev. Lett.}, 114:233602, 2015.

\bibitem{Busch1990}
P.~Busch.
\newblock {On the energy-time uncertainty relation. Part I: Dynamical time and
  time indeterminacy}.
\newblock {\em Found. Phys.}, 20:1, 1990.

\bibitem{Busch1990a}
P.~Busch.
\newblock {On the energy-time uncertainty relation. Part II: Pragmatic time
  versus energy indeterminacy}.
\newblock {\em Found.Phys.}, 20:33, 1990.

\bibitem{Pfeifer1995}
P.~Pfeifer and J.~Fr\"ohlich.
\newblock Generalized time-energy uncertainty relations and bounds on lifetimes
  of resonances.
\newblock {\em Rev. Mod. Phys.}, 67:759, 1995.

\bibitem{Muga2007}
G.~Muga, R.~S. Mayato, and I.~Egusquiza.
\newblock {\em Time in quantum mechanics}.
\newblock Springer, 2007.

\bibitem{Muga2009}
G.~Muga, A.~Ruschhaupt, and A.~del Campo.
\newblock {\em Time in Quantum Mechanics - Vol. 2}.
\newblock Springer, 2009.

\bibitem{Sen2014}
D.~Sen.
\newblock The uncertainty relations in quantum mechanics.
\newblock {\em Current Science}, 107:203, 2014.

\bibitem{Dodonov2015}
V.~V. Dodonov and A.~V. Dodonov.
\newblock Energy-time and frequency-time uncertainty relations: exact
  inequalities.
\newblock {\em Phys. Scr.}, 90:074049, 2015.

\bibitem{Frey2016}
M.~R. Frey.
\newblock Quantum speed limits---primer, perspectives, and potential future
  directions.
\newblock {\em Quantum Inf. Process.}, 15:3919, 2016.

\bibitem{Messiah1966}
A.~Messiah.
\newblock {\em Quantum Mechanics}, volume~II.
\newblock John Wiley \& Sons, Amsterdam, The Netherlands, 1966.

\bibitem{Hilgevoord1990}
J.~Hilgevoord and J.~Uffink.
\newblock {\em A New View on the Uncertainty Principle}, pages 121--137.
\newblock Springer US, Boston, MA, 1990.

\bibitem{Kholevo1974}
A.~S. Kholevo.
\newblock A generalization of the rao–cramer inequality.
\newblock {\em Theory of Probability \& Its Applications}, 18:359--362, 1974.

\bibitem{Frowis2015}
F.~Fr\"owis, R.~Schmied, and N.~Gisin.
\newblock Tighter quantum uncertainty relations following from a general
  probabilistic bound.
\newblock {\em Phys. Rev. A}, 92:012102, 2015.

\bibitem{Erker2016}
P.~{Erker}, M.~T. {Mitchison}, R.~{Silva}, M.~P. {Woods}, N.~{Brunner}, and
  M.~{Huber}.
\newblock {Autonomous quantum clocks: how thermodynamics limits our ability to
  measure time}.
\newblock {\em Phys. Rev. X}, 7:031022, 2017.

\bibitem{Aharonov1961}
Y.~Aharonov and D.~Bohm.
\newblock Time in the quantum theory and the uncertainty relation for time and
  energy.
\newblock {\em Phys. Rev.}, 122:1649, 1961.

\bibitem{Briggs2008}
J.~S. Briggs.
\newblock A derivation of the time-energy uncertainty relation.
\newblock {\em J. Phys.: Conf. Ser.}, 99:012002, 2008.

\bibitem{Fleming1973}
G.~N. Fleming.
\newblock A unitarity bound on the evolution of nonstationary states.
\newblock {\em Il Nuovo Cimento A (1965-1970)}, 16:232, 1973.

\bibitem{Bhattacharyya1983}
K.~Bhattacharyya.
\newblock Quantum decay and the {M}andelstam-{T}amm-energy inequality.
\newblock {\em J. Phys. A: Math. Gen.}, 16:2993, 1983.

\bibitem{Anandan1990}
J.~Anandan and Y.~Aharonov.
\newblock Geometry of quantum evolution.
\newblock {\em Phys. Rev. Lett.}, 65:1697, 1990.

\bibitem{Vaidman1992}
L.~Vaidman.
\newblock Minimum time for the evolution to an orthogonal quantum state.
\newblock {\em Am. J. Phys.}, 60:182, 1992.

\bibitem{Brody2003}
D.~C. Brody.
\newblock Elementary derivation for passage times.
\newblock {\em J. Phys. A: Math. Gen.}, 36:5587, 2003.

\bibitem{Andrecut2004}
M.~Andrecut and M.~K. Ali.
\newblock Maximum speed of quantum evolution.
\newblock {\em Int. J. Theo. Phys.}, 43:969, 2004.

\bibitem{Zych2006}
P.~Kosi\ifmmode~\acute{n}\else \'{n}\fi{}ski and M.~Zych.
\newblock Elementary proof of the bound on the speed of quantum evolution.
\newblock {\em Phys. Rev. A}, 73:024303, 2006.

\bibitem{Giovannetti2003a}
V.~Giovannetti, S.~Lloyd, and L.~Maccone.
\newblock The role of entanglement in dynamical evolution.
\newblock {\em EPL (Europhysics Letters)}, 62:615, 2003.

\bibitem{Svozil2005}
K.~Svozil, L.~B. Levitin, T.~Toffoli, and Z.~Walton.
\newblock Maximum speed of quantum gate operation.
\newblock {\em Int. J. Theo. Phys.}, 44:965, 2005.

\bibitem{Zander2007}
C.~Zander, A.~R. Plastino, A.~Plastino, and M.~Casas.
\newblock Entanglement and the speed of evolution of multi-partite quantum
  systems.
\newblock {\em J. Phys. A: Math. Theor.}, 40:2861, 2007.

\bibitem{Pfeifer1993}
P.~Pfeifer.
\newblock How fast can a quantum state change with time?
\newblock {\em Phys. Rev. Lett.}, 70:3365, 1993.

\bibitem{Wootters1981}
W.~K. Wootters.
\newblock Statistical distance and {Hilbert} space.
\newblock {\em Phys. Rev. D}, 23:357, 1981.

\bibitem{Uhlmann1976}
A.~Uhlmann.
\newblock The “transition probability” in the state space of a∗-algebra.
\newblock {\em Rep. Math. Phys.}, 9:273, 1976.

\bibitem{Jozsa1994}
R.~Jozsa.
\newblock Fidelity for mixed quantum states.
\newblock {\em J. Mod. Opt.}, 41:2315, 1994.

\bibitem{Kakutani1948}
S.~Kakutani.
\newblock On equivalence of infinite product measures.
\newblock {\em Ann. Math.}, 49:214, 1948.

\bibitem{Bures1969}
D.~Bures.
\newblock An extension of {Kakutani's} theorem on infinite product measures to
  the tensor product of semifinite w*-algebras.
\newblock {\em Trans. Am. Math. Soc.}, 135:199, 1969.

\bibitem{Uhlmann1992}
A.~Uhlmann.
\newblock An energy dispersion estimate.
\newblock {\em Phys. Lett. A}, 161:329, 1992.

\bibitem{Deffner2013}
S.~{Deffner} and E.~{Lutz}.
\newblock {Energy-time uncertainty relation for driven quantum systems}.
\newblock {\em J. Phys. A: Math. Theor.}, 46:G5302, 2013.

\bibitem{Braunstein1994}
S.~L. Braunstein and C.~M. Caves.
\newblock Statistical distance and the geometry of quantum states.
\newblock {\em Phys. Rev. Lett.}, 72:3439, 1994.

\bibitem{Braunstein1995}
S.~L. Braunstein and G.~J. Milburn.
\newblock Dynamics of statistical distance: Quantum limits for two-level
  clocks.
\newblock {\em Phys. Rev. A}, 51:1820, 1995.

\bibitem{Braunstein1996}
S.~L. Braunstein, C.~M. Caves, and G.~J. Milburn.
\newblock Generalized uncertainty relations: Theory, examples, and lorentz
  invariance.
\newblock {\em Ann. Phys.}, 247:135, 1996.

\bibitem{Giovannetti2003}
V.~Giovannetti, S.~Lloyd, and L.~Maccone.
\newblock Quantum limits to dynamical evolution.
\newblock {\em Phys. Rev. A}, 67:052109, 2003.

\bibitem{Giovannetti2003b}
V.~{Giovannetti}, S.~{Lloyd}, and L.~{Maccone}.
\newblock {The quantum speed limit}.
\newblock In D.~{Abbott}, J.~H. {Shapiro}, and Y.~{Yamamoto}, editors, {\em
  Fluctuations and Noise in Photonics and Quantum Optics}, volume 5111 of {\em
  Proc. SPIE}, pages 1--6, 2003.

\bibitem{Giovannetti2004}
V.~{Giovannetti}, S.~{Lloyd}, and L.~{Maccone}.
\newblock {The speed limit of quantum unitary evolution}.
\newblock {\em J. Opt. B}, 6:807, 2004.

\bibitem{Andrews2007}
M.~Andrews.
\newblock Bounds to unitary evolution.
\newblock {\em Phys. Rev. A}, 75:062112, 2007.

\bibitem{Jones2010}
P.~J. Jones and P.~Kok.
\newblock Geometric derivation of the quantum speed limit.
\newblock {\em Phys. Rev. A}, 82:022107, 2010.

\bibitem{Zwierz2012}
M.~Zwierz.
\newblock Comment on {``Geometric derivation of the quantum speed limit''}.
\newblock {\em Phys. Rev. A}, 86:016101, 2012.

\bibitem{Bremermann1967}
H.~J. Bremermann.
\newblock Quantum noise and information.
\newblock In {\em Proceedings of the Fifth Berkeley Symposium on Mathematical
  Statistics and Probability, Volume 4: Biology and Problems of Health}, pages
  15--20, Berkeley, Calif., 1967. University of California Press.

\bibitem{Bekenstein1974}
J.~D. Bekenstein.
\newblock Generalized second law of thermodynamics in black-hole physics.
\newblock {\em Phys. Rev. D}, 9:3292, 1974.

\bibitem{Bekenstein1990}
J.~D. Bekenstein and M.~Schiffer.
\newblock Quantum limitations on the storage and transmission of information.
\newblock {\em Int. J. Mod. Phys. C}, 1:355, 1990.

\bibitem{Goold2016}
J.~Goold, M.~Huber, A.~Riera, L.~del Rio, and P.~Skrzypczyk.
\newblock The role of quantum information in thermodynamics - a topical review.
\newblock {\em J. Phys. A: Math. Theor.}, 49:143001, 2016.

\bibitem{delCampo2013}
A.~del Campo, J.~Goold, and M.~Paternostro.
\newblock More bang for your buck: Super-adiabatic quantum engines.
\newblock {\em Sci. Rep.}, 4:6208, 2014.

\bibitem{Talkner2007}
Peter Talkner, Eric Lutz, and Peter Hanggi.
\newblock Fluctuation theorems: Work is not an observable.
\newblock {\em Phy. Rev. E}, 75:050102(R), 2007.

\bibitem{Campisi2011}
M.~{Campisi}, P.~{H{\"a}nggi}, and P.~{Talkner}.
\newblock {Colloquium: Quantum fluctuation relations: Foundations and
  applications}.
\newblock {\em Rev. Mod. Phys.}, 83:771, July 2011.

\bibitem{Deffner2016work}
S.~Deffner, J.~P. Paz, and W.~H. Zurek.
\newblock Quantum work and the thermodynamic cost of quantum measurements.
\newblock {\em Phys. Rev. E}, 94:010103, 2016.

\bibitem{Umegaki1962}
H.~Umegaki.
\newblock Conditional expectation in an operator algebra, iv (entropy and
  information).
\newblock {\em KODAI MATHEMATICAL SEMINAR REPORTS}, 14(2):59--85, 1962.

\bibitem{Deffner2013TL}
S.~Deffner and E.~Lutz.
\newblock Thermodynamic length for far-from-equilibrium quantum systems.
\newblock {\em Phys. Rev. E}, 87:022143, 2013.

\bibitem{Kosloff2017}
R.~Kosloff and Y.~Rezek.
\newblock The quantum harmonic {Otto} cycle.
\newblock {\em Entropy}, 19:136, 2017.

\bibitem{Zheng2016}
Y.~{Zheng}, S.~{Campbell}, G.~{De Chiara}, and D.~{Poletti}.
\newblock {Cost of counterdiabatic driving and work output}.
\newblock {\em Phys. Rev. A}, 94:042132, 2016.

\bibitem{Torrontegui2013}
E.~Torrontegui, S.~Ib\'{a}\~{n}ez, S.~Mart\'{\i}nez-Garaot, M.~Modugno, A.~del
  Campo, D.~Gu\'{e}ry-Odelin, A.~Ruschhaupt, X.~Chen, and J.~G. Muga.
\newblock {Shortcuts to Adiabaticity}.
\newblock {\em Adv. At. Mol. Opt. Phys.}, 62:117, 2013.

\bibitem{Demirplak2003}
M.~Demirplak and S.~A. Rice.
\newblock Adiabatic population transfer with control fields.
\newblock {\em J. Chem. Phys. A}, 107:9937, 2003.

\bibitem{Demirplak2005}
M.~Demirplak and S.~A Rice.
\newblock {Assisted adiabatic passage revisited}.
\newblock {\em J. Phys. Chem. B}, 109:6838, 2005.

\bibitem{Berry2009}
M.~Berry.
\newblock Transitionless quantum driving.
\newblock {\em J. Phys. A: Math. Theor.}, 42:365303, 2009.

\bibitem{Campbell2017}
S.~Campbell and S.~Deffner.
\newblock Trade-off between speed and cost in shortcuts to adiabaticity.
\newblock {\em Phys. Rev. Lett.}, 118:100601, 2017.

\bibitem{Abah2017}
O.~{Abah} and E.~{Lutz}.
\newblock Energy efficient quantum machines.
\newblock {\em EPL}, 118:40005, 2017.

\bibitem{Mukherjee2016}
V.~Mukherjee, W.~Niedenzu, A.~G. Kofman, and G.~Kurizki.
\newblock Speed and efficiency limits of multilevel incoherent heat engines.
\newblock {\em Phys. Rev. E}, 94:062109, 2016.

\bibitem{Woods2016}
M.~P. {Woods}, R.~{Silva}, and J.~{Oppenheim}.
\newblock {Autonomous quantum machines and finite sized clocks}.
\newblock {\em arXiv:1607.04591}, 2016.

\bibitem{SarandySciRep2015}
A.~C. Santos and M.~S. Sarandy.
\newblock Superadiabatic controlled evolutions and universal quantum
  computation.
\newblock {\em Sci. Rep.}, 5:15775, 2015.

\bibitem{Jordan2017}
S.~P. Jordan.
\newblock Fast quantum computation at arbitrarily low energy.
\newblock {\em Phys. Rev. A}, 95:032305, 2017.

\bibitem{Giovannetti2006}
V.~Giovannetti, S.~Lloyd, and L.~Maccone.
\newblock Quantum metrology.
\newblock {\em Phys. Rev. Lett.}, 96:010401, 2006.

\bibitem{Paris2009}
M.~G.~A. {Paris}.
\newblock {Quantum estimation for quantum technology}.
\newblock {\em Int. J. Quant. Inf.}, 7:125, 2009.

\bibitem{Toth2013}
G\'eza T\'oth and D\'enes Petz.
\newblock Extremal properties of the variance and the quantum fisher
  information.
\newblock {\em Phys. Rev. A}, 87:032324, 2013.

\bibitem{Yu2013}
S.~{Yu}.
\newblock {Quantum Fisher Information as the Convex Roof of Variance}.
\newblock {\em arXiv:1302.5311}, 2013.

\bibitem{Frowis2012}
F.~Fr\"owis.
\newblock Kind of entanglement that speeds up quantum evolution.
\newblock {\em Phys. Rev. A}, 85:052127, May 2012.

\bibitem{Zwierz2010}
M.~Zwierz, C.~A. P\'erez-Delgado, and P.~Kok.
\newblock General optimality of the heisenberg limit for quantum metrology.
\newblock {\em Phys. Rev. Lett.}, 105:180402, 2010.

\bibitem{Zhang2016}
C.~{Zhang}, B.~{Yadin}, Z.-B. {Hou}, H.~{Cao}, B.-H. {Liu}, Y.-F. {Huang},
  R.~{Maity}, V.~{Vedral}, C.-F. {Li}, G.-C. {Guo}, and D.~{Girolami}.
\newblock {Detecting metrologically useful asymmetry and entanglement by few
  local measurements}.
\newblock {\em arXiv:1611.02004}, 2016.

\bibitem{Beau2017}
M.~Beau and A.~del Campo.
\newblock Nonlinear quantum metrology of many-body open systems.
\newblock {\em Phys. Rev. Lett.}, 119:010403, 2017.

\bibitem{Krotov1996}
V.~F. Krotov.
\newblock {\em Global Methods in Optimal Control Theory}.
\newblock Mercel Dekker, New York, 1996.

\bibitem{Walmsley2003}
I.~Walmsley and H.~Rabitz.
\newblock Quantum physics under control.
\newblock {\em Phys. Today}, 56:43, 2003.

\bibitem{Bason2012}
M.~G. {Bason}, M.~{Viteau}, N.~{Malossi}, P.~{Huillery}, E.~{Arimondo},
  D.~{Ciampini}, R.~{Fazio}, V.~{Giovannetti}, R.~{Mannella}, and O.~{Morsch}.
\newblock {High-fidelity quantum driving}.
\newblock {\em Nat. Phys.}, 8:147, 2012.

\bibitem{Binder2015}
F.~C. Binder, S.~Vinjanampathy, K.~Modi, and J.~Goold.
\newblock Quantacell: powerful charging of quantum batteries.
\newblock {\em New J. Phys.}, 17:075015, 2015.

\bibitem{Campaioli2017}
F.~Campaioli, F.~A. Pollock, F.~C. Binder, L.~C\'eleri, J.~Goold,
  S.~Vinjanampathy, and K.~Modi.
\newblock Enhancing the charging power of quantum batteries.
\newblock {\em Phys. Rev. Lett.}, 118:150601, 2017.

\bibitem{Bose2003}
S.~Bose.
\newblock Quantum communication through an unmodulated spin chain.
\newblock {\em Phys. Rev. Lett.}, 91:207901, 2003.

\bibitem{Murphy2010}
M.~Murphy, S.~Montangero, V.~Giovannetti, and T.~Calarco.
\newblock Communication at the quantum speed limit along a spin chain.
\newblock {\em Phys. Rev. A}, 82:022318, 2010.

\bibitem{Lieb1972}
E.~H. Lieb and D.~W. Robinson.
\newblock The finite group velocity of quantum spin systems.
\newblock {\em Commun. Math. Phys.}, 28:251, 1972.

\bibitem{Caneva2011}
T.~Caneva, T.~Calarco, R.~Fazio, G.~E. Santoro, and S.~Montangero.
\newblock Speeding up critical system dynamics through optimized evolution.
\newblock {\em Phys. Rev. A}, 84:012312, 2011.

\bibitem{Hegerfeldt2013}
G.~C. Hegerfeldt.
\newblock {Driving at the quantum speed limit: Optimal control of a two-level
  system}.
\newblock {\em Phys. Rev. Lett.}, 111:260501, 2013.

\bibitem{Poggi2013}
P.~M. {Poggi}, F.~C. {Lombardo}, and D.~A. {Wisniacki}.
\newblock {Quantum speed limit and optimal evolution time in a two-level
  system}.
\newblock {\em EPL (Europhysics Letters)}, 104:40005, 2013.

\bibitem{richerme14}
P.~{Richerme}, Z.-X. {Gong}, A.~{Lee}, C.~{Senko}, J.~{Smith}, M.~{Foss-Feig},
  S.~{Michalakis}, A.~V. {Gorshkov}, and C.~{Monroe}.
\newblock {Non-local propagation of correlations in long-range interacting
  quantum systems}.
\newblock {\em Nature}, 511:198, 2014.

\bibitem{delCampo2015}
A.~del Campo and K.~Sengupta.
\newblock Controlling quantum critical dynamics of isolated systems.
\newblock {\em Eur. Phys. J. Spec. Top.}, 224:189, 2015.

\bibitem{Khalil2015}
T.~Khalil and J.~Richert.
\newblock Excitation of time-dependent quantum systems: An application of time
  energy uncertainty relations.
\newblock {\em Physica B: Cond. Matt.}, 457:78, 2015.

\bibitem{Mukherjee2013}
V.~{Mukherjee}, A.~{Carlini}, A.~{Mari}, T.~{Caneva}, S.~{Montangero},
  T.~{Calarco}, R.~{Fazio}, and V.~{Giovannetti}.
\newblock {Speeding up and slowing down the relaxation of a qubit by optimal
  control}.
\newblock {\em Phys. Rev. A}, 88:062326, 2013.

\bibitem{deffner14}
S.~{Deffner}.
\newblock {Optimal control of a qubit in an optical cavity}.
\newblock {\em J. Phys. B: At. Mol. Opt. Phys.}, 47:145502, 2014.

\bibitem{Caneva2013}
T.~Caneva, S.~Montangero, M.~D. Lukin, and T.~Calarco.
\newblock {Noise-resistant optimal spin squeezing via quantum control}.
\newblock {\em Phys. Rev. A}, 93:013851, 2016.

\bibitem{Mukherjee2015}
V.~Mukherjee, V.~Giovannetti, R.~Fazio, S.~F. Huelga, T.~Calarco, and
  S.~Montangero.
\newblock Efficiency of quantum controlled non-markovian thermalization.
\newblock {\em New J. Phys.}, 17:063031, 2015.

\bibitem{Reich2013}
D.~M. Reich and C.~P Koch.
\newblock {Cooling molecular vibrations with shaped laser pulses: optimal
  control theory exploiting the timescale separation between coherent
  excitation and spontaneous emission}.
\newblock {\em \NJP}, 15:125028, 2013.

\bibitem{Brody2015}
D.~C. Brody, G.~W. Gibbons, and D.~M Meier.
\newblock Time-optimal navigation through quantum wind.
\newblock {\em New J. Phys.}, 17:033048, 2015.

\bibitem{Gajdacz2015}
M.~Gajdacz, K.~K. Das, J.~Arlt, J.~F. Sherson, and T.~Opatrn\'y.
\newblock Time-limited optimal dynamics beyond the quantum speed limit.
\newblock {\em Phys. Rev. A}, 92:062106, 2015.

\bibitem{Baksic2016}
A.~Baksic, H.~Ribeiro, and A.~A. Clerk.
\newblock Speeding up adiabatic quantum state transfer by using dressed states.
\newblock {\em Phys. Rev. Lett.}, 116:230503, 2016.

\bibitem{Arenz2017}
C.~{Arenz}, B.~{Russell}, D.~{Burgarth}, and H.~{Rabitz}.
\newblock {The roles of drift and control field constraints upon quantum
  control speed limits}.
\newblock {\em arXiv:1704.06289}, 2017.

\bibitem{Barnes2013a}
E.~Barnes.
\newblock {Analytically solvable two-level quantum systems and Landau-Zener
  interferometry}.
\newblock {\em Phys. Rev. A}, 88:013818, 2013.

\bibitem{Boscain2012}
U.~Boscain and R.~Long.
\newblock {Time minimal trajectories for two-level quantum systems with two
  bounded controls}.
\newblock {\em 51st IEEE Conference on Decision and Control (CDC)}, page 3626,
  2012.

\bibitem{Hegertfeldt2014}
G.~C. Hegerfeldt.
\newblock High-speed driving of a two-level system.
\newblock {\em Phys. Rev. A}, 90:032110, 2014.

\bibitem{Bukov2017}
M.~{Bukov}, A.~G.~R. {Day}, D.~{Sels}, P.~{Weinberg}, A.~{Polkovnikov}, and
  P.~{Mehta}.
\newblock {Machine Learning Meets Quantum State Preparation. The Phase Diagram
  of Quantum Control}.
\newblock {\em arXiv:1705.00565}, 2017.

\bibitem{Mondal2016PLA}
D.~Mondal and A.~K. Pati.
\newblock Quantum speed limit for mixed states using an experimentally
  realizable metric.
\newblock {\em Phys. Lett. A}, 380:1395, 2016.

\bibitem{Pires2016}
D.~P. Pires, M.~Cianciaruso, L.~C. C\'eleri, G.~Adesso, and D.~O. Soares-Pinto.
\newblock Generalized geometric quantum speed limits.
\newblock {\em Phys. Rev. X}, 6:021031, 2016.

\bibitem{bengtsson2007}
I.~Bengtsson and K.~Zyczkowski.
\newblock {\em Geometry of quantum states: an introduction to quantum
  entanglement}.
\newblock Cambridge University Press, 2007.

\bibitem{Campbell1986}
L.~L. Campbell.
\newblock An extended {{\v{C}}encov} characterization of the information
  metric.
\newblock {\em Proc. Am. Math. Soc.}, 98:135, 1986.

\bibitem{Andersson2014b}
O.~Andersson and H.~Heydari.
\newblock Geometric uncertainty relation for mixed quantum states.
\newblock {\em J. Math. Phys.}, 55:042110, 2014.

\bibitem{Andersson2014}
O.~{Andersson} and H.~{Heydari}.
\newblock {Quantum speed limits and optimal Hamiltonians for driven systems in
  mixed states}.
\newblock {\em J. Phys. A: Math. Theor.}, 47:215301, 2014.

\bibitem{Andersson2014a}
O.~Andersson and H.~Heydari.
\newblock Geometry of quantum evolution for mixed quantum states.
\newblock {\em Phys. Scr.}, 2014:014004, 2014.

\bibitem{Mirsky1975}
L.~Mirsky.
\newblock A trace inequality of {John von Neumann}.
\newblock {\em Monatshefte f{\"u}r Mathematik}, 79:303, 1975.

\bibitem{Grigorieff1991}
R.~Grigorieff.
\newblock A note on von {Neumann}'s trace inequalitv.
\newblock {\em Mathematische Nachrichten}, 151:327, 1991.

\bibitem{Simon1979}
B.~Simon.
\newblock {\em Trace ideals and their applications}, volume~35.
\newblock Cambridge University Press Cambridge, 1979.

\bibitem{Garraway1997}
B.~M. Garraway.
\newblock Nonperturbative decay of an atomic system in a cavity.
\newblock {\em Phys. Rev. A}, 55:2290, 1997.

\bibitem{Breuer1999}
H.-P. Breuer, B.~Kappler, and F.~Petruccione.
\newblock Stochastic wave-function method for non-{Markovian} quantum master
  equations.
\newblock {\em Phys. Rev. A}, 59:1633, 1999.

\bibitem{Xu2014}
Z.-Y. {Xu}, S.~{Luo}, W.~L. {Yang}, C.~{Liu}, and S.~{Zhu}.
\newblock Quantum speedup in a memory environment.
\newblock {\em Phys. Rev. A}, 89:012307, 2014.

\bibitem{Breuer2009}
H.-P. Breuer, E.-M. Laine, and J.~Piilo.
\newblock Measure for the degree of non-{Markovian} behavior of quantum
  processes in open systems.
\newblock {\em Phys. Rev. Lett.}, 103:210401, 2009.

\bibitem{Brune1996}
M.~Brune, F.~Schmidt-Kaler, A.~Maali, J.~Dreyer, E.~Hagley, J.~M. Raimond, and
  S.~Haroche.
\newblock Quantum {Rabi} oscillation: A direct test of field quantization in a
  cavity.
\newblock {\em Phys. Rev. Lett.}, 76:1800, 1996.

\bibitem{Madsen2011}
K.~H. Madsen, S.~Ates, T.~Lund-Hansen, A.~L\"offler, S.~Reitzenstein,
  A.~Forchel, and P.~Lodahl.
\newblock Observation of non-{Markovian} dynamics of a single quantum dot in a
  micropillar cavity.
\newblock {\em Phys. Rev. Lett.}, 106:233601, 2011.

\bibitem{Zhang2014}
Y.-J. {Zhang}, W.~{Han}, Y.-J. {Xia}, J.-P. {Cao}, and H.~{Fan}.
\newblock {Quantum speed limit for arbitrary initial states}.
\newblock {\em Sci. Rep.}, 4:4890, 2014.

\bibitem{Cai2017}
X.~Cai and Y.~Zheng.
\newblock Quantum dynamical speedup in a nonequilibrium environment.
\newblock {\em Phys. Rev. A}, 95:052104, 2017.

\bibitem{Xu2014a}
Z.-Y. {Xu} and S.-Q. {Zhu}.
\newblock Quantum speed limit of a photon under non-{M}arkovian dynamics.
\newblock {\em Chin. Phys. Lett.}, 31:020301, 2014.

\bibitem{Zhu2014}
S.~Zhu and Z.-Y. Xu.
\newblock Quantum speed limit with non-{Markovian} noise.
\newblock {\em Research in Optical Sciences}, page JW2A.48, 2014.

\bibitem{Zhang2015}
Y.-J. Zhang, W.~Han, Y.-J. Xia, J.-P. Cao, and H.~Fan.
\newblock Classical-driving-assisted quantum speed-up.
\newblock {\em Phys. Rev. A}, 91:032112, 2015.

\bibitem{Liu2016}
H.-B. Liu, W.~L. Yang, J.-H. An, and Z.-Y. Xu.
\newblock Mechanism for quantum speedup in open quantum systems.
\newblock {\em Phys. Rev. A}, 93:020105, 2016.

\bibitem{Chenu2017}
A.~Chenu, M.~Beau, J.~Cao, and A.~del Campo.
\newblock Quantum simulation of generic many-body open system dynamics using
  classical noise.
\newblock {\em Phys. Rev. Lett.}, 118:140403, 2017.

\bibitem{Beau2017arXiv}
M.~{Beau}, J.~{Kiukas}, I.~L. {Egusquiza}, and A.~{del Campo}.
\newblock {Nonexponential quantum decay under environmental decoherence}.
\newblock {\em arXiv:1706.06943}, June 2017.

\bibitem{Xu2016}
Z.-Y. Xu.
\newblock Detecting quantum speedup in closed and open systems.
\newblock {\em New J. Phys.}, 18:073005, 2016.

\bibitem{Mo2017}
M.~Mo, J.~Wang, and Y.~Wu.
\newblock Quantum speedup via engineering multiple environments.
\newblock {\em Ann. Phys.}, 529, 2017.

\bibitem{Hou2015}
S.~C. Hou, S.~L. Liang, and X.~X. Yi.
\newblock Non-{Markovianity} and memory effects in quantum open systems.
\newblock {\em Phys. Rev. A}, 91:012109, 2015.

\bibitem{Vacchini2016}
Heinz-Peter Breuer, Elsi-Mari Laine, Jyrki Piilo, and Bassano Vacchini.
\newblock Colloquium.
\newblock {\em Rev. Mod. Phys.}, 88:021002, Apr 2016.

\bibitem{Liu2015}
C.~Liu, Z.-Y. Xu, and S.~Zhu.
\newblock Quantum-speed-limit time for multiqubit open systems.
\newblock {\em Phys. Rev. A}, 91:022102, 2015.

\bibitem{Poggi2015}
P.~M. Poggi, F.~C. Lombardo, and D.~A. Wisniacki.
\newblock Enhancement of quantum speed limit time due to cooperative effects in
  multilevel systems.
\newblock {\em J. Phys. A: Math. Theor.}, 48:35FT02, 2015.

\bibitem{Poggi2015a}
P.~M. Poggi, F.~C. Lombardo, and D.~A. Wisniacki.
\newblock Time-optimal control fields for quantum systems with multiple avoided
  crossings.
\newblock {\em Phys. Rev. A}, 92:053411, 2015.

\bibitem{Hou2015JPA}
L.~Hou, B.~Shao, Y.-B. Wei, and J.~Zou.
\newblock Quantum speed limit in a qubit-spin-bath system.
\newblock {\em J. Phys. A: Math. Theor.}, 48:495302, 2015.

\bibitem{Hou2017}
L.~Hou, B.~Shao, Y.-B. Wei, and J.~Zou.
\newblock Quantum speed limit of the evolution of the qubits in a finite {XY}
  spin chain.
\newblock {\em Euro. Phys. J. D}, 71:22, 2017.

\bibitem{Song2016}
Y.-J. Song, L.-M. Kuang, and Q.-S. Tan.
\newblock Quantum speedup of uncoupled multiqubit open system via dynamical
  decoupling pulses.
\newblock {\em Quantum Inf. Process.}, 15:2325, 2016.

\bibitem{Wei2016}
Y.-B. Wei, J.~Zou, Z.-M. Wang, and B.~Shao.
\newblock Quantum speed limit and a signal of quantum criticality.
\newblock {\em Sci. Rep.}, 6:19308, 2016.

\bibitem{Heyl2017}
M.~Heyl.
\newblock Quenching a quantum critical state by the order parameter: Dynamical
  quantum phase transitions and quantum speed limits.
\newblock {\em Phys. Rev. B}, 95:060504, 2017.

\bibitem{Yu2016}
W.-J. Yu, Y.-B. Wei, H.~Li, L.~Li, J.~Zou, and B.~Shao.
\newblock Quantum speed limit for dynamics of central spin model with initial
  system-bath correlations.
\newblock {\em Int. J. Theor. Phys.}, 55:4785, 2016.

\bibitem{Wu2015}
N.~Wu, A.~Nanduri, and H.~Rabitz.
\newblock Optimal suppression of defect generation during a passage across a
  quantum critical point.
\newblock {\em Phys. Rev. B}, 91:041115, 2015.

\bibitem{Francuz2016}
A.~Francuz, J.~Dziarmaga, B.~Gardas, and W.~H. Zurek.
\newblock Space and time renormalization in phase transition dynamics.
\newblock {\em Phys. Rev. B}, 93:075134, 2016.

\bibitem{Hou2016}
L.~Hou, B.~Shao, and J.~Zou.
\newblock Quantum speed limit for a central system in {Lipkin-Meshkov-Glick}
  bath.
\newblock {\em Eur. Phys. J. D}, 70:35, 2016.

\bibitem{Lipkin1965}
H.~J. Lipkin, N.~Meshkov, and A.~J. Glick.
\newblock Validity of many-body approximation methods for a solvable model.
\newblock {\em Nuclear Physics}, 62:188, 1965.

\bibitem{Campbell2015LMG}
S.~Campbell, G.~de~Chiara, M.~Paternostro, G.~M. Palma, and R.~Fazio.
\newblock Shortcut to adiabaticity in the {Lipkin-Meshkov-Glick} model.
\newblock {\em Phys. Rev. Lett.}, 114:177206, 2015.

\bibitem{Campbell2016LMG}
S.~Campbell.
\newblock Criticality revealed through quench dynamics in the
  {Lipkin-Meshkov-Glick} model.
\newblock {\em Phys. Rev. B}, 94:184403, 2016.

\bibitem{Hutter2012}
A.~Hutter and S.~Wehner.
\newblock Almost all quantum states have low entropy rates for any coupling to
  the environment.
\newblock {\em Phys. Rev. Lett.}, 108:070501, 2012.

\bibitem{Gardas2016}
B.~Gardas, S.~Deffner, and A.~Saxena.
\newblock $\mathcal{PT}$-symmetric slowing down of decoherence.
\newblock {\em Phys. Rev. A}, 94:040101, 2016.

\bibitem{Uzdin2016}
R.~Uzdin and R.~Kosloff.
\newblock Speed limits in {Liouville} space for open quantum systems.
\newblock {\em EPL (Europhysics Letters)}, 115:40003, 2016.

\bibitem{Machnes2014}
S.~{Machnes} and M.~B. {Plenio}.
\newblock {Surprising Interactions of Markovian noise and Coherent Driving}.
\newblock {\em arXiv:1408.3056}, 2014.

\bibitem{Dehdashti2015}
Sh. Dehdashti, M.~Bagheri Harouni, B.~Mirza, and H.~Chen.
\newblock Decoherence speed limit in the spin-deformed boson model.
\newblock {\em Phys. Rev. A}, 91:022116, 2015.

\bibitem{Jing2016}
J.~Jing, L.-A. Wu, and A.~del Campo.
\newblock Fundamental speed limits to the generation of quantumness.
\newblock {\em Sci. Rep.}, 6:38149, 2016.

\bibitem{Mondal2016}
D.~Mondal, C.~Datta, and S.~Sazim.
\newblock Quantum coherence sets the quantum speed limit for mixed states.
\newblock {\em Phys. Lett. A}, 380:689, 2016.

\bibitem{Pires2015}
D.~P. Pires, L.~C. C\'eleri, and D.~O. Soares-Pinto.
\newblock Geometric lower bound for a quantum coherence measure.
\newblock {\em Phys. Rev. A}, 91:042330, 2015.

\bibitem{Marvian2015}
I.~Marvian and D.~A. Lidar.
\newblock Quantum speed limits for leakage and decoherence.
\newblock {\em Phys. Rev. Lett.}, 115:210402, 2015.

\bibitem{Marvian2016}
I.~Marvian, R.~W. Spekkens, and P.~Zanardi.
\newblock Quantum speed limits, coherence, and asymmetry.
\newblock {\em Phys. Rev. A}, 93:052331, 2016.

\bibitem{Sun2015}
Z.~Sun, J.~Liu, J.~Ma, and X.~Wang.
\newblock Quantum speed limits in open systems: Non-markovian dynamics without
  rotating-wave approximation.
\newblock {\em Sci. Rep.}, 5:8444, 2015.

\bibitem{Deffner2017}
S.~{Deffner}.
\newblock {Geometric quantum speed limits: a case for Wigner phase space}.
\newblock {\em New J. Phys. (2017) DOI:10.1088/1367-2630/aa83dc}.

\bibitem{Holder1889}
O.~H\"older.
\newblock Ueber einen {Mittelwerthssatz}.
\newblock {\em {Nachrichten von der K\"oniglichen Gesellschaft der
  Wissenschaften und der Georg-Augusts-Universit\"at zu G\"ottingen}}, 1889:38,
  1889.

\bibitem{Chen2010}
X.~Chen, A.~Ruschhaupt, S.~Schmidt, A.~del Campo, D.~Gu\'ery-Odelin, and J.~G.
  Muga.
\newblock Fast optimal frictionless atom cooling in harmonic traps: Shortcut to
  adiabaticity.
\newblock {\em Phys. Rev. Lett.}, 104:063002, 2010.

\bibitem{Campo2012}
A.~del Campo and M.~G. Boshier.
\newblock Shortcuts to adiabaticity in a time-dependent box.
\newblock {\em Sci. Rep.}, 2:648, 2012.

\bibitem{Masuda2010}
S.~Masuda and K.~Nakamura.
\newblock Fast-forward of adiabatic dynamics in quantum mechanics.
\newblock {\em Proc. R. Soc. A}, 466:1135, 2010.

\bibitem{Masuda2011}
S.~Masuda and K.~Nakamura.
\newblock Acceleration of adiabatic quantum dynamics in electromagnetic fields.
\newblock {\em Phys. Rev. A}, 84:043434, 2011.

\bibitem{Torrontegui2012}
E.~Torrontegui, S.~Mart\'{\i}nez-Garaot, A.~Ruschhaupt, and J.~G. Muga.
\newblock Shortcuts to adiabaticity: Fast-forward approach.
\newblock {\em Phys. Rev. A}, 86:013601, 2012.

\bibitem{Torrontegui2012a}
E.~Torrontegui, X.~Chen, M.~Modugno, S.~Schmidt, A.~Ruschhaupt, and J.~G. Muga.
\newblock {Fast transport of Bose–-Einstein condensates}.
\newblock {\em New J. Phys.}, 14:013031, 2012.

\bibitem{Masuda2014a}
S.~Masuda, K.~Nakamura, and A.~del Campo.
\newblock High-fidelity rapid ground-state loading of an ultracold gas into an
  optical lattice.
\newblock {\em Phys. Rev. Lett.}, 113:063003, 2014.

\bibitem{Kiely2015}
A.~Kiely, J.~P.~L. McGuinness, J.~G. Muga, and A.~Ruschhaupt.
\newblock Fast and stable manipulation of a charged particle in a penning trap.
\newblock {\em J. Phys. B: At. Mol. Opt. Phys.}, 48:075503, 2015.

\bibitem{Deffner2016}
S.~Deffner.
\newblock Shortcuts to adiabaticity: suppression of pair production in driven
  {Dirac} dynamics.
\newblock {\em New J. Phys.}, 18:012001, 2016.

\bibitem{Deffner2014}
S.~Deffner, C.~Jarzynski, and A.~del Campo.
\newblock {Classical and quantum shortcuts to adiabaticity for scale-invariant
  driving}.
\newblock {\em Phys. Rev. X}, 4:021013, 2014.

\bibitem{Jarzysnki2013}
C.~{Jarzynski}.
\newblock {Generating shortcuts to adiabaticity in quantum and classical
  dynamics}.
\newblock {\em Phys. Rev. A}, 88:040101, 2013.

\bibitem{Patra2016}
A.~Patra and C.~Jarzynski.
\newblock Classical and quantum shortcuts to adiabaticity in a tilted piston.
\newblock {\em J. Phys. Chem. B}, 121:3403, 2016.

\bibitem{Chen2011a}
X.~Chen, E.~Torrontegui, D.~Stefanatos, J.~Li, and J.~G. Muga.
\newblock Optimal trajectories for efficient atomic transport without final
  excitation.
\newblock {\em Phys. Rev. A}, 84:043415, 2011.

\bibitem{Stefanatos2013}
D.~Stefanatos.
\newblock Optimal shortcuts to adiabaticity for a quantum piston.
\newblock {\em Automatica}, 49:3079, 2013.

\bibitem{Saberi2014}
H.~Saberi, T.~Opatrn\'y, K.~M\o{}lmer, and A.~del Campo.
\newblock Adiabatic tracking of quantum many-body dynamics.
\newblock {\em Phys. Rev. A}, 90:060301, 2014.

\bibitem{PolkovnikovArXiv}
D.~{Sels} and A.~{Polkovnikov}.
\newblock {Minimizing irreversible losses in quantum systems by local
  counter-diabatic driving}.
\newblock {\em PNAS}, 114:20, 2017.

\bibitem{Xiao2014}
G.~Xiao and J.~Gong.
\newblock {Suppression of work fluctuations by optimal control: An approach
  based on Jarzynski's equality}.
\newblock {\em Phys. Rev. E}, 90:052132, 2014.

\bibitem{Masuda2014}
S.~{Masuda} and S.~A. {Rice}.
\newblock {Fast-Forward Assisted STIRAP}.
\newblock {\em J. Phys. Chem. A}, 119:3479, 2015.

\bibitem{Torrontegui2014}
E.~Torrontegui, S.~Mart\'{\i}nez-Garaot, and J.~G. Muga.
\newblock Hamiltonian engineering via invariants and dynamical algebra.
\newblock {\em Phys. Rev. A}, 89:043408, 2014.

\bibitem{Acconcia2015}
T.~V. Acconcia, M.~V.~S. Bonan\ifmmode~\mbox{\c{c}}\else \c{c}\fi{}a, and
  S.~Deffner.
\newblock Shortcuts to adiabaticity from linear response theory.
\newblock {\em Phys. Rev. E}, 92:042148, 2015.

\bibitem{Garaot2015}
S.~Mart\'{\i}nez-Garaot, A.~Ruschhaupt, J.~Gillet, T.~Busch, and J.~G. Muga.
\newblock Fast quasiadiabatic dynamics.
\newblock {\em Phys. Rev. A}, 92:043406, 2015.

\bibitem{Kiely2016}
A.~Kiely, A.~Benseny, T.~Busch, and A.~Ruschhaupt.
\newblock Shaken not stirred: creating exotic angular momentum states by
  shaking an optical lattice.
\newblock {\em J. Phys B: At., Mol. Opt. Phys.}, 49:215003, 2016.

\bibitem{Chen2010PRL}
X.~Chen, I.~Lizuain, A.~Ruschhaupt, D.~Gu\'ery-Odelin, and J.~G. Muga.
\newblock Shortcut to adiabatic passage in two- and three-level atoms.
\newblock {\em Phys. Rev. Lett.}, 105:123003, 2010.

\bibitem{Santos2016}
A.~C. Santos, R.~D. Silva, and M.~S. Sarandy.
\newblock Shortcut to adiabatic gate teleportation.
\newblock {\em Phys. Rev. A}, 93:012311, 2016.

\bibitem{Coulamy2016}
I.~B. {Coulamy}, A.~C. {Santos}, I.~{Hen}, and M.~S. {Sarandy}.
\newblock {Energetic cost of superadiabatic quantum computation}.
\newblock {\em Front. ICT}, 3:19, 2016.

\bibitem{Funo2017}
K.~Funo, J.-N. Zhang, C.~Chatou, K.~Kim, M.~Ueda, and A.~del Campo.
\newblock Universal work fluctuations during shortcuts to adiabaticity by
  counterdiabatic driving.
\newblock {\em Phys. Rev. Lett.}, 118:100602, 2017.

\bibitem{Peskin1995}
M.~E. Peskin and D.~V. Schroeder.
\newblock {\em An Introduction To Quantum Field Theory (Frontiers in Physics)}.
\newblock Westview Press, Boulder, CO, USA, 1995.

\bibitem{Thaller1956}
B.~Thaller.
\newblock {\em {The Dirac equation}}.
\newblock Springer, Berlin, Germany, 1956.

\bibitem{Dirac1928}
P.~A.~M Dirac.
\newblock The quantum theory of the electron.
\newblock {\em Proc. R. Soc. A}, 117:778, 1928.

\bibitem{Pickl2008}
P.~Pickl and D.~D\"urr.
\newblock On adiabatic pair creation.
\newblock {\em Commun. Math. Phys.}, 282:161, 2008.

\bibitem{Fillion2012}
F.~Fillion-Gourdeau, E.~Lorin, and A.~D. Bandrauk.
\newblock {L}andau-{Z}ener-{S}t\"uckelberg interferometry in pair production
  from counterpropagating lasers.
\newblock {\em Phys. Rev. A}, 86:032118, 2012.

\bibitem{Fillion2013}
F.~Fillion-Gourdeau, E.~Lorin, and A.~D. Bandrauk.
\newblock Resonantly enhanced pair production in a simple diatomic model.
\newblock {\em Phys. Rev. Lett.}, 110:013002, 2013.

\bibitem{Fillion2013a}
F.~Fillion-Gourdeau, E.~Lorin, and A.~D. Bandrauk.
\newblock {Enhanced Schwinger pair production in many-centre systems}.
\newblock {\em J. Phys. B: At. Mol. Opt. Phys.}, 46:175002, 2013.

\bibitem{Schmidt2015}
M.~Schmidt, V.~Peano, and F.~Marquardt.
\newblock {Optomechanical Dirac physics}.
\newblock {\em New J. Phys.}, 17:023025, 2015.

\bibitem{Deffner2015}
S.~Deffner and A.~Saxena.
\newblock {Quantum work statistics of charged {D}irac particles in
  time-dependent fields}.
\newblock {\em Phys. Rev. E}, 92:032137, 2015.

\bibitem{Wehling2014}
T.~O. {Wehling}, A.~M. {Black-Schaffer}, and A.~V. {Balatsky}.
\newblock {Dirac materials}.
\newblock {\em Adv. Phys.}, 63:1, 2014.

\bibitem{Fillion2015}
F.~Fillion-Gourdeau and S.~MacLean.
\newblock {Time-dependent pair creation and the Schwinger mechanism in
  graphene}.
\newblock {\em Phys. Rev. B}, 92:035401, 2015.

\bibitem{Villamizar2015}
D.~V. Villamizar and E.~I. Duzzioni.
\newblock Quantum speed limit for a relativistic electron in a uniform magnetic
  field.
\newblock {\em Phys. Rev. A}, 92:042106, 2015.

\bibitem{Tian2015}
Z.~Tian, J.~Wang, H.~Fan, and J.~Jing.
\newblock Relativistic quantum metrology in open system dynamics.
\newblock {\em Sci. Rep.}, 5:7946, 2015.

\bibitem{Khan2015}
S.~Khan and N.~A. Khan.
\newblock Relativistic quantum speed limit time in dephasing noise.
\newblock {\em Eur. Phys. J. Plus}, 130:216, 2015.

\bibitem{Wang2017}
K.~{Wang}, Y.~F. {Zhang}, Q.~{Wang}, Z.~W. {Long}, and J.~{Jing}.
\newblock {Quantum speed limit for relativistic spin-0 and spin-1 bosons on
  commutative and noncommutative planes}.
\newblock {\em arXiv:1703.01063}, 2017.

\bibitem{Wang2017a}
K.~{Wang}, Y.~F. {Zhang}, Q.~{Wang}, Z.~W. {Long}, and J.~{Jing}.
\newblock {Quantum speed limit for a relativistic electron in the
  noncommutative phase space}.
\newblock {\em arXiv:1702.03167}, 2017.

\bibitem{Mostafazadeh2010}
A.~Mostafazadeh.
\newblock Pseudo-hermitian representation of quantum mechanics.
\newblock {\em Int. J. Geo. Meth. Mod. Phys.}, 7:1191, 2010.

\bibitem{Berry2011JO}
M.~V. Berry.
\newblock Optical polarization evolution near a non-hermitian degeneracy.
\newblock {\em J. Opt.}, 13:115701, 2011.

\bibitem{Moiseyev2011}
N.~Moiseyev.
\newblock {\em Non-Hermitian quantum mechanics}.
\newblock Cambridge University Press, 2011.

\bibitem{Gardas2016Coriolis}
B.~Gardas, S.~Deffner, and A.~Saxena.
\newblock Repeatability of measurements: Non-hermitian observables and quantum
  coriolis force.
\newblock {\em Phys. Rev. A}, 94:022121, 2016.

\bibitem{Gardas2016thermo}
B.~Gardas, S.~Deffner, and A.~Saxena.
\newblock Non-hermitian quantum thermodynamics.
\newblock {\em Sci. Rep.}, 6:23408, 2016.

\bibitem{Bender1998}
C.~M. Bender and S.~Boettcher.
\newblock Real spectra in non-hermitian hamiltonians having
  $\mathcal{P}\mathcal{T}$ symmetry.
\newblock {\em Phys. Rev. Lett.}, 80:5243, 1998.

\bibitem{Bender2007}
C.~M. Bender.
\newblock {Making sense of non-Hermitian Hamiltonians}.
\newblock {\em Reports Prog. Phys.}, 70:947--1018, 2007.

\bibitem{Deffner2015PRL}
S.~Deffner and A.~Saxena.
\newblock Jarzynski equality in $\mathcal{P}\mathcal{T}$-symmetric quantum
  mechanics.
\newblock {\em Phys. Rev. Lett.}, 114:150601, 2015.

\bibitem{Bender2017}
C.~M. Bender, D.~C. Brody, and M.~P. M\"uller.
\newblock Hamiltonian for the zeros of the riemann zeta function.
\newblock {\em Phys. Rev. Lett.}, 118:130201, 2017.

\bibitem{Bender2013}
C.~M. Bender, M.~Dekieviet, and S.~P. Klevansky.
\newblock $\mc{PT}$-quantum mechanics.
\newblock {\em Philos. Trans. R. Soc. A Math. Phys. Eng. Sci.}, 371:20120523,
  2013.

\bibitem{Bender2003}
C.~M. Bender, D.~C. Brody, and H.~F. Jones.
\newblock {Must a Hamiltonian be Hermitian?}
\newblock {\em Am. J. Phys.}, 71:1095, 2003.

\bibitem{Rushhaupt2005}
A.~Ruschhaupt, F.~Delgado, and J.~G. Muga.
\newblock Physical realization of {{\cal{PT}}} -symmetric potential scattering
  in a planar slab waveguide.
\newblock {\em J. Phys. A: Math. Gen.}, 38:L171, 2005.

\bibitem{Klaiman2008}
S.~Klaiman, U.~G\"unther, and N.~Moiseyev.
\newblock Visualization of branch points in $\mathcal{P}\mathcal{T}$-symmetric
  waveguides.
\newblock {\em Phys. Rev. Lett.}, 101:080402, 2008.

\bibitem{Makris2008}
K.~G. Makris, R.~El-Ganainy, D.~N. Christodoulides, and Z.~H. Musslimani.
\newblock Beam dynamics in $\mathcal{P}\mathcal{T}$ symmetric optical lattices.
\newblock {\em Phys. Rev. Lett.}, 100:103904, 2008.

\bibitem{Musslimani2008}
Z.~H. Musslimani, K.~G. Makris, R.~El-Ganainy, and D.~N. Christodoulides.
\newblock Optical solitons in $\mathcal{P}\mathcal{T}$ periodic potentials.
\newblock {\em Phys. Rev. Lett.}, 100:030402, 2008.

\bibitem{Ruter2010}
C.~E. R\"{u}ter, K.~G. Makris, R.~El-Ganainy, D.~N. Christodoulides, M.~Segev,
  and D.~Kip.
\newblock {Observation of parity–time symmetry in optics}.
\newblock {\em Nat. Phys.}, 6:192, 2010.

\bibitem{Bender2002}
C.~M. Bender, D.~C. Brody, and H.~F. Jones.
\newblock Complex extension of quantum mechanics.
\newblock {\em Phys. Rev. Lett.}, 89:270401, 2002.

\bibitem{Bender2007a}
C.~M. Bender, D.~C. Brody, H.~F Jones, and B.~K Meister.
\newblock Faster than hermitian quantum mechanics.
\newblock {\em Phys. Rev. Lett.}, 98:040403, 2007.

\bibitem{Uzdin2012}
R.~Uzdin, U.~G{\"u}nther, S.~Rahav, and N.~Moiseyev.
\newblock Time-dependent hamiltonians with 100\% evolution speed efficiency.
\newblock {\em J. Phys. A}, 45:415304, 2012.

\bibitem{Frydryszak2008}
A.~M. Frydryszak and V.~M. Tkachuk.
\newblock Quantum brachistochrone problem for a spin-1 system in a magnetic
  field.
\newblock {\em Phys. Rev. A}, 77:014103, 2008.

\bibitem{Borras2008}
A.~Borras, C.~Zander, A.~R. Plastino, M.~Casas, and A.~Plastino.
\newblock Entanglement and the quantum brachistochrone problem.
\newblock {\em EPL (Europhysics Letters)}, 81:30007, 2008.

\bibitem{Kuzmak2013}
A.~R. Kuzmak and V.~M. Tkachuk.
\newblock The quantum brachistochrone problem for two spins-{$\frac{1}{2}$}
  with anisotropic heisenberg interaction.
\newblock {\em J. Phys. A: Math. Theor.}, 46:155305, 2013.

\bibitem{Kuzmak2015}
A.~R. Kuzmak and V.~M. Tkachuk.
\newblock The quantum brachistochrone problem for an arbitrary spin in a
  magnetic field.
\newblock {\em Phys. Lett. A}, 379:1233, 2015.

\bibitem{Russel2014}
B.~Russell and S.~Stepney.
\newblock Zermelo navigation and a speed limit to quantum information
  processing.
\newblock {\em Phys. Rev. A}, 90:012303, 2014.

\bibitem{Russel2015}
B.~Russell and S.~Stepney.
\newblock Zermelo navigation in the quantum brachistochrone.
\newblock {\em J. Phys. A: Math. Theo.}, 48:115303, 2015.

\bibitem{Villamizar2017}
D.~V. {Villamizar}, A.~C.~S. {Leal}, R.~{Auccaise}, and E.~I. {Duzzioni}.
\newblock {Estimating the time evolution of NMR systems via quantum speed
  limit}.
\newblock {\em arXiv:1705.06137}, 2017.

\bibitem{Mostafazadeh2007}
A.~Mostafazadeh.
\newblock Quantum brachistochrone problem and the geometry of the state space
  in pseudo-hermitian quantum mechanics.
\newblock {\em Phys. Rev. Lett.}, 99:130502, 2007.

\bibitem{Assis2008}
P.~E.~G. Assis and A.~Fring.
\newblock The quantum brachistochrone problem for non-hermitian hamiltonians.
\newblock {\em J. Phys. A: Math. Theor.}, 41:244002, 2008.

\bibitem{Giri2008}
P.~R. Giri.
\newblock Lower bound of minimal time evolution in quantum mechanics.
\newblock {\em Int. J. Theo. Phys.}, 47:2095, 2008.

\bibitem{Gunther2008}
U.~G\"unther and B.~F. Samsonov.
\newblock $\mathcal{P}\mathcal{T}$-symmetric brachistochrone problem, {Lorentz}
  boosts, and nonunitary operator equivalence classes.
\newblock {\em Phys. Rev. A}, 78:042115, 2008.

\bibitem{Gunther2008PRL}
U.~G\"unther and B.~F. Samsonov.
\newblock Naimark-dilated $\mathcal{P}\mathcal{T}$-symmetric brachistochrone.
\newblock {\em Phys. Rev. Lett.}, 101:230404, 2008.

\bibitem{Mostafazadeh2009}
A.~Mostafazadeh.
\newblock Hamiltonians generating optimal-speed evolutions.
\newblock {\em Phys. Rev. A}, 79:014101, 2009.

\bibitem{Bender2009}
C.~M. Bender and D.~C. Brody.
\newblock {\em Optimal Time Evolution for Hermitian and Non-Hermitian
  Hamiltonians}, pages 341--361.
\newblock Springer, Berlin, Heidelberg, 2009.

\bibitem{Huang2009}
K.~Huang.
\newblock {\em Introduction to statistical physics}.
\newblock CRC Press, 2009.

\bibitem{Shen1984}
Y.-R. Shen.
\newblock The principles of nonlinear optics.
\newblock {\em New York, Wiley-Interscience, 1984, 575 p.}, 1, 1984.

\bibitem{Xi2016}
X.~Chen, Y.~Ban, and G.~C. Hegerfeldt.
\newblock Time-optimal quantum control of nonlinear two-level systems.
\newblock {\em Phys. Rev. A}, 94:023624, 2016.

\bibitem{Dou2014}
F.~Q. Dou, L.~B. Fu, and J.~Liu.
\newblock High-fidelity fast quantum driving in nonlinear systems.
\newblock {\em Phys. Rev. A}, 89:012123, 2014.

\bibitem{Giovannetti2012}
V.~Giovannetti, S.~Lloyd, and L.~Maccone.
\newblock Quantum measurement bounds beyond the uncertainty relations.
\newblock {\em Phys. Rev. Lett.}, 108:260405, 2012.

\bibitem{Landauer1961}
R.~Landauer.
\newblock Irreversibility and heat generation in the computing process.
\newblock {\em IBM J. Res. Dev.}, 5:183, 1961.

\bibitem{Deffner2013PRX}
S.~Deffner and C.~Jarzynski.
\newblock Information processing and the second law of thermodynamics: An
  inclusive, hamiltonian approach.
\newblock {\em Phys. Rev. X}, 3:041003, 2013.

\bibitem{Berut2012}
A.~Berut, A.~Arakelyan, A.~Petrosyan, S.~Ciliberto, R.~Dillenschneider, and
  E.~Lutz.
\newblock Experimental verification of landauer's principle linking information
  and thermodynamics.
\newblock {\em Nature}, 483:187, 2012.

\bibitem{Jun2014}
Y.~Jun, M.~Gavrilov, and J.~Bechhoefer.
\newblock High-precision test of {Landauer}'s principle in a feedback trap.
\newblock {\em Phys. Rev. Lett.}, 113:190601, 2014.

\bibitem{PetersonPRSA}
J.~P.~S. Peterson, R.~S. Sarthour, A.~M. Souza, I.~S. Oliveira, J.~Goold,
  K.~Modi, D.~O. Soares-Pinto, and L.~C. C{\'e}leri.
\newblock Experimental demonstration of information to energy conversion in a
  quantum system at the landauer limit.
\newblock {\em Proc. Royal Soc. A}, 472(2188), 2016.

\bibitem{Ciampini2017}
M.~A. Ciampini, L.~Mancino, A.~Orieux, C.~Vigliar, P.~Mataloni, M.~Paternostro,
  and M.~Barbieri.
\newblock Experimental extractable work-based multipartite separability
  criteria.
\newblock {\em NPJ Quantum Information}, 3:10, 2017.

\bibitem{Gaudenzi2017}
R.~{Gaudenzi}, E.~{Burzur{\'{\i}}}, S.~{Maegawa}, H.~S.~J. {van der Zant}, and
  F.~{Luis}.
\newblock {Quantum-enhanced Landauer erasure and storage}.
\newblock {\em arXiv:1703.04607}, 2017.

\bibitem{Kliesch2014}
M.~Kliesch, C.~Gogolin, and J.~Eisert.
\newblock {\em Many-Electron Approaches in Physics, Chemistry and Mathematics:
  Lieb-Robinson bounds and the simulation of time evolution of local
  observables in lattice systems}.
\newblock Springer, 2014.

\bibitem{Cheneau2012}
M.~{Cheneau}, P.~{Barmettler}, D.~{Poletti}, M.~{Endres}, P.~{Schau{\ss}},
  T.~{Fukuhara}, C.~{Gross}, I.~{Bloch}, C.~{Kollath}, and S.~{Kuhr}.
\newblock {Light-cone-like spreading of correlations in a quantum many-body
  system}.
\newblock {\em Nature}, 481:484, 2012.

\bibitem{Groenewold1971}
H.~J. Groenewold.
\newblock A problem of information gain by quantal measurements.
\newblock {\em Int. J. Theo. Phys.}, 4:327, 1971.

\bibitem{Nielsen2002}
M.~A. Nielsen and I.~Chuang.
\newblock {\em Quantum computation and quantum information}.
\newblock AAPT, 2002.

\bibitem{Acconcia2017}
T.~V. Acconcia and S.~Deffner.
\newblock Quantum speed limits and the maximal rate of quantum learning.
\newblock {\em arXiv:1706.03826}, 2017.

\bibitem{Lockwood1996}
M.~Lockwood.
\newblock {'Many Minds': Interpretations of Quantum Mechanics}.
\newblock {\em BJPS}, 47:159, 1996.

\end{thebibliography}

\end{document}